\documentclass{an}

\usepackage{times}
\usepackage{graphicx}
\graphicspath{{./figs/}{./png/}}
\usepackage{url}
\newcommand{\EQ}{\begin{equation}}
\newcommand{\EN}{\end{equation}}

\newcommand{\Eq}[1]{Eq.~(\ref{#1})}
\newcommand{\rr}{{\vec{r}}}
\newcommand{\uu}{{\vec{u}}}
\newcommand{\BB}{{\vec{B}}}
\newcommand{\JJ}{{\vec{J}}}

\newcommand{\FF}{{\vec{F}}}
\newcommand{\AAA}{{\vec{A}}}
\newcommand{\SSS}{{\vec{S}}}
\newcommand{\ttau}{{\vec{\tau}}}

\newcommand{\dd}{{\rm d} {}}
\newcommand{\DD}{{\rm D} {}}
\newcommand{\kms}{\,{\rm km/s}}
\newcommand{\km}{\,{\rm km}}
\newcommand{\s}{\,{\rm s}}
\newcommand{\K}{\,{\rm K}}
\newcommand{\G}{\,{\rm G}}
\newcommand{\kG}{\,{\rm kG}}
\newcommand{\AU}{\,{\rm AU}}
\newcommand{\g}{\,{\rm g}}
\newcommand{\cm}{\,{\rm cm}}
\newcommand{\yr}{\,{\rm yr}}

\newcommand{\erg}{\,{\rm erg}}
\begin{document}

%%%\thesaurus{ 06         % Section 6 -- Formation, structure and evolution
%%%                       % of stars
%%%           (09.10.1;  % ISM: jets and outflows
%%%            02.01.2;  % Accretion, accretion disks
%%%            02.13.1;  % Magnetic fields
%%%            02.13.2)} % Magnetohydrodynamics (MHD)

\titlerunning{Stellar dynamo driven wind braking versus disc coupling}
\title{Stellar dynamo driven wind braking versus disc coupling}

\author{
  B.~von~Rekowski\inst{1,2}
  \and A.~Brandenburg\inst{3}
}

\institute{
  Department of Astronomy \& Space Physics,
  Uppsala University, Box 515, 751 20 Uppsala, Sweden
  \and School of Mathematics and Statistics, University of St.\ Andrews,
  North Haugh, St.\ Andrews, Fife KY16 9SS, UK
  \and NORDITA, Blegdamsvej 17, DK-2100 Copenhagen \O, Denmark
}

%%%\offprints{brigitta@mcs.st-and.ac.uk}

\date{Received; accepted; published online}
%\date{\today,~ $ $Revision: 1.120 $ $}

%-------------------------------------------------------------------------------

\abstract{
Star-disc coupling is considered in numerical models where the stellar field is
not an imposed perfect dipole, but instead a more irregular self-adjusting
dynamo-generated field.
Using axisymmetric simulations of the hydromagnetic mean-field equations,
it is shown that the resulting stellar field configuration is more complex,
but significantly better suited for driving a stellar wind.
In agreement with recent findings by a number of people,
star-disc coupling is less efficient in braking the star
than previously thought. Moreover, stellar wind braking becomes
equally important.
In contrast to a perfect stellar dipole field, 
dynamo-generated stellar fields favor field-aligned accretion
with considerably higher velocity at low latitudes,
where the field is weaker and originating in the disc.
Accretion is no longer nearly periodic (as it is in the case of a
stellar dipole), but it is more irregular and episodic.
}
\correspondence{brigitta@mcs.st-and.ac.uk}
\maketitle

%-------------------------------------------------------------------------------
\section{Introduction}

Over the past decade, numerical simulations have been used to address
the problem of star--disc coupling by a stellar dipole magnetosphere
(Hayashi et al.\ 1996,
Hirose et al.\ 1997, Miller \& Stone 1997, Goodson \& Winglee 1999,
von Rekowski \& Brandenburg 2004, hereafter referred to as vRB04,
Romanova et al.\ 2004).
One of the perhaps most striking differences compared with the standard
pictures of star--disc coupling (K\"onigl 1991, Cameron \& Campbell 1993,
Shu et al.\ 1994, Hartmann et al.\ 1994) is the realization that the stellar
magnetospheric field may not
permeate the surrounding disc over broad enough a range in radius
to ensure sufficient coupling between the two.
Instead, differential rotation between disc and star leads to magnetic helicity
injection that in turn inflates the closed magnetic loops which then open up
close to the inner
disc edge (Lovelace et al.\ 1995,
Agapitou \& Papaloizou 2000, Uzdensky et al.\ 2002).
This led Matt \& Pudritz (2004, 2005) to the conclusion that the star
cannot be braked by the magnetic interaction with the disc, 
as was assumed so far.
Instead, stellar winds may be the more important agent causing stellar
spin-down.

One of the important new features used in the simulations of vRB04
is the fact that the disc itself can produce a magnetic field.
This feature was introduced by von Rekowski et al.\ (2003) to study the
possibility of collimated outflows in the absence of an ambient magnetic
field.
This idea seems now quite reasonable also from an observational point of view.
M\'enard \& Duch\^ene (2004) report observations of the Taurus-Auriga
molecular cloud which is a nearby star-forming region of low-mass stars.
They find that jets and outflows from protostellar star--disc systems
in this cloud are not always aligned with the local magnetic field
of the cloud where they are born. This suggests that jets are not likely
to be driven by the cloud's own field.
However, theoretical models have shown that magnetic fields are important
for jet acceleration and collimation.
Further, there is now also observational evidence that jets and outflows
are essentially hydromagnetic in nature. Observations of CO outflows exclude
purely
radiative and thermal driving of jets and outflows from low-mass protostellar
star--disc systems (Pudritz 2004).
Thermal pressure might still be dynamically important to lift up the winds.
Moreover, there are observational indications of warm wind regions
with temperatures of a few times $10^4\K$ (G\'omez de Castro \& Verdugo 2003).
These winds cannot be due to stellar radiation because the star is too cool,
suggesting that the warm winds are magnetohydrodynamically (MHD) driven.
So it is quite plausible that a disc dynamo might be
necessary for the launching of jets and outflows.

Also the protostar's own magnetic field could contribute to the launching of
jets and outflows. From observations it is not clear whether the jets and
outflows originate from the central star or from the surface of the inner part
of the circumstellar disc.
In the present paper we extend the simulations of von Rekowski et al.\ (2003)
and vRB04 by allowing the
stellar magnetic field to be self-generated by its own dynamo and to
react to its environment.
One of the immediate consequences of a stellar dynamo
is that the stellar field is
no longer an ideal dipole, but it has a range of higher multipoles.
Indeed, observations of solar-type low-mass protostars surrounded by a disc
(classical T Tauri stars, hereafter CTTS) suggest kilogauss fields that are
rather irregular and in general nonaxisymmetric
(Johns-Krull et al.\ 1999a,b, Johns-Krull \& Valenti 2001).
Valenti \& Johns-Krull (2004) made measurements of the magnetic field strength
and geometry at the surface of CTTS. They find a mean surface field strength
of about $1\kG$ to $3\kG$ and a top surface field strength of as large as
about $6\kG$. Their measurements indicate that the standard magnetospheric
accretion model (where a stellar dipole is threading the disc leading to a
funnel-shaped polar accretion flow along the magnetospheric field lines) seems
far too simple and idealized. Their observations show that not a fixed fraction
of the stellar surface is coupled to the disc by a magnetic field, and that
the magnetic star--disc coupling is also not periodic in time,
both opposed to the standard magnetospheric accretion models.
Accretion and
the stellar magnetosphere are rather irregular and nonaxisymmetric, and they
rule out a global dipolar field at the stellar surface of at least some CTTS.
Although CTTS seem to have complex magnetic fields that are not
simple dipoles, the stellar fields still seem to
have a very large scale component in and around the surface of at least a few
hundred gauss.

Observations (e.g. with the Hubble Space Telescope) show that outflows
from protostellar star--disc systems can have different shapes. They can be
weakly collimated gas bubbles or well-collimated jets extending far
into space.
Outflow speeds range between 50 and 500 $\kms$. Interestingly,
high-resolution observations with the VLT show that also the accretion flow
velocity is rather high, of the order of $100\kms$ (Stempels \& Piskunov
2002). Protostellar systems are very dynamic, with changes in the disc
brightness, and knots and gaps in the jet. Stempels \& Piskunov (2002)
find that the time variability related to accretion and outflow processes
occurs on time scales ranging from years to minutes.

In this paper, we investigate the effects of a stellar dynamo on star--disc
coupling and outflows. The accretion flow geometry is not known
from observations other than it seems to be complex and highly time-dependent.
Our stellar dynamo models suggest irregular and loose
magnetic star--disc coupling by the {\it disc} dynamo-generated
field at low latitudes.
Accretion is episodic but fast, and along field lines in those
equatorial regions, where the field is very weak.

We further suggest that the observed jets and outflows
could originate from the star and/or from the inner disc.
In the presence of stellar and disc dynamos, stellar and disc wind
mass loss rates are comparable, and stellar wind speeds even exceed
disc wind speeds.
The fast stellar wind significantly contributes in braking the star.

%%%%%%
%!\begin{figure}[t!]
%!%GB: fig created with ....pro
%!\centerline{
%!   \includegraphics[width=8.5cm]{....ps}
%!}
%!\caption[]{
%!...
%!}\label{...}
%!\end{figure}
%%%%%%

%-------------------------------------------------------------------------------
\section{The models} \label{TDM}

A detailed description of the models without stellar dynamo and of the models
with an anchored stellar dipole magnetosphere can be found
in Sect.~2 of von Rekowski et al.\ (2003) and vRB04, respectively.
We begin the description of the present models by reviewing the basic setup
of the earlier models.
The implementation of the stellar dynamo is described in Sect.~\ref{ADDSD}.

\subsection{Basic equations} \label{BE}

We solve the set of MHD equations
using cylindrical polar coordinates $(\varpi,\varphi,z)$.
In all our models, we solve the continuity equation,
\begin{equation}
   {\partial\varrho\over\partial t}+\vec{\nabla}\cdot(\varrho\uu)
   =q_\varrho^{\rm disc}+q_\varrho^{\rm star},
\label{Cont}
\end{equation}
the Navier--Stokes equation,
\begin{equation}
   \frac{\DD\uu}{\DD t}
   = - {1\over\varrho}\vec{\nabla}p - \vec{\nabla}\Phi
     + \frac{1}{\varrho}\left[\vec{F} + (\vec{\uu}_{\rm Kep}-\vec{\uu})
       q_\varrho^{\rm disc}\right],
\label{Momentum}
\end{equation}
and the mean-field induction equation,
\begin{equation}
   {\partial\AAA\over\partial t}=\uu\times\BB+\alpha\BB-\eta\mu_0\JJ.
\label{Ind}
\end{equation}
Here, $\uu$ is the velocity field, $\varrho$ is the gas density,
$p$ is the gas pressure, $\Phi$ is the gravitational potential,
$t$ is time, ${\DD}/{\DD}t=\partial/\partial t+\uu\cdot\vec{\nabla}$
is the advective derivative,
$\FF=\JJ\times\BB+\vec{\nabla}\cdot\ttau_{\rm visc}$ is the sum of the Lorentz
and viscous forces, $\JJ=\vec{\nabla}\times\BB/\mu_0$ is the current
density due to the mean magnetic field $\BB$, $\mu_0$ is the magnetic
permeability, and $\ttau_{\rm visc}$ is the viscous stress tensor.
To ensure that $\BB$ is solenoidal, we solve the induction equation in terms of
the magnetic vector potential $\AAA$, where $\BB=\vec{\nabla}\times\AAA$.

In the disc, we assume a turbulent (Shakura--Sunyaev) kinematic viscosity,
$\nu_{\rm t}=\alpha_{\rm SS}c_{\rm s}z_0$,
where $\alpha_{\rm SS}$ is the Shakura--Sunyaev coefficient (less than unity),
$c_{\rm s}=(\gamma p/\varrho)^{1/2}$ is the sound speed, $\gamma = c_p/c_v$
is the ratio of the specific heat at constant pressure, $c_p$, and
the specific heat at constant volume, $c_v$,
and $z_0$ is the disc half-thickness.

The mass in the star and the disc that is explicitly accounted for in the
simulations, cannot include the contributions from the most dense inner
parts.
For this reason we have, as described in vRB04, modeled this by a
self-regulatory mass source in the disc and a self-regulatory mass sink in the
star,
$q_\varrho^{\rm disc}$ and $q_\varrho^{\rm star}$, respectively.
Matter is injected with Keplerian speed $\vec{\uu}_{\rm Kep}$ in the
disc and removed from the central parts of the star without affecting
its angular velocity.

Star, disc, and corona are represented as regions with three different
polytropes, but the same polytropic index.
The method has been described in detail by von Rekowski et al.\ (2003)
and vRB04, and is summarized in Appendix~\ref{Sec-cool-hot}.
Axisymmetry is assumed throughout this paper.

\subsection{Disc dynamo and stellar dynamo} \label{ADDSD}

To date, almost all models of the formation and collimation
of winds and jets from protostellar accretion discs rely on externally
imposed poloidal magnetic fields and ignore any dynamo-generated poloidal
field produced in the disc or in the star.
This is not the case in the models developed in our earlier papers, which also
form the basis of the models used in the present paper.
In von Rekowski et al.\ (2003) and vRB04, we have studied outflows and accretion
in connection with magnetic fields that are
produced by a disc dynamo,
resolving the disc and the star at the same time as well as assuming a non-ideal
corona.
The seed magnetic field is large scale, poloidal,
of mixed parity, weak and confined to the disc.
The mixed-parity choice allows for the symmetry to be violated also in the
(statistically) saturated magnetic fields.

Our present models are similar to those of von Rekowski et al.\ (2003) and
vRB04, except for the
modeling of the stellar magnetic field.
In von Rekowski et al.\ (2003), the magnetic field in the star--disc system is
solely generated and maintained by the disc dynamo so that the stellar field is
produced entirely by advection of the disc field. In vRB04, we also model
-- in addition to the disc dynamo -- a stellar magnetosphere that is initially
a dipole threading the disc but can freely evolve with time outside the
anchoring region starting at about 1.5 stellar radii. In the following we
usually refer to this type of model as ``stellar dipole'' model (or similar)
although it is important to note that the dipole is imposed only in the
anchoring region (which includes the star and extends up to 1.5 stellar radii).

In this paper we assume in our main models that both the disc field and
the stellar field be generated
by disc and stellar dynamos, respectively.
These dynamos are considered to be the only mechanism for the generation and
maintenance of the entire magnetic field in our system.
We study the interaction of both stellar and disc fields, the resulting
field configuration as well as the associated outflow and accretion processes.
We assume that the magnetic fields in the disc and in the star be generated
by standard $\alpha^2\Omega$ dynamos (e.g.,\ Krause \& R\"adler 1980),
where $\alpha$ is the mean-field $\alpha$ effect and
$\Omega$ is the angular velocity of the plasma.
This implies an extra electromotive force, $\alpha\BB$, in the induction
equation for the mean magnetic field, $\BB$, that is restricted to the disc
and to the star.
As usual, the $\alpha$ effect is antisymmetric about the midplane.
We take
\begin{equation}
\alpha={z\over z_0}\,
{\alpha_0^{\rm disc}\,\xi_{\rm disc}(\varpi,z)
\over1+v_{\rm A}^2/(c_{\rm s}^2{\rm Ma}^2_{\rm disc})}
+ {z\over r_*}\,
{\alpha_0^{\rm star}\,\xi_{\rm star}(\varpi,z)
\over1+v_{\rm A}^2/(c_{\rm s}^2{\rm Ma}^2_{\rm star})},     \label{alpha}
\end{equation}
where $v_{\rm A}$ is the Alfv\'en speed based on the total magnetic field
(i.e.\ $v_{\rm A} = |B|/\sqrt{\mu_0 \varrho}$),
$\xi_{\rm disc}$ and $\xi_{\rm star}$ are time-independent profiles specifying
the shapes of the disc and the star, respectively
(see Appendix~\ref{Sec-cool-hot}), $r_*$ is the stellar radius, and
$\alpha_0^{\rm disc}$ and $\alpha_0^{\rm star}$ are $\alpha$ effect
coefficients that control the intensity of dynamo action, whereas
${\rm Ma}_{\rm disc}$ and ${\rm Ma}_{\rm star}$ are parameters that control
the strength of the saturated magnetic field.

In the disc, we choose the $\alpha$ effect to be negative in the upper half
(i.e.\ $\alpha_0^{\rm disc}<0$),
consistent with results from simulations of accretion disc turbulence driven by
the magneto-rotational instability (Brandenburg et al.\ 1995).
The resulting dynamo-generated magnetic field symmetry is roughly dipolar
because the accretion disc is embedded in a (not perfectly) conducting corona
rather than a vacuum (see vRB04 for more references).
In the star, we choose the $\alpha$ effect to be positive in the upper
hemisphere
(i.e.\ $\alpha_0^{\rm star}>0$),
as is expected if turbulence in the star is due to convection.
Further, we include $\alpha$ quenching which leads to the stellar and
disc dynamos saturating at a level close to equipartition between
magnetic and thermal energies,
depending on ${\rm Ma}_{\rm disc}$ and ${\rm Ma}_{\rm star}$.

Note that by letting the $\alpha$ effect operate in the whole star,
we implicitly assume our protostar to be fully convective which can indeed be the case
for classical T Tauri stars.
Corresponding three-dimensional turbulent
MHD simulations by Dobler et al.\ (2005) have shown that
outside a fully convective star, the stellar dynamo-generated magnetic field
is dominated by the large scale field that has a strong poloidal component
with a significant dipole moment.
They also discuss the effect of having no stably stratified overshoot
layer, that is normally thought important for producing large scale fields
(Durney et al.\ 1993, Hawley et al.\ 2000).
However, buoyant magnetic flux tubes can remain within the convective star due to
the effect of turbulent pumping, as is seen in simulations of stratified
convective dynamos (Nordlund et al.\ 1992, Tobias et al.\ 1998).

We recall that the star is rotating
differentially with respect to both radius and latitude, as a consequence
of the initial setup and determined mainly by the magnetic feed-back
as time evolves.
The star belongs entirely to the computational domain. In our stellar
dynamo models,
no boundary conditions are imposed on the velocity or
magnetic field at the stellar surface. The resulting dynamo effect due to the
differential rotation ($\Omega$ effect) is difficult to quantify in and
around the star.
The magnetic diffusivity $\eta$ is finite in the whole domain,
and dependent on position.

On the outer boundaries of the computational domain,
on $\varpi=\varpi_{\max}$ and $z=\pm z_{\max}$,
we impose outflow conditions (see also Sect.~\ref{NMP}).
On the axis ($\varpi=0$) we impose regularity conditions.

\begin{table*}[t!]
\caption[]{Varying model parameters of the runs discussed in this paper.
Note that Model~N of vRB04 is different from Model~N of the
present paper in that in vRB04 the gap between star and disc is much smaller
and the magnetic diffusivity is as large as in Model~S there. In Model~N,
there is no stellar dynamo or anchored magnetosphere but there is a disc
dynamo. Model~S is very similar
to Model~S of vRB04 (with an anchored stellar magnetosphere).
The difference is
that in this paper the angular velocity profile in the star is not fixed,
whereas in vRB04 it was fixed to rather small values.
(The asterisk in Model~S indicates that
the magnetosphere is anchored up to a radius of about 0.275,
whereas the stellar radius is located at 0.15.)
The main differences
between the models of the runs discussed in this paper are marked in bold
in this table. Model~Ref is the reference model of this paper, with a
stellar dynamo as well as a disc dynamo.
Model~SmS has a smaller star, and Model~SDd has a stronger
disc dynamo, both compared to the reference model. Model~NDd has no
disc dynamo, and Model~ND has no disc at all. In order to produce a setup
without a disc,
the following model parameters are changed (marked with $**$ in the table):
$\beta_{\rm disc}=1$ for the specific entropy,
$\alpha_0^{\rm disc}=0$ and no seed magnetic field in the original disc for the
dynamo, $\alpha_{\rm SS}=0$ and $\eta_{\rm disc,turb}=0$ for the turbulent
diffusivities, and no mass source.
}
\label{tab1}
\vspace{12pt}
\centerline{
\begin{tabular}{ll|c|c|cc|ccc}
Model & Description & $r_*$ &
$\beta_{\rm disc}$ &
$\alpha_0^{\rm disc}$ & ${\rm Ma}_{\rm disc}$ &
$\alpha_0^{\rm star}$ & ${\rm Ma}_{\rm star}$ &
$A_{\rm star}$
\\
\hline
\hline
%Run1a: No stellar dynamo or anchored magnetosphere but disc dynamo
N   & {\bf N}o stellar dynamo, no magnetosphere, but disc dynamo
& .15  & .005 & -.15 &   1 &  0   & $-$ &  0 \\
\hline
%Run1b: Smaller Star compared to reference model
SmS & {\bf Sm}aller {\bf S}tar compared to reference model
& .15  & .005 & -.15 &   1 &{\bf
                             +1.5}&  10 &  0 \\
%
%Run3: Reference model with stellar dynamo and disc dynamo
Ref & {\bf Ref}erence model with stellar and disc dynamos
&{\bf
  .25} & .005 & -.15 &   1 & +1.5 &   5 &  0 \\
%
%strong: Strong anchored magnetosphere and disc dynamo
S   & {\bf S}trong anchored magnetosphere and disc dynamo
& .15$-$
  .275*& .005 & -.1  &   3 &  0   & $-$ &{\bf
                                          25}\\
%
%Run5: Stronger Disc dynamo compared to reference model
SDd & {\bf S}tronger {\bf D}isc {\bf d}ynamo compared to Model~Ref
& .3   & .005 &{\bf
                -.3} &   2 & +1   &   5 &  0 \\
\hline
%Run6b: No Disc dynamo but stellar dynamo
NDd & {\bf N}o {\bf D}isc {\bf d}ynamo but stellar dynamo
& .3   & .005 &{\bf
                 0}  & $-$ & +1   &   5 &  0 \\
\hline
%nodisc-big: No Disc but stellar dynamo
ND  & {\bf N}o {\bf D}isc but stellar dynamo
& .25  &{\bf
         1}** &  0** & $-$ & +1.5 &   5 &  0 \\
\end{tabular}
}
\end{table*}

\subsection{Normalization and model parameters} \label{NMP}

In all our simulations we use dimensionless variables.
Our models can be rescaled easily,
and thus they can be applied to a range of different astrophysical objects.
Here, we consider values for our normalization parameters that are
typical of a protostellar star--disc system.
As in vRB04, we scale the sound speed with a typical coronal sound speed of
$c_{\rm s0} = 10^2\km\s^{-1}$,
which corresponds to a temperature of $T_0 \approx 4\times10^5\K$,
and we scale the surface density with a typical
disc surface density of $\Sigma_0=1\g\cm^{-2}$.
Further, we assume $M_*=1\,M_\odot$ (where $M_*$ and $M_\odot$ are the
stellar and solar mass, respectively), a mean specific weight of $\mu=0.6$,
and $\gamma=5/3$. Fixing our normalization parameters ($c_{\rm s0}$ or $T_0$,
and $\Sigma_0$, $M_*$, $\mu$ and $\gamma$) defines the units for all quantities.
[In the following, $G\approx6.67\times10^{-8}\cm^3\g^{-1}\s^{-2}$ is the
gravitational constant and ${\cal R}\approx8.3\times10^7\cm^2\s^{-2}\K^{-1}$
is the universal gas constant. With $M_\odot\approx2\times10^{33}\g$,
it follows that $GM_*=GM_\odot\approx1.3\times10^{26}\cm^3\s^{-2}$.]
The resulting units for our dimensionless quantities are given in
Appendix~\ref{NonDim}.

The axisymmetric computations have been carried out in the domain
$(\varpi,z)\in[0,2]\times[-1,+1]$, with the mesh sizes
$\delta \varpi=\delta z=0.01$. In our units, the domain then
extends to $\pm 0.1\AU$ in $z$ and $0.2\AU$ in $\varpi$.
[In Model~ND, the domain is larger in $z$, with $-2 \le z \le +2$;
the mesh sizes are the same.]
We impose regularity conditions on the axis ($\varpi=0$) and
do not allow for inflows at the outer boundaries of the computational domain,
on $\varpi=\varpi_{\max}$ and $z=\pm z_{\max}$.
The equations are solved also on the boundaries;
the derivatives are calculated with single-sided schemes
of the same order (sixth order) as in the domain.

We choose a set of values for our model parameters such that
the resulting dimensions of our star--disc system and the resulting initial
profiles of the physical quantities are close to those for a standard
accretion disc around a protostellar object.
Our (main) model parameters are the stellar radius, disc inner radius
and disc height describing the geometry of the star--disc system,
the entropy contrast parameters $\beta_{\rm disc}$ and
$\beta_{\rm star}$, the (turbulent) diffusivity coefficients, and
the $\alpha$ effect parameters for the disc and for the star.

We vary the stellar radius
between $r_*=0.15$ and $r_*=0.3$, corresponding to $3$ to $6$ solar radii.
The disc geometry remains unchanged, with the disc inner radius
$\varpi_{\rm in}=0.6$ (which is therefore located at $4$ to $2$ stellar radii)
and a disc semi-thickness of $z_0=0.15$ (i.e.\ $1$ to $0.5$ stellar radii).
The disc outer radius $\varpi_{\rm out}$ is close to the outer domain boundary.

In our piecewise polytropic model,
we need to choose values for the polytrope parameters in the disc and
in the star (see Appendix~\ref{Sec-cool-hot}).
Our choice for the entropy contrast between disc and corona,
$\beta_{\rm disc} = 0.005$, leads to an initial disc temperature
ranging between $9000\K$ in the inner part and $900\K$ in the outer part.
Real protostellar discs have typical temperatures
of about a few thousand Kelvin (e.g.,\ Papaloizou \& Terquem 1999).
As turns out from our simulations, the disc temperature increases by less than
a factor of $2$ with time, except for the inner disc edge where the increase
is much higher.
The low disc temperature corresponds to a relatively high disc density.
In the inner part of the disc, it is about $3\times10^{-10}\g\cm^{-3}$
initially, but increases to about $10^{-9}\g\cm^{-3}$ at later stages.
In the outer part it is a few times $10^{-11}\g\cm^{-3}$.
For the star--corona entropy contrast we choose always $\beta_{\rm star}=0.02$.

The Shakura--Sunyaev coefficient of the turbulent disc kinematic viscosity,
$\nu_{\rm t} = \alpha_{\rm SS} c_{\rm s} z_0 \xi_{\rm disc}(\varpi,z)$, is
$\alpha_{\rm SS} = 0.004$ in all models with disc, whereas the turbulent disc
magnetic diffusivity $\eta_{\rm t} = \eta_{\rm disc} \xi_{\rm disc}(\varpi,z)$
is also dependent on position but varies in the different models. In the stellar
dynamo models,
$\alpha_{\rm SS}^{(\eta)} \equiv {\eta_{\rm disc}/(c_{\rm s}z_0})$ is typically
a few times higher than $\alpha_{\rm SS}$, but $\alpha_{\rm SS}^{(\eta)}$ can
also be as much as an order of magnitude higher than $\alpha_{\rm SS}$. Values
of the effective magnetic diffusivity in the disc range from around $0.0003$
in the outer parts to top values of around $0.003$ toward the inner disc edge so
that in the disc, the diffusion time $t_{\rm diff} \equiv z_0^2/\eta_{\rm disc}$
ranges between $75$ and $7.5$ in the stellar dynamo models.
In the stellar dipole models, the effective magnetic diffusivity in the midplane
is a few times $10^{-3}$, leading to $t_{\rm diff} \approx 5 \dots 10$.

The dimensionless dynamo numbers related to
the $\alpha$ effect in the disc and in the star,
${\cal R}_\alpha^{\rm disc} \equiv \alpha_0^{\rm disc}z_0/\eta_{\rm disc}$ and
${\cal R}_\alpha^{\rm star} \equiv \alpha_0^{\rm star}r_*/\eta_{\rm star}$, are
${\cal R}_\alpha^{\rm disc} \approx -75 \dots -7.5$ and
${\cal R}_\alpha^{\rm star} \approx +100 \dots +200$ in Model~Ref, and
${\cal R}_\alpha^{\rm disc} \approx -3 \dots -7$ in Model~S.
Note that
$\alpha_0^{\rm disc} \approx (1 \dots 3) \, c_{\rm s}$ in Model~Ref and Model~S,
and
$\alpha_0^{\rm star} \approx 2 \, c_{\rm s}$ (near the stellar surface)
in Model~Ref.
Since CTTS can be fully convective, the $\alpha$ effect would operate
over a comparatively large volume so that
in these protostars the dynamo may be operating in a
highly supercritical regime.

\begin{figure*}[t!]
\centerline{
   \includegraphics[width=8.5cm]{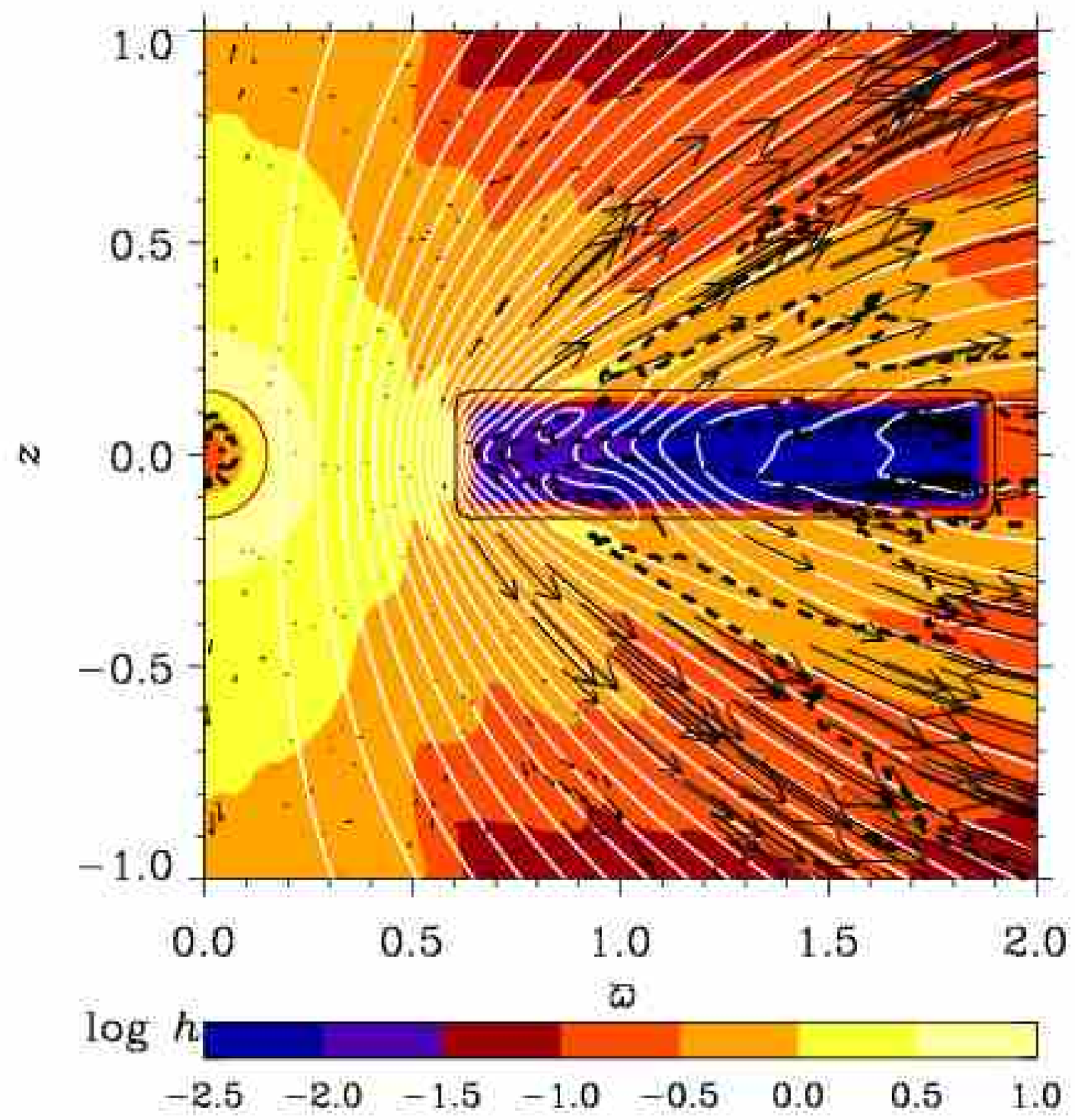}
   \includegraphics[width=8.5cm]{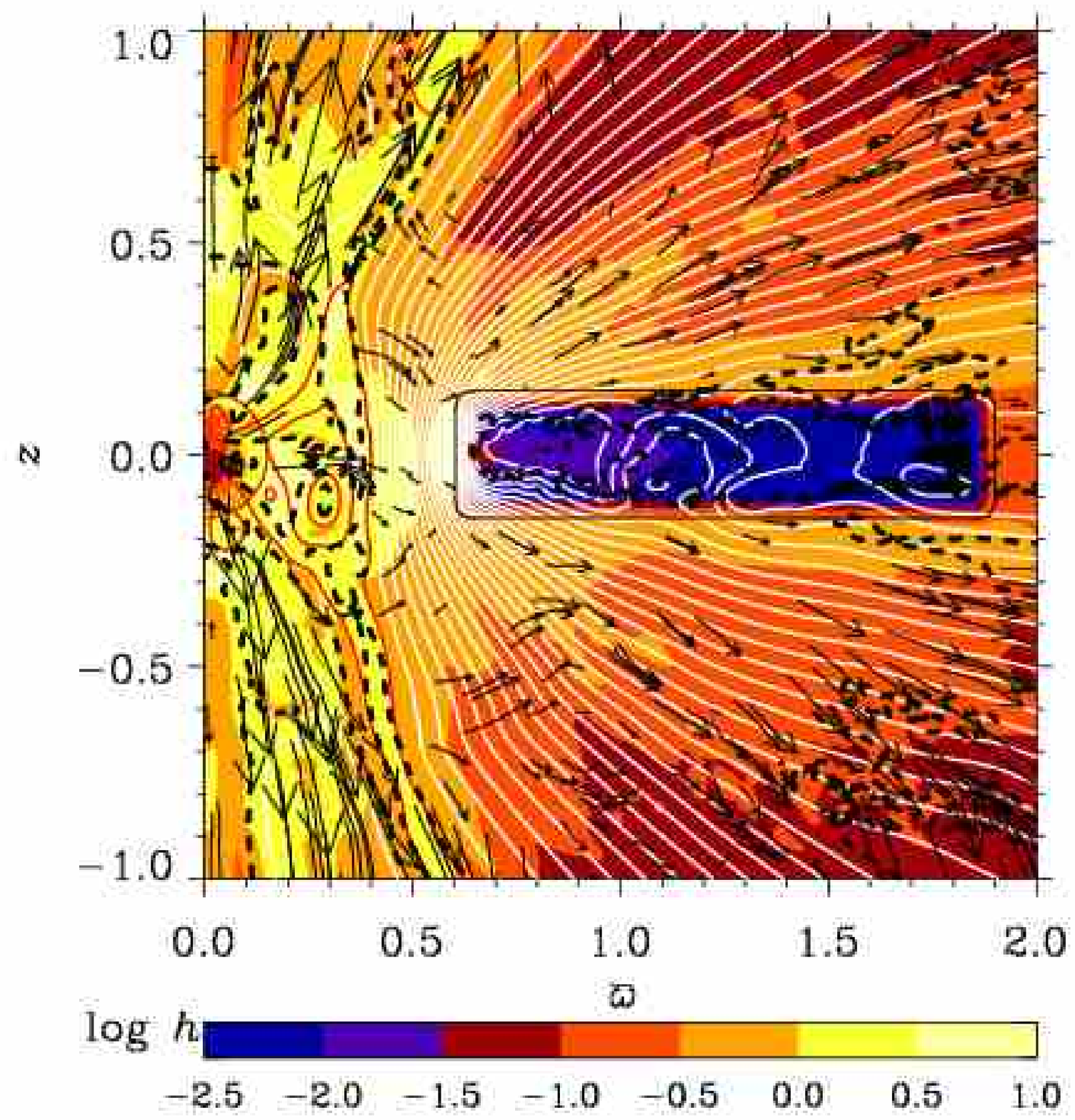}
}
\caption[]{
The emergence of a fast stellar wind in the presence of
a stellar dynamo ({\it right hand panel}) compared to the case without
stellar dynamo ({\it left hand panel}) where the stellar wind is very weak.
Black arrows: poloidal velocity vectors;
solid lines: poloidal magnetic field lines
(white means right-handed and red/gray means left-handed);
colors/gray shades: temperature $T$ [$\log_{10}h=(-2,-1,0,1)$ corresponds to
$T\approx
(3{\times}10^{3},3{\times}10^{4},3{\times}10^{5},3{\times}10^{6})\,\mbox{K}\,$];
black dashed line: poloidal Alfv\'en surface.
{\it Left}: Model~N (where $\alpha_0^{\rm star}=0$)
at time $t \approx 940$ days (steady state).
{\it Right}: Model~SmS (where $\alpha_0^{\rm star}=+1.5$)
at time $t \approx 1644$ days (steady state for the disc dynamo).
}\label{R1aR1b}
\end{figure*}

With our initial setup as the hydrostatic equilibrium of a piecewise polytropic
model with smoothed gravitational potential,
the initial stellar surface angular velocity depends on the chosen stellar
radius. In our models, this angular velocity varies at the equator
between 1 (for $r_*=0.15$) and 0.1 (for $r_*=0.3$),
corresponding to rotation periods of 10~days to 100~days.
In the case of a rotation period of around 10~days, the corotation radius is
$\varpi_{\rm co}\approx 1$ in nondimensional form.
The inner disc radius (which is fixed in our models) is rotating with
roughly Keplerian angular velocity, resulting in a rotation period of the inner
disc edge of around 4.5~days in all our models. The Keplerian speed of the inner
disc edge is about $130\kms$, and the free-fall velocity is
$v_{\rm ff} = \sqrt{2GM_*/r} \approx 183\kms \dots 283\kms$ in the gap,
between the inner disc radius and the stellar radius of Model~Ref.
The Keplerian orbital period of the disc radius close to our outer boundary
is about 25~days.
For an accretion velocity of $10\kms$, the advection time through the disc is
$t_{\rm accr} \equiv
\left(\varpi_{\rm out}-\varpi_{\rm in}\right)/u_{\rm accr}^{\rm disc}
\approx$ 20~days.
We ran the simulations discussed in this paper over many rotation periods,
between 165 and 1650 days.

For the stellar and accretion disc dynamos to work, we have weak seed
magnetic fields in both the disc and the star.
There is no externally imposed magnetic field in the system
in all our new models presented in this paper.
The initial hydrostatic solution is an unstable
equilibrium because of the vertical
shear between the disc, star and corona (Urpin \& Brandenburg 1998).
In addition, angular momentum transfer by viscous and magnetic
stresses -- the latter from the stellar and disc dynamos --
drives the solution immediately away from the initial state.
The initial state should certainly not be regarded as an approximation
to the final time-dependent state.
In fact,
all the results presented in the next section are representative of a
statistically steady state, reached after initial transients have died out.

%-------------------------------------------------------------------------------
\section{Results} \label{RES}

A summary of the varying input parameters of the models discussed
in this paper is given in Table~\ref{tab1}.
They will be explained in more detail below as the models are being
motivated.
We begin with Model~N which has no stellar dynamo and no anchored magnetosphere,
but it has a disc dynamo.

\subsection{Stellar dynamo versus no own stellar field}

In Fig.~\ref{R1aR1b} we show the resulting flow and field lines of
Model~N
and compare with those of Model~SmS where there is a stellar dynamo.
In both models, the stellar radius (about 3 solar radii) and
the inner disc edge (about 12 solar radii) are the same.
Obviously, this comparison is somewhat artificial, because in a
model without a stellar dynamo the disc would in reality be truncated
at a smaller radius.
Here we have chosen the same gap, because we want to see
the effect of a stellar dynamo while keeping the other parameters the same.
We will discuss a smaller gap in the case without a star's own magnetic
field at the end of this subsection.

\begin{figure*}[t!]
\centerline{
   \includegraphics[width=8.5cm]{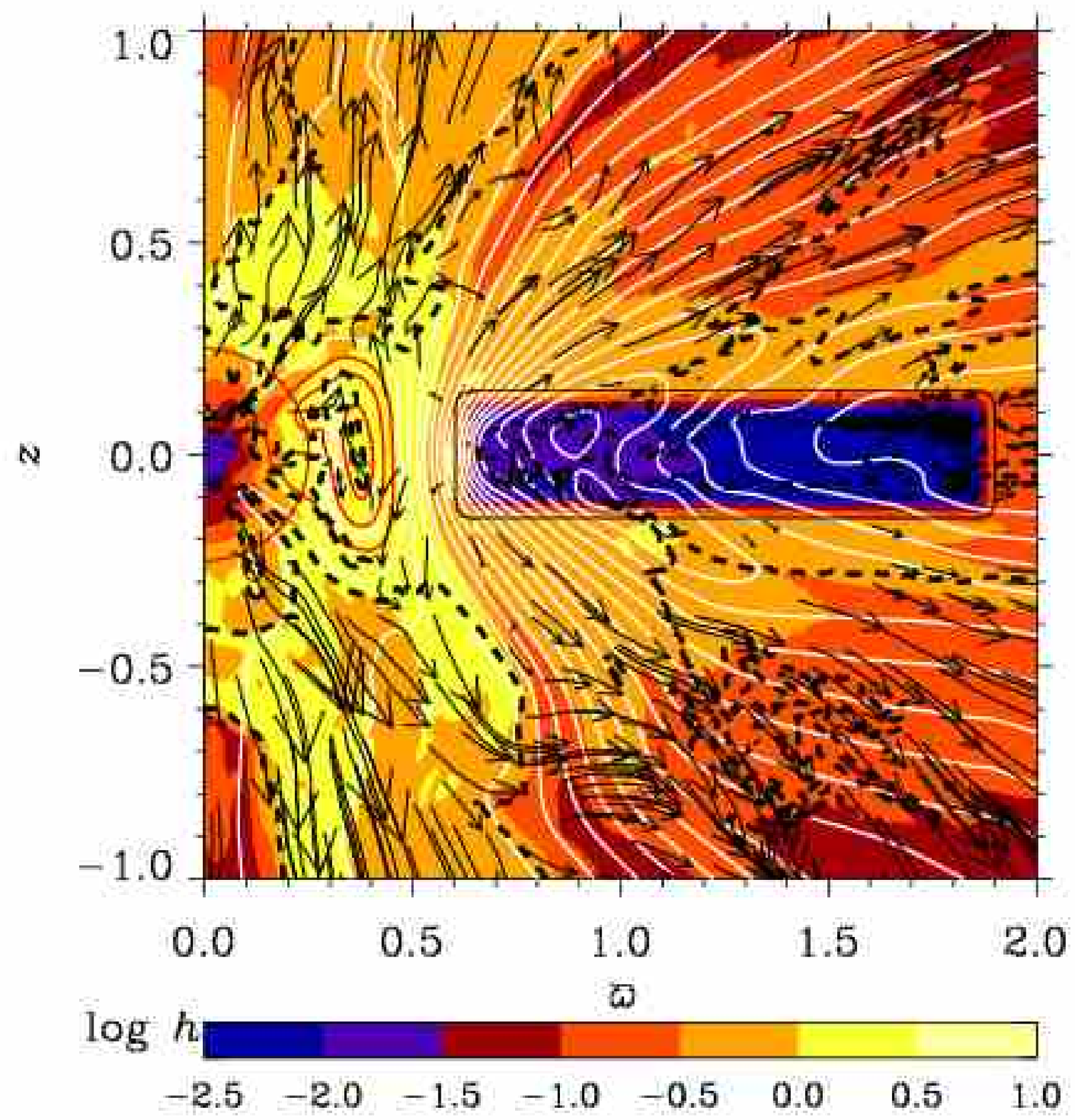}
   \includegraphics[width=8.5cm]{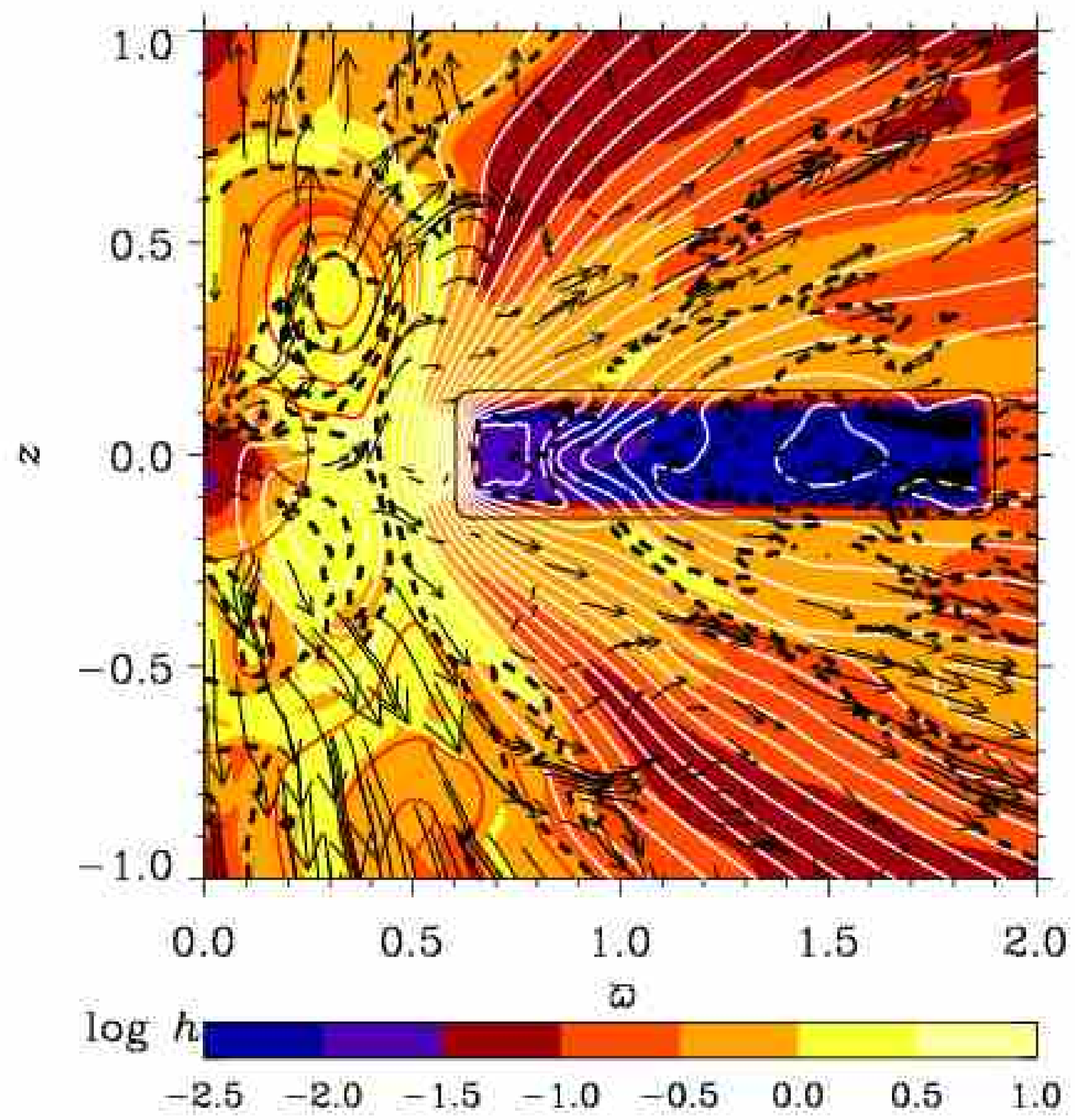}
}
\caption[]{
The stellar dynamo-generated magnetic field switches between dipolar
({\it left hand panel}) and quadrupolar ({\it right hand panel}) symmetry.
Shown is the same as in Fig.~\ref{R1aR1b}, but for Model~Ref where
the stellar radius is larger.
{\it Left}: at an earlier time ($t \approx 1171$ days) when the stellar field
configuration turns out to be close to dipolar.
{\it Right}: at a later time ($t \approx 1183$ days) when the stellar field
configuration turns out to be close to quadrupolar.
}\label{pnew_R3}
\end{figure*}

Comparing the left and right hand panels of Fig.~\ref{R1aR1b}, it becomes
clear that a stellar dynamo affects the dynamics of a star--disc system
significantly.
In Model~SmS, the stellar dynamo generates a magnetosphere that
is mostly of roughly dipolar symmetry and has a magnetic field strength of
$B_{\rm star,max} \approx 300$ G. As a result, the star is rotating fast
with an equatorial surface rotation period of $P_{\rm rot,surf} \approx 2$ days,
and a fast stellar wind is launched with a terminal speed of about
500 km/s. This fast wind from the stellar dynamo is supersonic and
super-Alfv\'enic and tends to collimate at low heights; it is not present
in the absence of a stellar dynamo (Model~N)
where the stellar wind reaches 20 km/s
along the rotation axis and 5 km/s further away from the axis,
and where the star rotates about 50 times more slowly than in Model~SmS.
In Model~N, the stellar wind is subsonic and
sub-Alfv\'enic. When there is no dynamo in the star, the stellar field results
entirely from advection of the dynamo-generated disc field, which is only
a few tens of gauss strong. In both cases, the compressed magnetic field
at the inner disc edge acts like a shield and prevents any pronounced accretion.

The disc wind is robust and its overall structure is largely
independent of the origin of the stellar magnetic field.
The disc wind terminal speed is about twice the Keplerian speed of the inner
disc edge.
However, the stellar dynamo and induced stellar wind distort the disc dynamo
and disc wind in places, and the terminal speed of about 350 km/s in the
undistorted inner disc wind of Model~N drops to about 250 km/s in Model~SmS.
Therefore, the total
wind mass loss rate $\dot{M}_{\rm wind}$ is dominated by the dense high-speed
stellar wind in the model with stellar dynamo, where $\dot{M}_{\rm wind}
\approx(2 \dots 6)\times10^{-7}M_\odot\yr^{-1}$, and by the less dense but fast
disc wind in the model without stellar dynamo, where $\dot{M}_{\rm wind}
\approx(1.5 \dots 2)\times10^{-7}M_\odot\yr^{-1}$.

We now want to discuss the effect of
a smaller gap in the case without the star's own magnetic
field. To this end, we compare Model~N of this paper and Model~N of vRB04.
The smaller gap in Model~N of vRB04 has significant consequences.
The smaller gap leads to an initially much faster star, with a rotation period
of a few days. Further, due to the small gap, the advected disc magnetic field
is now penetrating the star and it thus creates a stellar field strength of
more than 100\G. This fast magnetized star is the reason why the stellar wind
velocity and mass loss rate are higher in Model~N of vRB04.
Also the inner disc edge -- where the strongest disc wind originates from --
is rotating faster as it is further inward.
Therefore, also the disc wind velocity and mass loss rate are higher in vRB04.
Finally, the mass accretion rate is significantly larger in Model~N of vRB04.
This is because there, the build-up of a strong magnetic field advected from the disc
occurs only in the star, whereas in Model~N of this paper there is a field
in the gap, where it is built up to about $250\G$ and acts as a barrier for the
disc matter.
(The field strength in Model~N of this paper decreases then to a few tens of
gauss in the star.)

\begin{figure}[t!]
\centerline{
   \includegraphics[width=8.5cm]{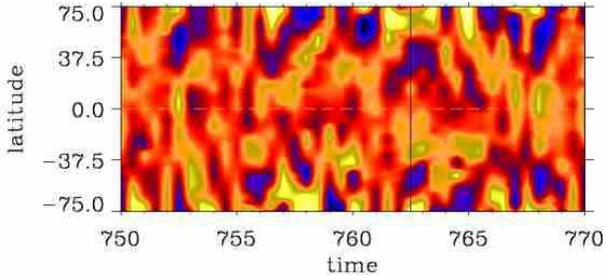}
}
\caption[]{
Space-time diagram for Model~Ref in a color/gray scale representation
of the radial magnetic field component.
No clear latitudinal migration pattern of the stellar surface field is
seen, although some high-latitude structures appear to show traces of
poleward migration.
Bright colors or light shades indicate positive values;
dark colors or dark shades indicate negative values.
The maximum strength of $B_r$ is about $180\G$.
The vertical lines at times 762.5 and 770 mark the corresponding times $t \approx$ 1171 days
and 1183 days, respectively, i.e.\ the times of Figs~\ref{pnew_R3} and
\ref{Runs_strong-103_VAR1525-VAR1540-R3}.
}\label{pbutter_R3}
\end{figure}

\subsection{Dipolar and quadrupolar stellar dynamos}

As our reference model, we choose a model where the stellar radius is
increased to about 5 solar radii.
The inner disc radius remains at the same
position (about 12 solar radii) so that it is now located at 2.4 stellar radii,
whereas in the models of Fig.~\ref{R1aR1b} it is located at 4 stellar radii.
By increasing the stellar radius without changing the disc location or geometry,
this results in a more realistic aspect ratio between the stellar radius and the
height of the inner disc edge and, furthermore, the gap and dynamo-free region
are decreased so that star--disc interaction is facilitated.
In Fig.~\ref{pnew_R3} we show the resulting magnetic field and flow patterns.
The accretion disc dynamo-generated magnetic field is -- as usual --
of dipolar symmetry with respect to the disc midplane (within the disc
as well as above and below it), with $B_\varphi<0$ in the conical shell above
the disc and $B_\varphi>0$ in the conical shell below the disc.
The stellar dynamo generates a magnetosphere that switches
between `dipolar' (Fig.~\ref{pnew_R3}, left hand panel)
and `quadrupolar' (Fig.~\ref{pnew_R3}, right hand panel) symmetries
(see also Fig.~\ref{pbutter_R3}).
Here, the quotation marks on `dipolar' and `quadrupolar' indicate that,
within the star, both poloidal and toroidal fields have dipolar/quadrupolar
symmetry in Fig.~\ref{pnew_R3}, left/right hand panel, respectively.
However, the stellar
dynamo-generated magnetic field {\it outside} the star develops in such a way
that the azimuthal component of the field tends to be of one sign within each
closed loop:
within almost the whole `dipolar' loop of Fig.~\ref{pnew_R3} (left hand panel),
it is $B_\varphi > 0$ {\it outside} the star,
and within the `quadrupolar' loops of Fig.~\ref{pnew_R3} (right hand panel),
it is $B_\varphi < 0$ {\it outside} the star.
This behavior of the toroidal field is possible because we do not impose any
northern-southern hemisphere symmetry for the rotation, but the angular velocity
evolves in a dynamical way within the star, at the stellar surface and outside
the star.

The `oscillation' of the magnetosphere is not periodic and results in
quick temporal changes of the magnetic field geometry with many different
field configurations of mixed parity.
Polarity reversals of the magnetic field occur mostly
at time scales of less than a day.
The irregular behavior of the magnetic field geometry with time, including the
magnetic polarity reversals, is likely to be due to the assumption of
strong convection in our model of the star, which led us to choosing
supercritical values of the dimensionless $\alpha$ dynamo number
in the star: ${\cal R}_\alpha^{\rm star} > 100$ with an effective resistivity of
a few times $10^{-3}$.
Since we are in a highly nonlinear regime, the magnetic field can then be
very irregular in the regime of an axisymmetric spherical $\alpha^2$ dynamo
(Meinel \& Brandenburg 1990). Tavakol et al.\ (1995) studied the structural
stability of axisymmetric $\alpha \Omega$ dynamos in spheres subject to
the increase of ${\cal R}_\alpha$. For negative radial rotational shear
(as is also the case in some of our models; see Sect.~\ref{Collimation} below),
their stable solution switches from antisymmetric
to mixed parity to oscillating symmetric to symmetric.
Our simulations have been carried out with $\alpha^2\Omega$ dynamos with large
${\cal R}_\alpha^{\rm star}$ in the star.
The resulting irregular behavior of the magnetic field is in accordance with
observations of CTTS, which show that CTTS have highly time-dependent fields
suggesting the existence of irregular magnetic cycles
(cf.\ Johns-Krull et al.\ 1999a,b).

The maximum stellar field strength varies also with time,
ranging between $250\G$ and $750\G$ inside the star.
The rotation period of the stellar surface is oscillating around a few days
around the equator, and the stellar wind reaches terminal speeds between
300 km/s and 400 km/s; it becomes supersonic and super-Alfv\'enic very quickly.
The disc field is less than $50\G$ strong in the bulk of our disc,
but it can grow to over $100\G$ around the inner disc edge.
The disc wind is little affected. It is fastest within a conical shell
originating from the inner disc edge and reaches a maximum terminal speed
of up to $250$ km/s.
The combined stellar and disc wind mass loss rate is $\dot{M}_{\rm wind}
\approx(2 \dots 4)\times10^{-7}M_\odot\yr^{-1}$, while
the mass accretion rate varies over a large range, between
$\dot{M}_{\rm accr} \approx 10^{-10}M_\odot\yr^{-1}$ and
$\dot{M}_{\rm accr} \approx 10^{-7}M_\odot\yr^{-1}$
(see Appendix~\ref{NonDim} for the unit for $\dot{M}$).

In Fig.~\ref{pbutter_R3}, a space-time diagram is shown for Model~Ref.
No clear latitudinal migration pattern of the radial magnetic field
at the stellar surface can be seen.
(The same is true for Model~SmS, Model~SDd and Model~NDd.)

\subsection{Stellar dynamo versus anchored stellar dipole}

\begin{figure*}[t!]
\centerline{
   \includegraphics[width=8.5cm]{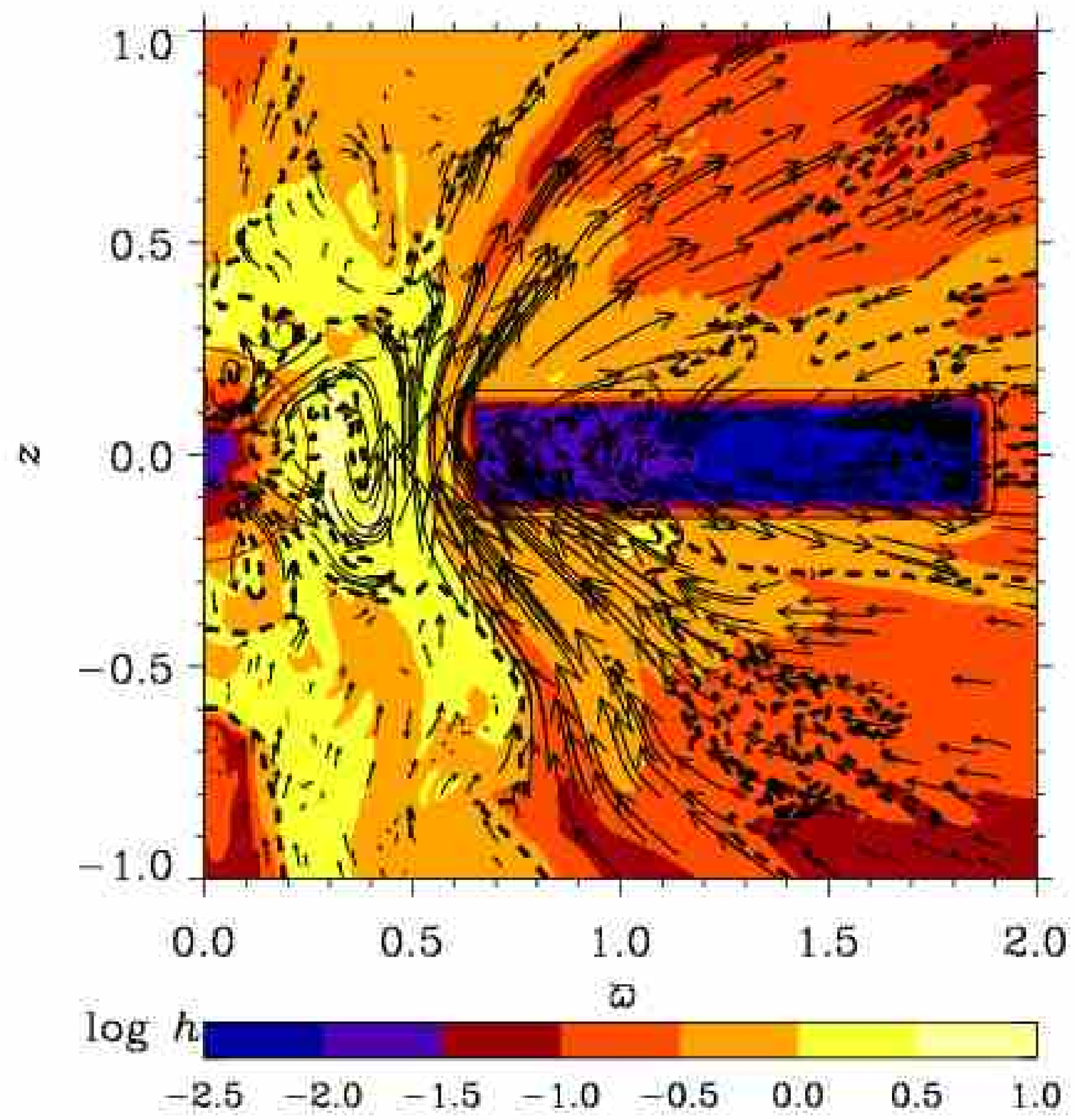}
   \includegraphics[width=8.5cm]{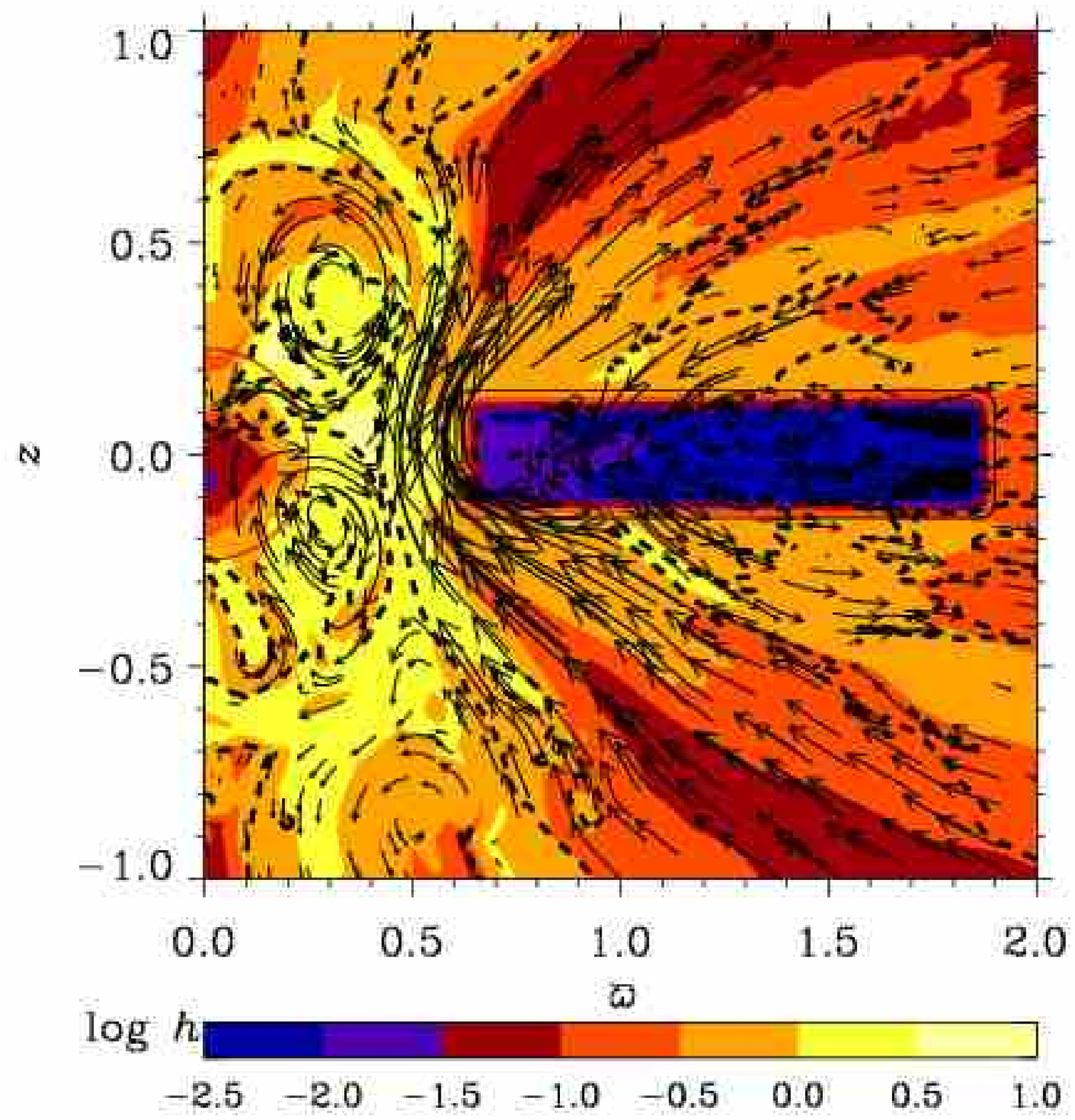}
}
\centerline{
   \includegraphics[width=8.5cm]{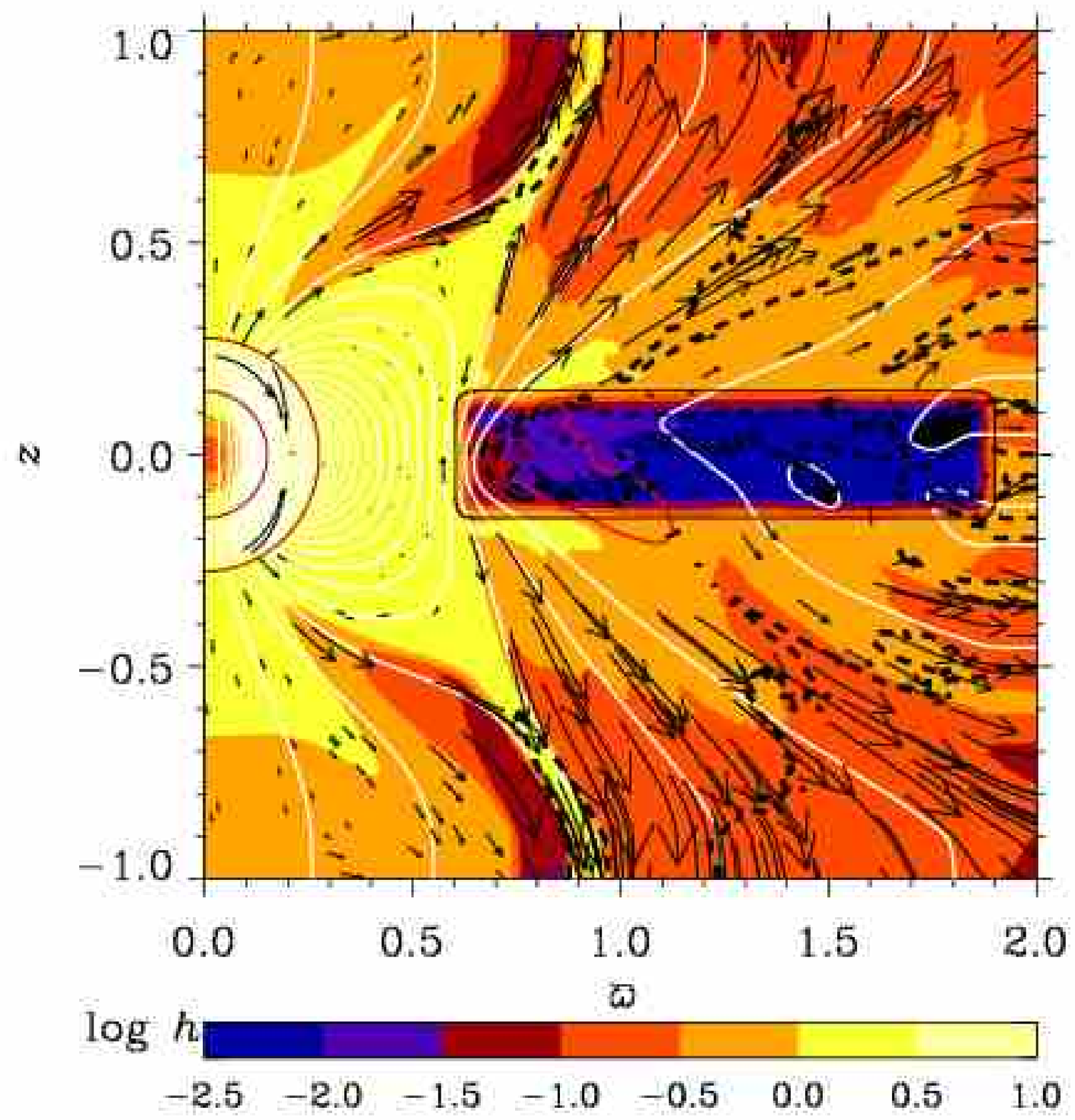}
   \includegraphics[width=8.5cm]{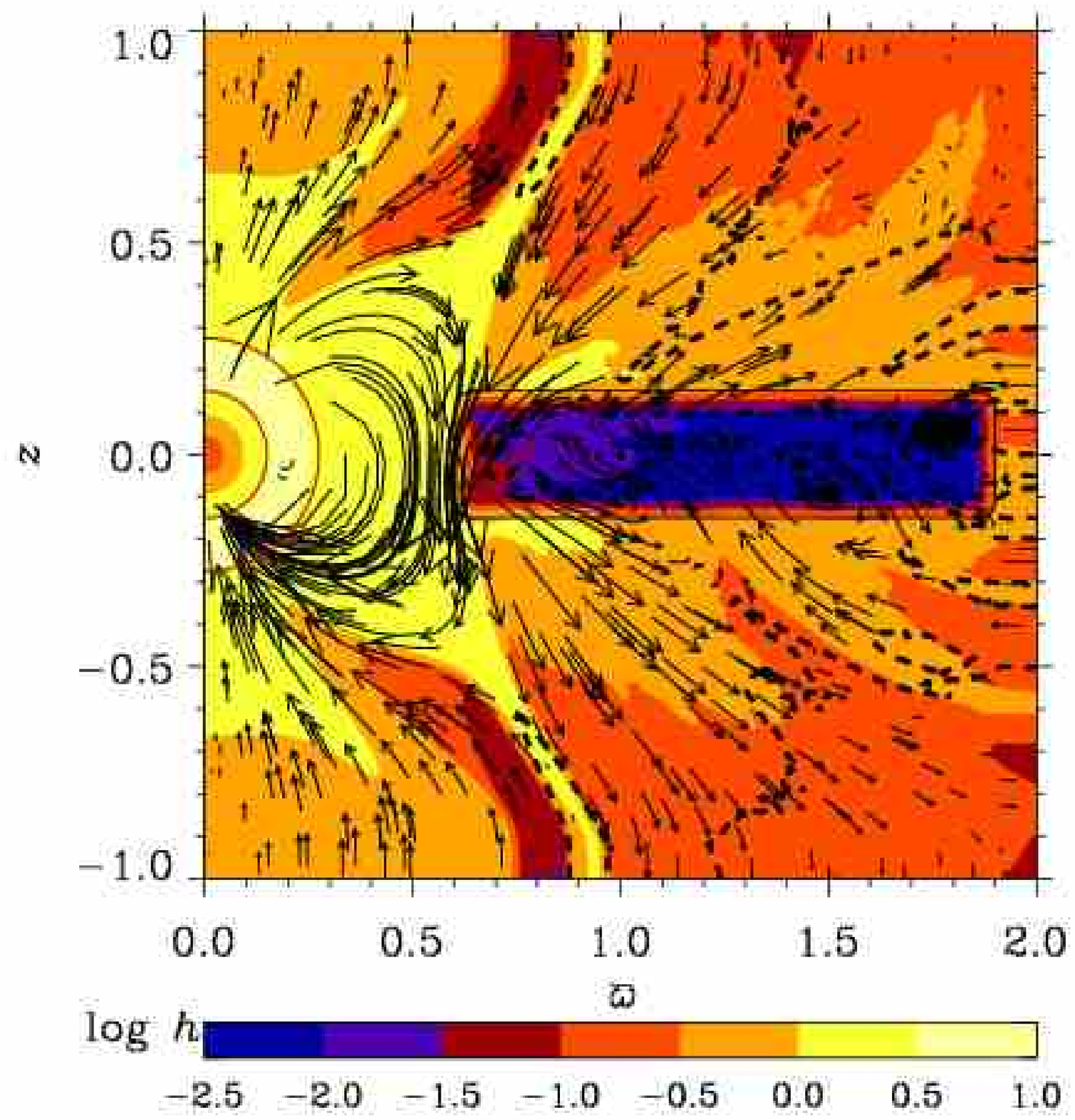}
}
\caption[]{
The absence of pronounced current sheets in a model with
stellar dynamo ({\it top row})
in contrast to the case of an imposed stellar dipole field ({\it bottom row}).
Colors/gray shades: temperature as in Figs~\ref{R1aR1b} and \ref{pnew_R3};
black dashed line: poloidal Alfv\'en surface;
black arrows: poloidal magnetic field vectors ({\it top row and bottom right};
              length of vectors is weighed with $\varpi$) and
              poloidal velocity vectors ({\it bottom left}), respectively.
{\it Top}: Model~Ref at times $t\approx$ 1171 days ({\it left})
           and $t\approx$ 1183 days ({\it right}).
{\it Bottom}: Model~S at time $t\approx 158$ days (cf.\ Figs~12 and 16 of
vRB04.)
Note that in Model~S, collimation at the upper and lower boundaries
($z>0.8$ and $z<-0.8$) is an artifact due to boundary conditions.
}\label{Runs_strong-103_VAR1525-VAR1540-R3}
\end{figure*}

\begin{figure*}[t!]
\centerline{
   \includegraphics[width=8.5cm]{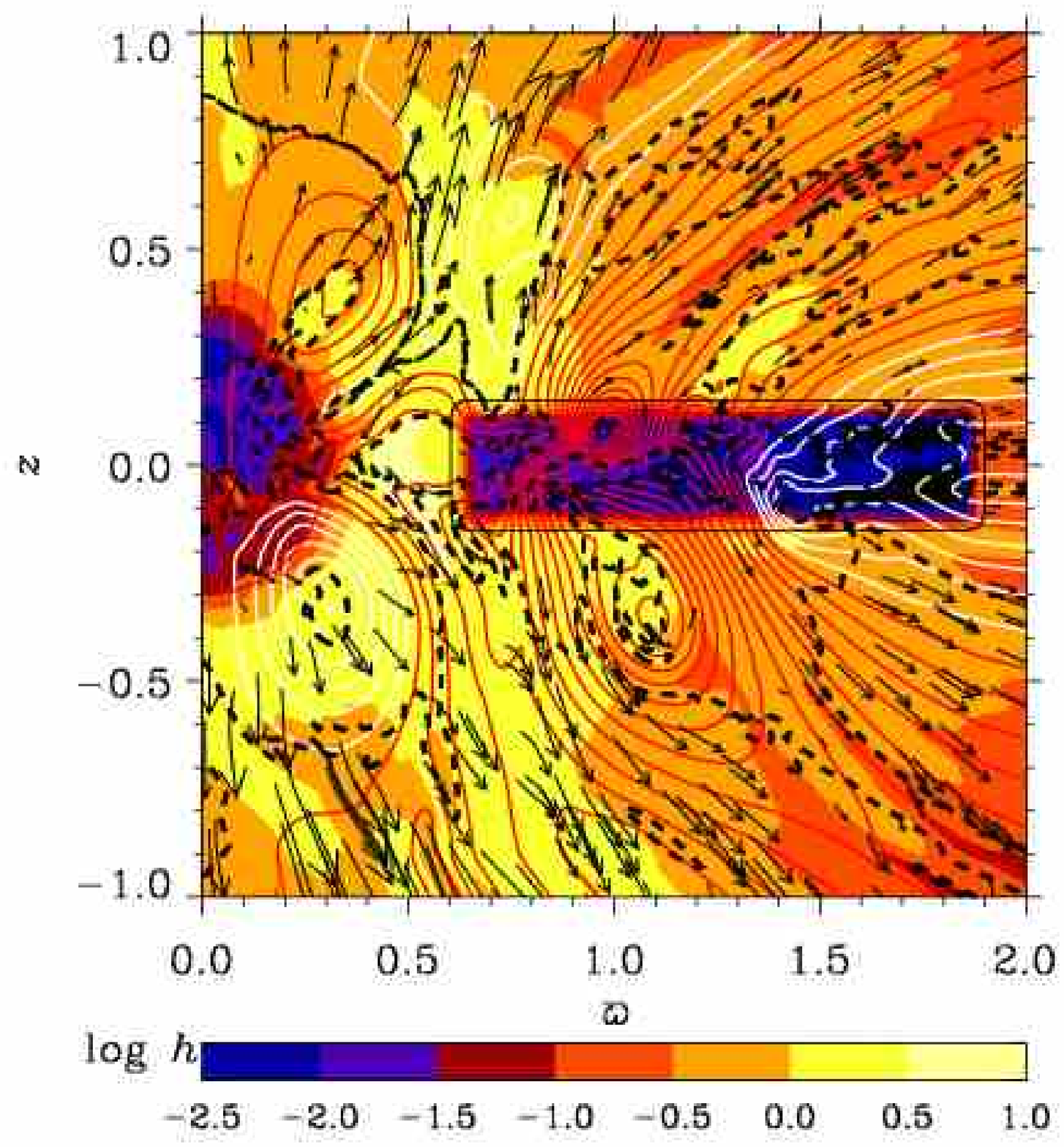}
   \includegraphics[width=8.5cm]{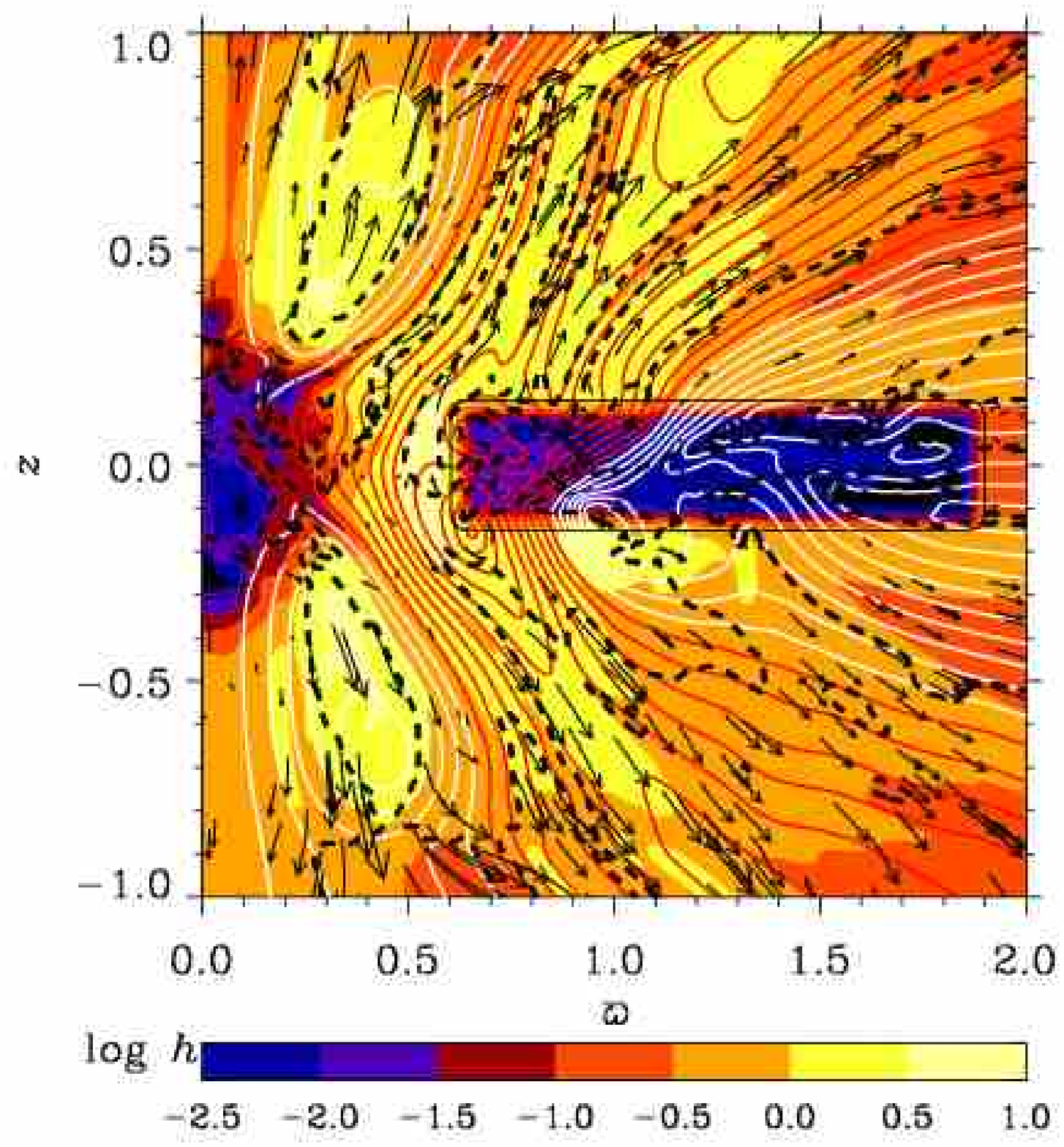}
}
\centerline{
%%%\resizebox{}{}{
   \includegraphics[width=8.5cm]{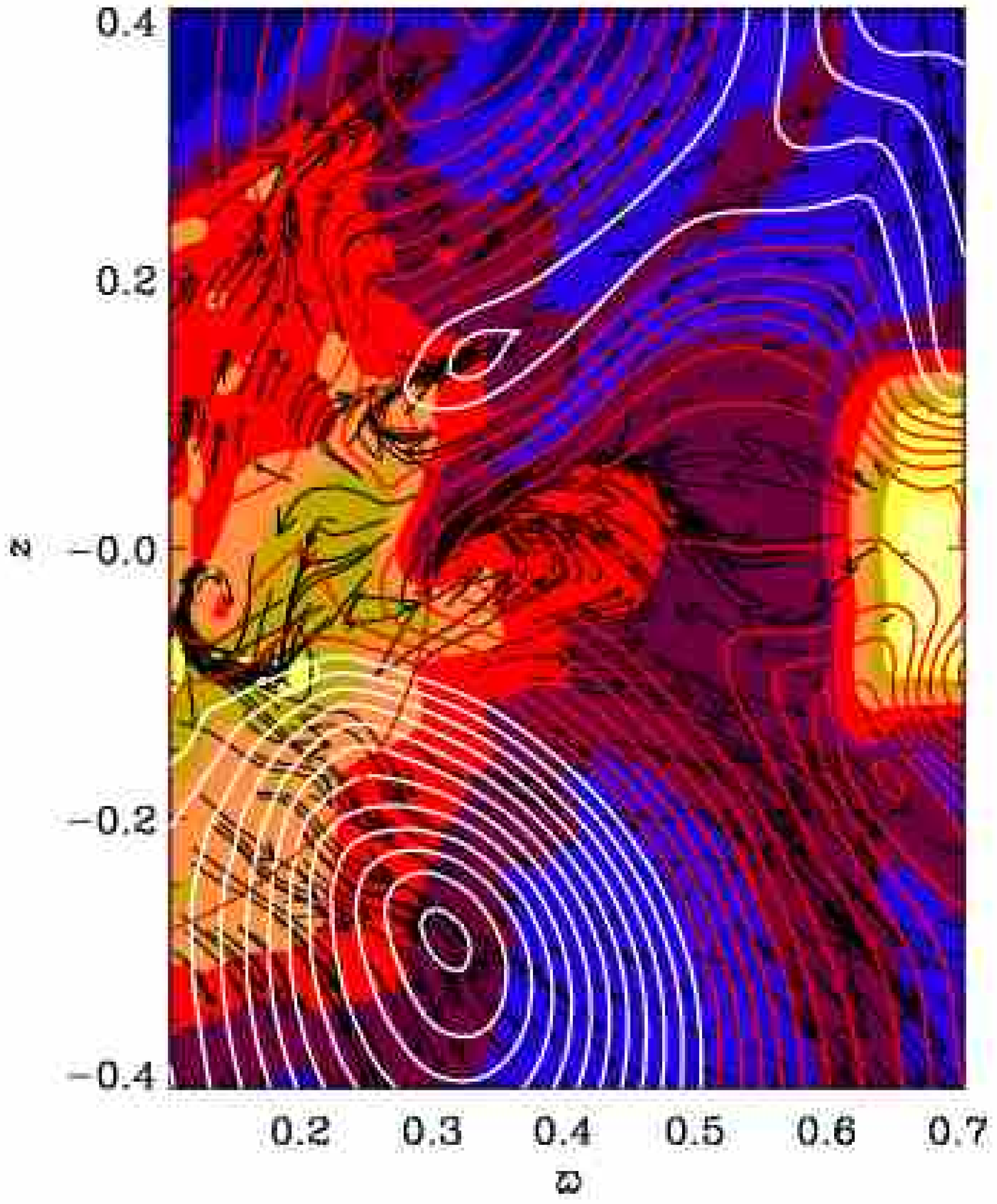}
%%%}
   \includegraphics[width=8.5cm]{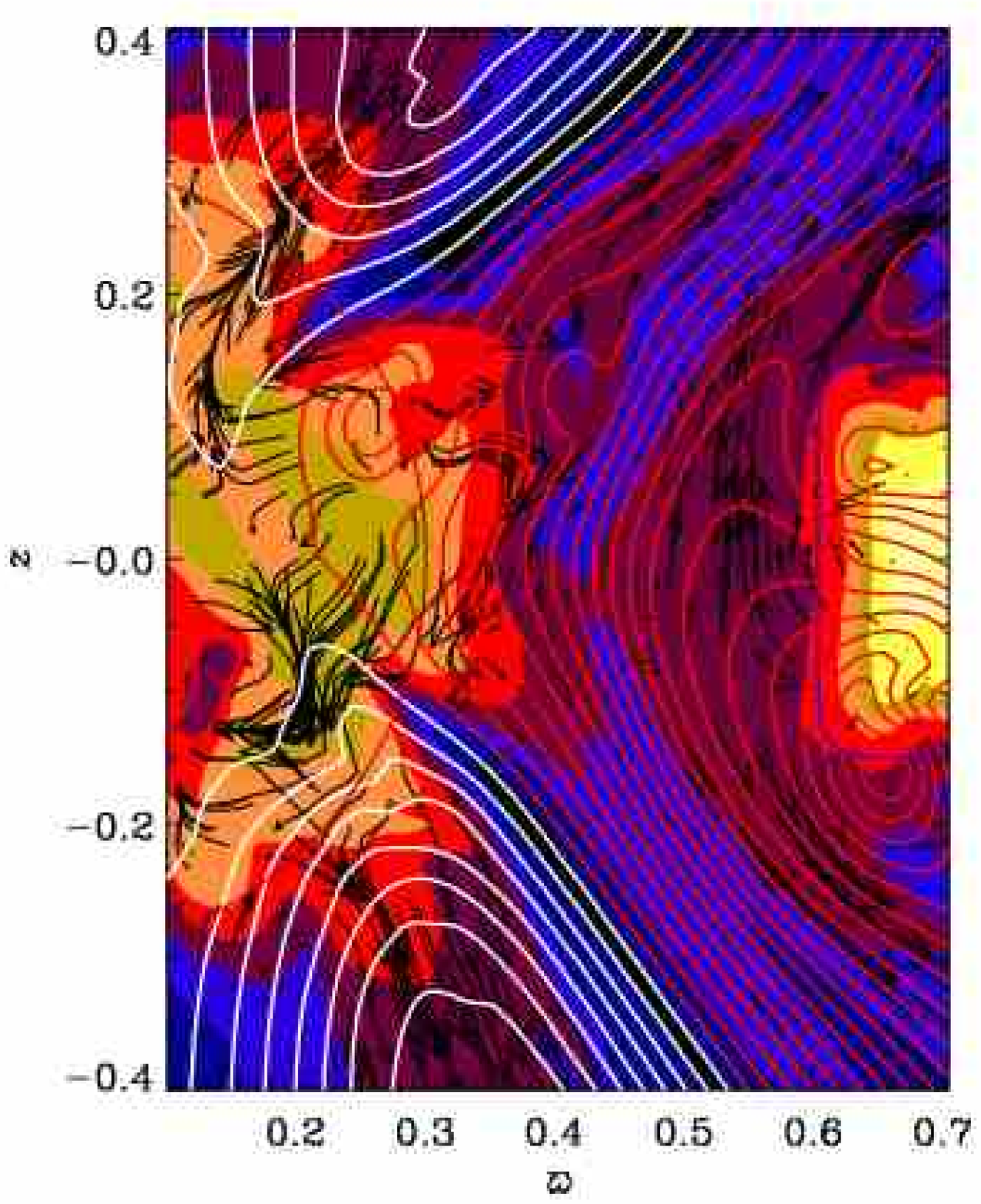}
}
\caption[]{
Magnetic star--disc coupling and mass accretion onto the star in Model~SDd
with stellar and disc dynamos, showing two distinct states:
when the star is magnetically connected to the disc
($t\approx 464$ days, {\it left column}) and
when the star is magnetically disconnected from the disc
($t\approx 469$ days, {\it right column}).
The {\it upper two panels} show temperature color/gray scale coded, together
with poloidal magnetic field lines and poloidal velocity vectors.
The {\it lower two panels} show gas density color/gray scale coded, together
with poloidal magnetic field lines and azimuthally integrated poloidal momentum
density vectors. Note that in this stellar dynamo model, star--disc coupling
is realized by the disc dynamo-generated magnetic field.
}\label{pnew_p_R5}
\end{figure*}

\begin{figure*}[t!]
%t=302,305,308,312,315,318
\centerline{
   \includegraphics[width=8.5cm]{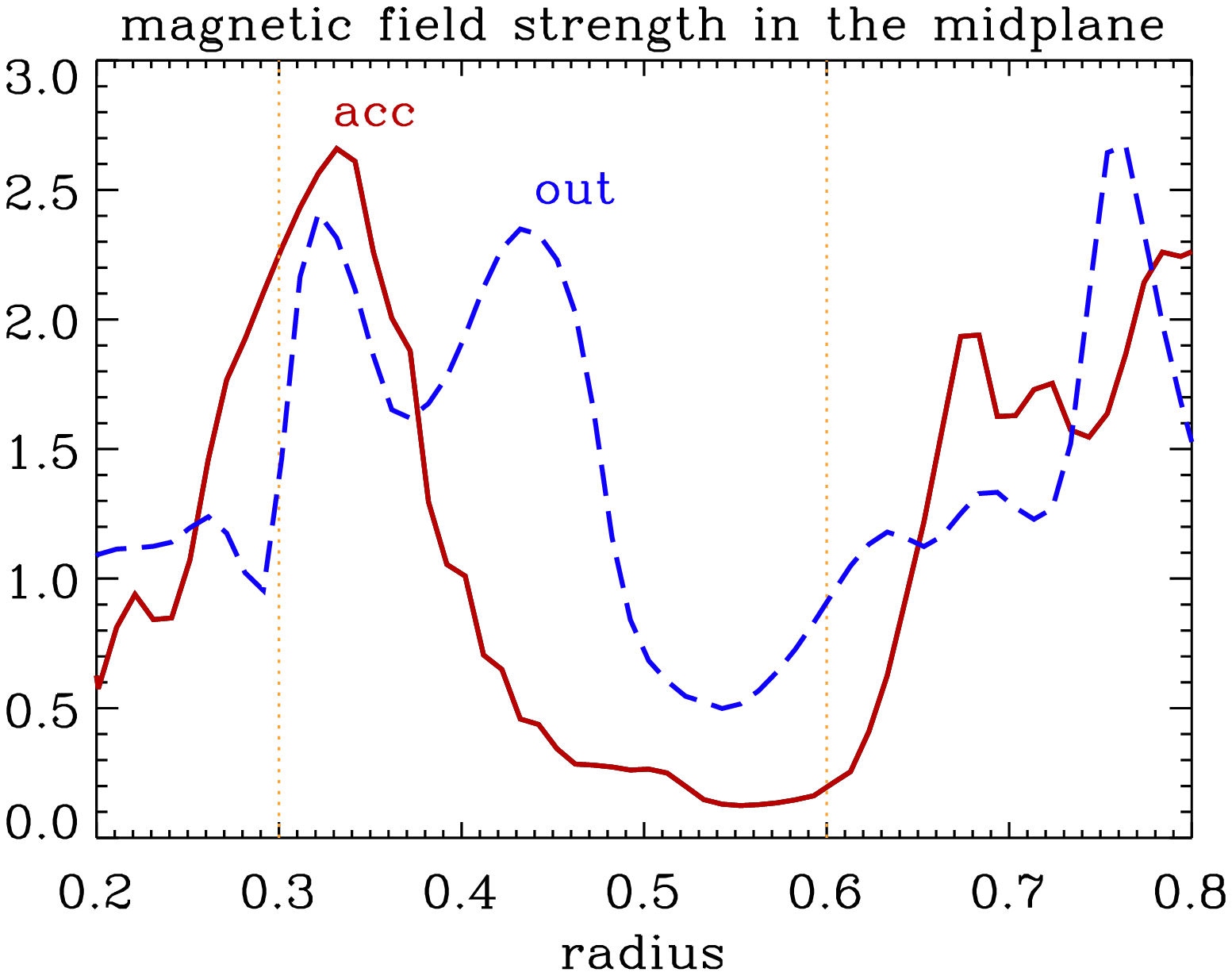}
   \includegraphics[width=8.5cm]{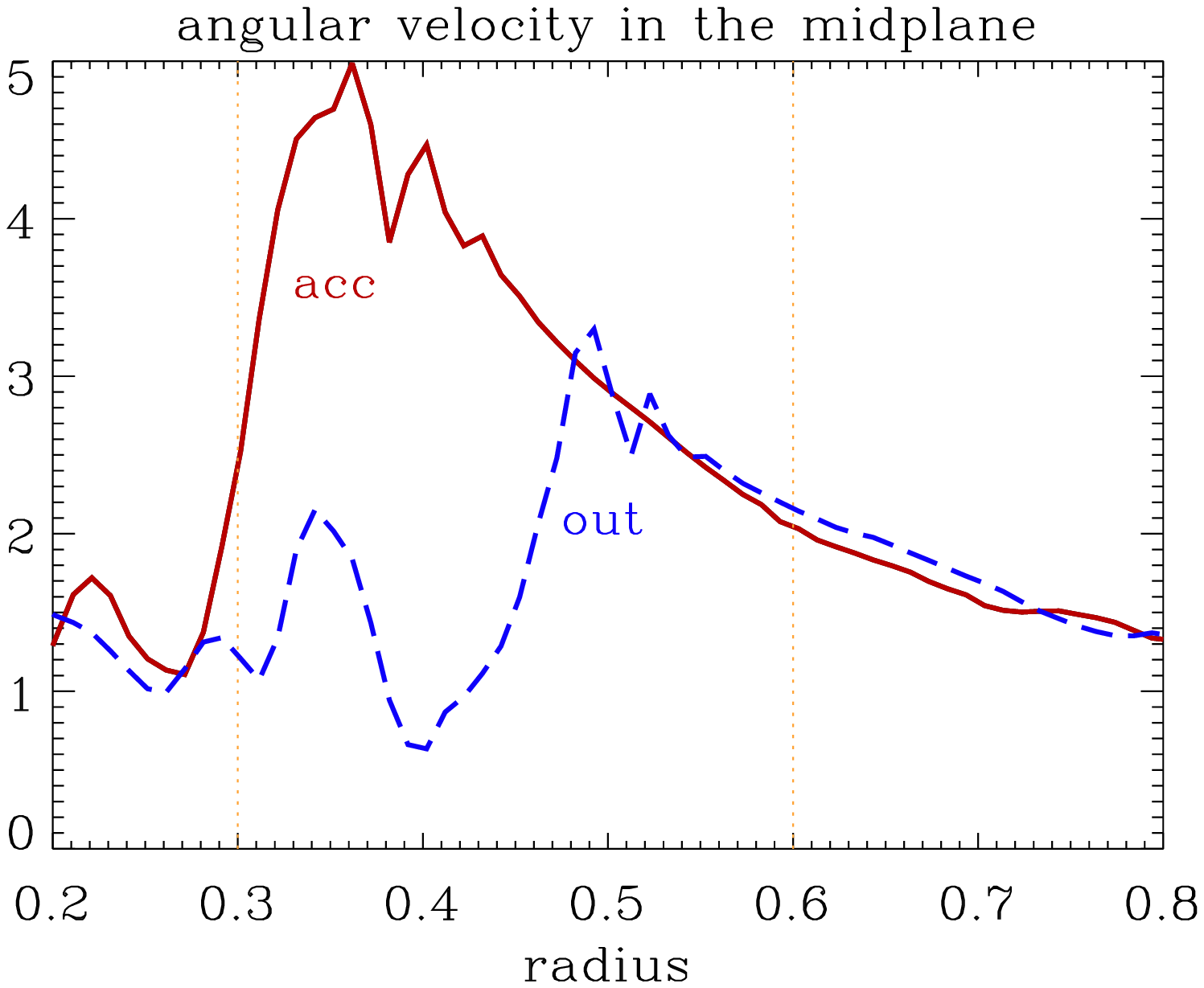}
}
\caption[]{
Radial profiles of the magnetic field strength ({\it left hand panel})
and the angular velocity ({\it right hand panel}) in the
midplane at the times of Fig.~\ref{pnew_p_R5} (Model~SDd),
$t\approx 464$ days (solid lines) and $t\approx 469$ days (dashed lines).
The vertical lines mark the stellar and inner disc radii.
The solid curves correspond to a time when the field in most parts of the gap
between the star and the disc is very weak, and hence disc matter -- carrying
Keplerian angular velocity from the disc throughout large parts of the gap --
can accrete onto the star (connected state).
The dashed curves correspond to the other extreme, a time when the field
in the gap is strong, so that accretion and rotation in the gap are suppressed
(disconnected state).
The stellar surface rotation period changes from 4 to 10 days at the equator;
the disc remains in Keplerian rotation.
}\label{Om_mag_R5}
\end{figure*}

We recall that the star is a part of the computational domain.
Therefore, in our stellar dynamo models no boundary conditions
are imposed in or around the star other than the prescription of a mean-field
dynamo in the star with a positive $\alpha$ effect in the upper hemisphere.
This leads to a dynamical evolution of the magnetic field, angular velocity and
poloidal velocity also within the star and at the stellar surface in the stellar
dynamo models, whereby the stellar magnetic field strength and geometry are
governed by the stellar dynamo parameters $\alpha_0^{\rm star}$ and
${\rm Ma}_{\rm star}$.
The resulting configuration of the dynamo-generated stellar magnetosphere
in our reference model (Model~Ref; cf.\ Fig.~\ref{pnew_R3} and
the top row of Fig.~\ref{Runs_strong-103_VAR1525-VAR1540-R3}) is highly
time-dependent on rather short time scales of usually less than a day,
including change of symmetry and polarity reversals, and varying magnetic field
strength. (In order to achieve a stellar field strength as high as $1\kG$ and
higher, one would need to increase the dynamo parameters such that the field
oscillations would probably have unrealistically high frequencies.)
Without restricting boundary conditions in the star, the rotation period
of the star is rather low, having values of a few days at the surface
around the equator.
As a consequence, an overall fast stellar wind develops in the stellar dynamo
models.
This wind is supersonic and super-Alfv\'enic, with terminal outflow speeds of
a few hundreds of km/s.

In the magnetospheric models of vRB04, where a dipole magnetosphere is
anchored in the star, the rotation profile in the star is fixed in time to
rather large rotation periods; in Model~S of vRB04, the minimum rotation period
in the star is about 10 days.
As a consequence, the whole stellar wind is slow in these magnetospheric models
of vRB04, namely of the order of a few tens of km/s.
Model~S of the present paper differs in that the angular velocity as well as
the azimuthal magnetic field
can now freely evolve also within the anchoring region (i.e. within the star and
around the stellar surface.) As a result, parts of the stellar surface
have a rotation period of only a few days; they are located at mid-latitudes.
These are the latitudes where
a fast wind is launched from the star, with terminal speeds comparable to those
of the fast inner disc wind (around 200 km/s).
High latitudes rotate more slowly and
the inner stellar wind from these latitudes remains slow (about 30 km/s);
see the lower left panel of Fig.~\ref{Runs_strong-103_VAR1525-VAR1540-R3}.
We will discuss the launching and acceleration mechanisms of
the winds and the angular momentum transport in more detail in
Sect.~\ref{DrivingMechanisms}.

We checked that in Model~S, the closed stellar field lines
rotate nearly rigidly with the star.
We also note that these field lines tend to show small temporal
oscillations in the angular velocity.
Whereas the angular velocity in the gap is around $1$ (so the corotating
disc radius is also around $1$), the outermost closed field line that just
penetrates the inner disc edge is rotating fastest
with the Keplerian angular velocity of the inner disc edge ($\Omega \approx 3$);
we return to this below in Sect.~\ref{Collimation}.
In contrast to the stellar dynamo models, the imposed dipole magnetosphere
is oscillating close to periodically with a period of roughly 20 days. The
oscillation period could be related to the advection time through the disc,
which is around 20 days because the disc accretion velocity is around 10 km/s.

Another important difference between the models with stellar dynamo and
with anchored stellar dipole is the way in which the stellar field connects
with the field in the disc.
In both models, the fields from star and disc arrange themselves
in such a way that 
no {\sf X}-point forms. In the anchored stellar dipole model this
leads necessarily to the formation of pronounced hot current sheets above and
below the midplane (vRB04).
This is not the case in the stellar dynamo models, where the
stellar field geometry is sufficiently more complex and able to adjust itself
such that no current sheets form (cf.\ upper and lower rows of
Fig.~\ref{Runs_strong-103_VAR1525-VAR1540-R3}).

The disc field and disc wind properties, like field strength and geometry,
as well as wind speed and mass loss rate, are fairly similar. This is expected
because the disc is modeled in the same way except for small changes in
the disc dynamo parameters.

\subsection{Magnetic star--disc coupling and accretion}

\begin{table*}[t!]
\caption[]{Results for accretion and wind properties of the main models
discussed in this paper. Note that we are interested in matter that is accreted
{\it onto the star}, so the mass accretion rate
$\dot{M}_{\rm accr}$ is given near the stellar radius.
Since accretion is highly episodic and irregular in time, the values of
$\dot{M}_{\rm accr}$ are peak values.
Typical values for the accretion flow velocity are given in the disc
as well as
in the gap.
Wind speeds are terminal speeds (in our computational domain, excluding possible
boundary effects).
$\dot{M}_{\rm wind}=\dot{M}_{\rm stellar \, wind} + \dot{M}_{\rm disc \, wind}$
is the total wind mass loss rate.
The accretion rate $\dot{M}_{\rm accr}$ and the wind mass loss rates
$\dot{M}_{\rm stellar \, wind}$ and $\dot{M}_{\rm disc \, wind}$ are in units
of $M_\odot\yr^{-1}$, whereas the accretion flow velocity $u_{\rm accr}$
and the wind speeds $u_{\rm stellar \, wind}$ and $u_{\rm disc \, wind}$
are in units of $\kms$.
The asterisk indicates that the
values are for inner and outer stellar winds in Model~S.
}
\label{tab2}
\vspace{12pt}
\centerline{
\begin{tabular}{l|ccc|cc|cc|c}
Model &
$\dot{M}_{\rm accr}$ &
$u_{\rm accr}^{\rm disc}$ & $u_{\rm accr}^{\rm gap}$ &
$\dot{M}_{\rm stellar \, wind}$ & $u_{\rm stellar \, wind}$ &
$\dot{M}_{\rm disc \, wind}$ & $u_{\rm disc \, wind}$ &
$\dot{M}_{\rm accr}/\dot{M}_{\rm wind}$
\\
\hline
\hline
%Run1a: No stellar dynamo or anchored magnetosphere but disc dynamo
N   & $2           \times 10^{-9}$ & $10$        & $<1$ &
      $4           \times 10^{-9}$ &  10 &
      $(1.5\dots2) \times 10^{-7}$ & 350 &
      $1:100$ \\
%
%strong: Strong anchored magnetosphere and disc dynamo
S   & $(1\dots2.5) \times 10^{-8}$ & $ 5\dots15$ & $15$ &
      $2           \times 10^{-8}$ & 30 / 200 * &
      $(2\dots3)   \times 10^{-7}$ & 250 &
      $1:10$ \\
%
%Run3: Reference model with stellar dynamo and disc dynamo
Ref & $(0.5\dots1) \times 10^{-7}$ & $10$        & $60$ &
      $(1\dots2)   \times 10^{-7}$ & 400 &
      $(1\dots2)   \times 10^{-7}$ & 250 &
      $1:4$ \\
%
%Run5: Stronger Disc dynamo compared to reference model
SDd & $(2\dots 6)  \times 10^{-7}$ & $10\dots20$ & $75\dots150$ &
      $(2\dots 6)  \times 10^{-7}$ & 450 &
      $(4\dots12)  \times 10^{-7}$ & 250 &
      $1:3$ \\
%
%Run6b: No Disc dynamo but stellar dynamo
NDd & negligible                 & $ 1\dots 3$ & negligible &
      $3         \times 10^{-7}$ & 300 &
      $2         \times 10^{-7}$ & 200 &
      negligible \\
\end{tabular}
}
\end{table*}

In our stellar dynamo models, for the stellar magnetic field to be strong enough
to penetrate the disc in the presence of a disc dynamo,
one needs a very strong stellar dynamo to create field strengths
of more than $1\kG$. In this case, however, the stellar wind is too strong to
allow for accretion, in particular to allow for a funnel flow along the stellar
field lines leading to polar accretion. Therefore, in our stellar dynamo models
an efficient accretion flow is only possible in mainly equatorial direction.
Also in the models with imposed dipole magnetosphere of vRB04, star--disc
coupling by the stellar magnetosphere requires a stellar surface field that is
stronger than $1\kG$. But in contrast to the stellar dynamo models, most of
the stellar wind remains weak so that polar accretion along the magnetospheric
field lines is possible in the stellar dipole models.

Even in the absence of a disc field,
a stellar field strength of at least about $500\G$ is necessary
in order to produce a coupling between star and disc by the stellar dynamo
field;
this is discussed further in Sect.~\ref{DrivingMechanisms}.
This leads to sufficient stellar wind speeds of a few hundred $\kms$, but
suppresses the already weak accretion.
Weaker stellar fields produce slower stellar winds.
However, the magnetic coupling to the disc is then essentially absent.

In order to get a coupling of star and disc, we made the disc dynamo
stronger.
In Model~SDd, the disc dynamo parameters are doubled compared to our reference
model (Model~Ref), leading to rms disc field strengths that are about 1.5 times
larger (around $40\G$). At the same time, the $\alpha$ effect
in the star is reduced by a third, which leads to somewhat smaller rms stellar
field strengths ($\le 200\G$).

Figure~\ref{pnew_p_R5} shows two different states regarding the star--disc
coupling. On the left hand side, star and disc are magnetically connected by
field lines generated by the {\it disc} dynamo, which allows for equatorial
accretion of disc material along those lines onto the star; these field lines
are very weak in the gap between the disc and the star,
where the plasma beta is above 100.
The top accretion rates are around
$\dot{M}_{\rm accr} \approx (2\dots6) \times10^{-7}M_\odot\yr^{-1}$,
which is comparable to the minimum of the total stellar and disc wind mass loss
rates, the latter assuming values of around
$\dot{M}_{\rm wind} \approx (6 \dots 18) \times10^{-7}M_\odot\yr^{-1}$.
The accretion flow velocity in the gap is rather high, up to $75\kms$ and
occasionally up to $150\kms$, but still considerably smaller than free-fall.

On the right hand side of Fig.~\ref{pnew_p_R5}, the magnetic field in the
gap -- again originating from the disc dynamo -- is almost vertical and
strong enough
to act as a shield for the star, not allowing disc matter to cross field lines.
This causes star and disc
to be disconnected in any way. In this state, disc matter leaving the inner edge
is diverted into the inner disc wind, and there is no accretion.

In Fig.~\ref{Om_mag_R5} we show
radial profiles of the magnetic field strength and the angular velocity in the
midplane of Model~SDd at the same two times as in Fig.~\ref{pnew_p_R5}.
The solid (red) curves correspond to a state when the field in the gap between
the star and the disc is weak and hence the Keplerian angular velocity
from the disc can extend almost all the way to the star.
The other extreme corresponds to the dashed (blue) curves where the field in the
gap is strong and the angular velocity in the gap is suppressed.

Table~\ref{tab2} summarizes the results for the accretion and wind properties
of the main models discussed in this paper. The main points are the following:
First, in our models with stellar as well as disc dynamo, especially in
Model~SDd, the accretion flow velocity reaches -- for the first time in our set
of models -- values that agree much better with observations of around $100\kms$
or higher.
The larger accretion flow velocity might be due to the much larger plasma beta
in the accretion region in the gap.
Second, in these models, the mass accretion rate becomes comparable to the total
wind mass loss rate.
Third, the stellar wind mass loss rate becomes comparable to the disc wind
mass loss rate in the presence of a stellar dynamo.
Forth, the accretion flow through the disc is clearly slower
when there is no disc field. This shows that a magnetic torque
in the disc is important as a driver of angular momentum transport outward,
for the accretion flow through the disc to be efficient.
Indeed, in the absence of a disc dynamo,
the disc wind does not carry much angular
momentum (see Sect.~\ref{DrivingMechanisms}),
and therefore there is less accretion through the disc.
Accretion flow in the gap
and mass accretion rate onto the star are negligible in Model~NDd,
because the stellar wind
is already too strong.

\subsection{Driving and structure of stellar and disc winds}
\label{DrivingMechanisms}

\begin{figure*}[t!]
\centerline{
   \includegraphics[width=8.5cm]{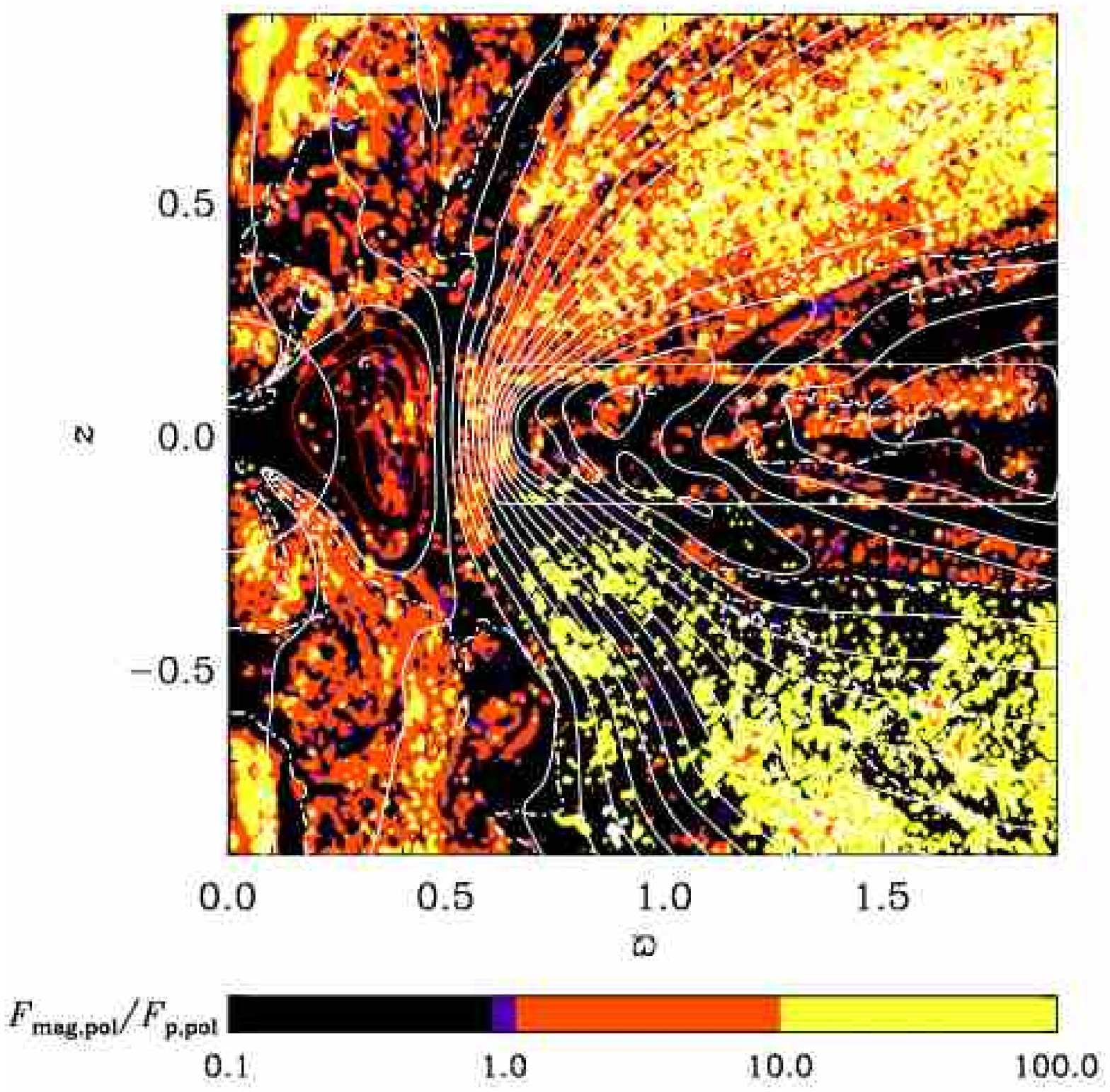}
   \includegraphics[width=8.5cm]{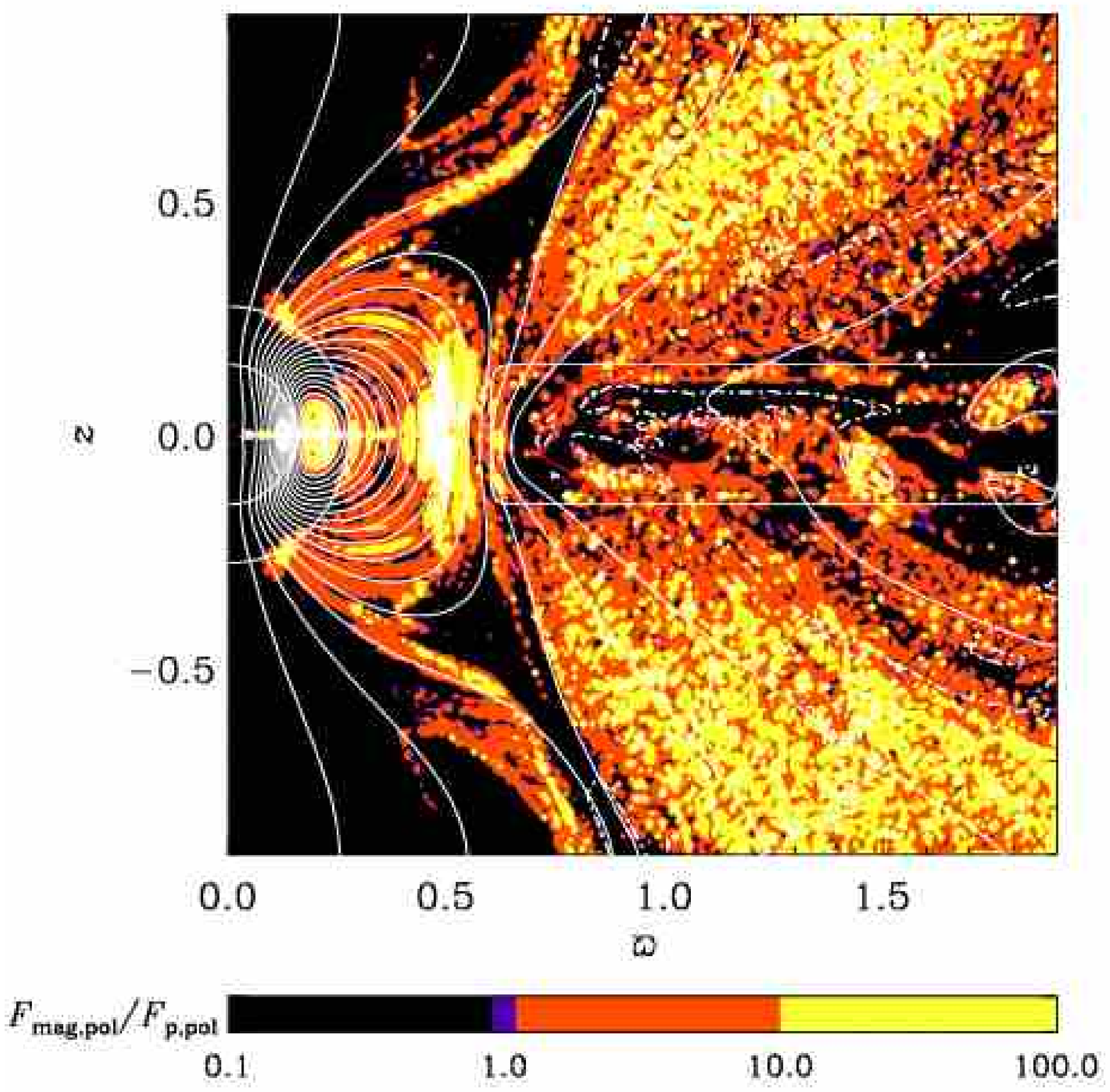}
}
\caption[]{
The force ratio, $|F^{\rm(mag)}_{\rm pol}|/|F_{\rm pol}^{\rm(p)}|$,
is shown everywhere, including the disc,
but not in the inner part of the star.
Regions with largest Lorentz force along the poloidal field
correspond to regions of highest angular momentum transport outward,
and thus to regions of magneto-centrifugal acceleration.
These regions are also those with lowest temperature and lowest density.
Plasma beta is low and wind speeds are high in these regions, as well.
White dash-dotted: fast magnetosonic surface.
{\it Left}: Model~Ref.
{\it Right}: Model~S, where the outer stellar wind is very similar to
the inner disc wind.
}\label{pEBforce_R3_strong}
\end{figure*}

As we have seen in Fig.~\ref{pnew_R3}, in the case with a stellar
dynamo there is a strong stellar wind.
In order to clarify its origin we need to look at the force balance.
For the magnetic acceleration,
we consider the Lorentz force and the gas pressure force, both along
the poloidal field whose unit vector is
$\hat{\BB}_{\rm pol}=\BB_{\rm pol}/|\BB_{\rm pol}|$, i.e.\
we consider the forces
\begin{eqnarray}
F^{\rm(mag)}_{\rm pol}=\hat{\BB}_{\rm pol}\cdot(\JJ\times\BB),\quad
F^{\rm(p)}_{\rm pol}=-\hat{\BB}_{\rm pol}\cdot\vec{\nabla}p.
\end{eqnarray}
In an axisymmetric system, if the magnetic field were purely poloidal,
the Lorentz force would point in the poloidal direction that is perpendicular
to $\BB_{\rm pol}$, and therefore it would be $F^{\rm(mag)}_{\rm pol}=0$.
This can be seen explicitly by writing down the detailed expression
for $F^{\rm(mag)}_{\rm pol}$ in axisymmetry,
\begin{eqnarray}
F^{\rm(mag)}_{\rm pol}&=&{\hat{\BB}_{\rm pol} \over \mu_0} \cdot
                       \left[
                       \left(
                       \vec{\nabla}\times B_\varphi\vec{e}_\varphi
                       \right)
                       \times B_\varphi\vec{e}_\varphi
                       \right] \\
&=& -{B_\varphi \over \mu_0 \sqrt{B_\varpi^2+B_z^2}}
                          \left(
                          {B_\varpi \over \varpi}
                          {\partial \over \partial\varpi}
                          \left(\varpi B_\varphi\right)
                          + B_z {\partial B_\varphi \over \partial z}
                          \right) \nonumber
\label{EBforce}
\end{eqnarray}
which vanishes for $B_\varphi=0$. Here, $\vec{e}_\varphi$ is the unit vector
in azimuthal direction.
However, in a rotating magnetized star--disc system there is of course
a strong toroidal component of the magnetic field in certain regions, leading
to $F^{\rm(mag)}_{\rm pol}$ being different from zero in those regions.
In Fig.~\ref{pEBforce_R3_strong} we plot the ratio
$|F^{\rm(mag)}_{\rm pol}|/|F_{\rm pol}^{\rm(p)}|$.
Because of how this ratio is defined, regions where this ratio is large
indicate regions where magnetic acceleration is strong.

\begin{figure*}[t!]
\centerline{
   \includegraphics[width=8.5cm]{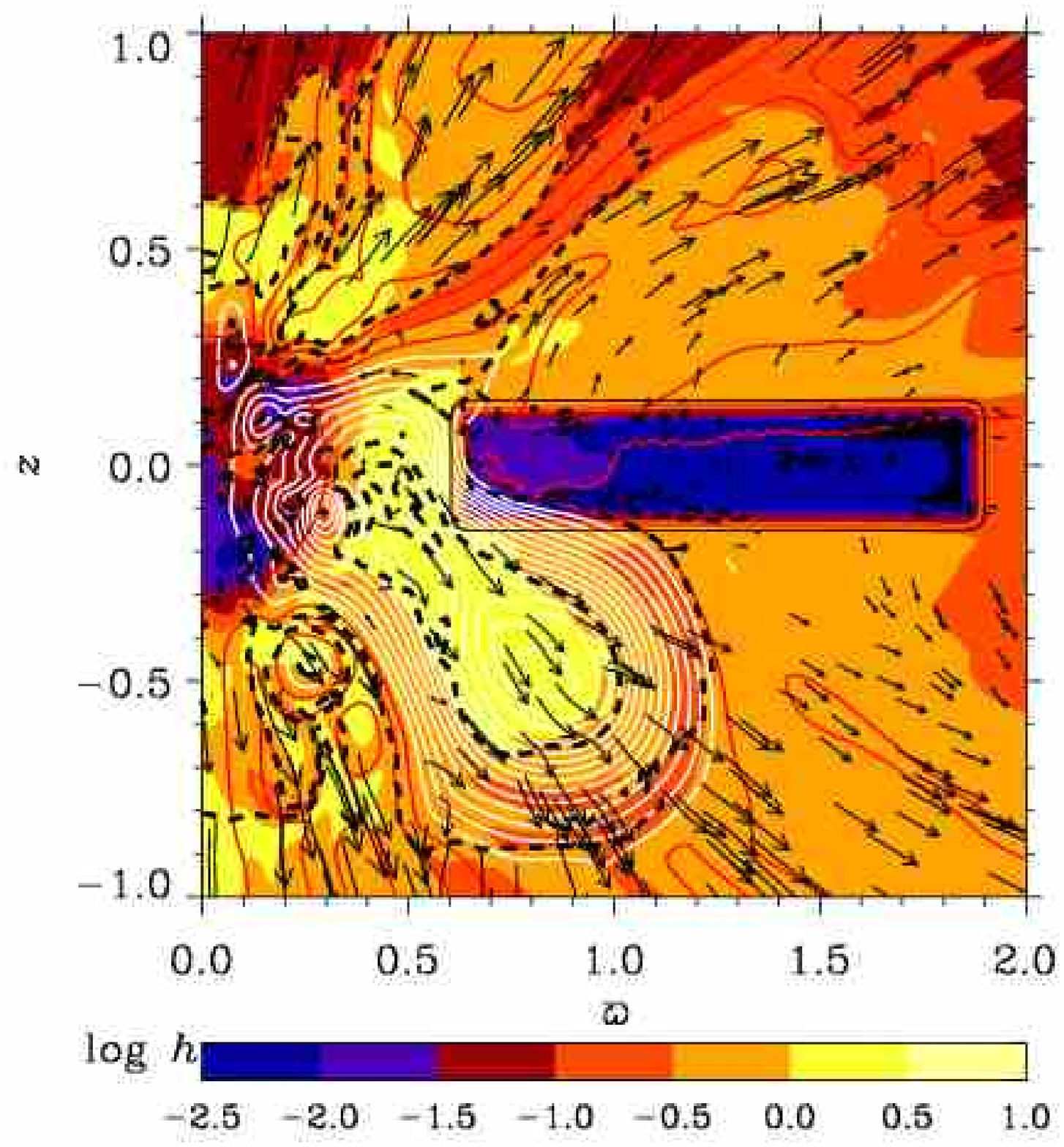}
   \includegraphics[width=8.5cm]{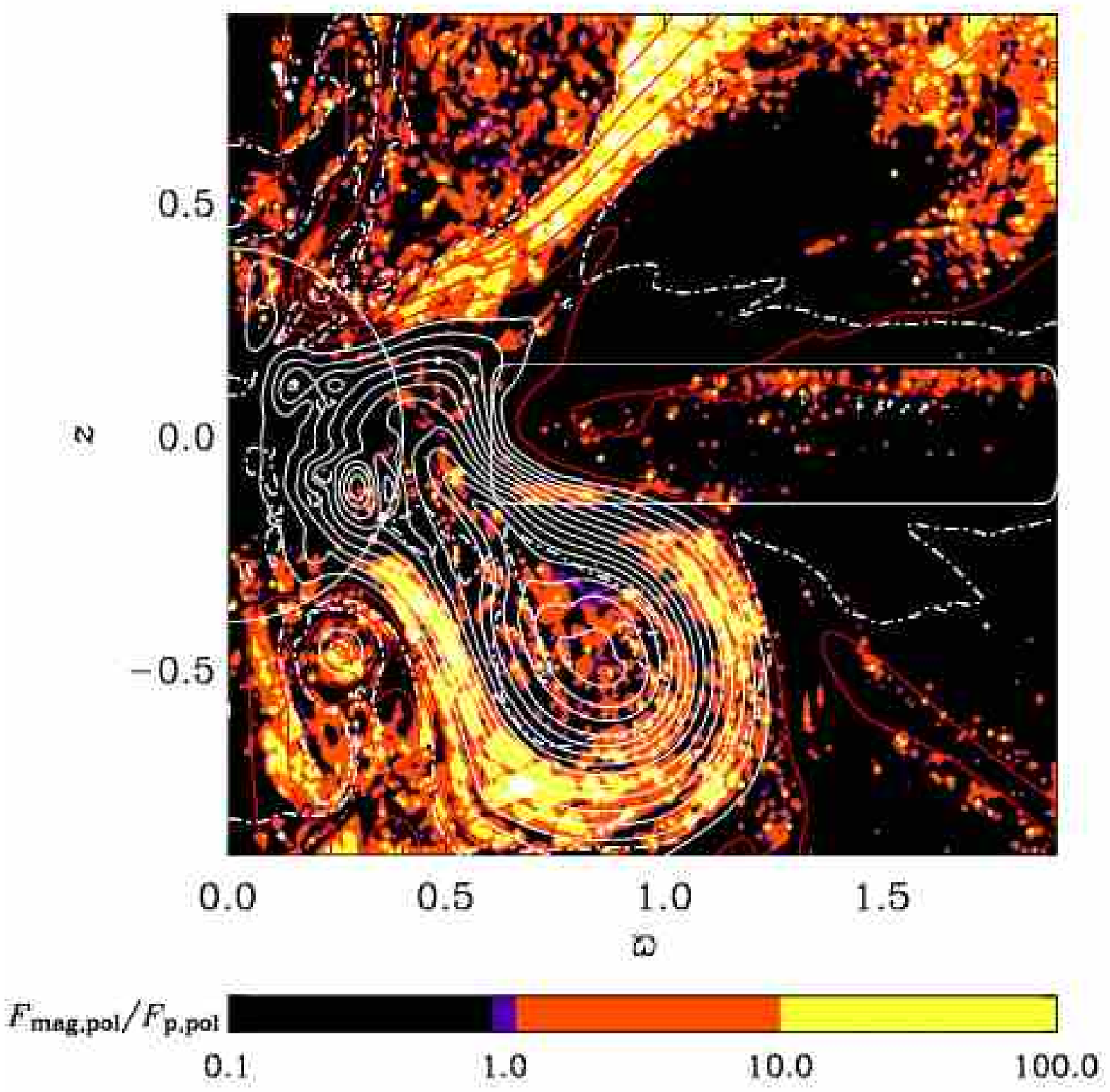}
}
\caption[]{
Model~NDd (no disc dynamo) at a time $t \approx 246$ days.
Note that if there is no ``disturbing'' disc field, then
star--disc coupling is possible for a stellar field strength of a few hundred
gauss, but there is virtually no accretion of disc matter onto the star.
{\it Left}:
Cf.\ Fig.~\ref{R1aR1b}.
{\it Right}:
Color/gray scale representation of the force ratio (also shown in the disc
but not in the inner part of the star),
together with poloidal magnetic field lines (white/red [gray] means
right/left-handed).
The white dash-dotted line is the fast magnetosonic surface.
As in Hayashi et al.\ (1996), hot plasmoids are ejected with a stellar wind
speed of about $300 \kms$ which is more than twice the Keplerian speed at
the inner disc edge.
}\label{pnew_pEBforce-159.85860-ori-R6b}
\end{figure*}

Concerning the disc wind,
note the light shades especially above and below the disc
in the region of field lines originating from the inner disc edge,
indicating that magnetic acceleration is strong there.
This ``conical shell'' corresponds also to the region of largest
angular momentum transport away from the disc, so that
the conical inner disc wind is magneto-centrifugally accelerated
along the field lines.
This conical structure
requires a disc dynamo that is saturated so that the field
above and below the disc is well-organized.
The structured disc wind due to the disc dynamo has been discussed
in detail in von Rekowski et al.\ (2003);
it is essentially independent of the modeling of the stellar field.
No magneto-centrifugal driving of the inner disc wind occurs when
the disc is virtually non-magnetized, i.e.\ the disc dynamo is switched off
($\alpha_0^{\rm disc}=0$); see
Fig.~\ref{pnew_pEBforce-159.85860-ori-R6b}, right hand panel.
Here, the regions above and below the disc are predominantly black
and correspond to the regions with basically no angular momentum transport,
which is suggestive of mostly pressure driving
of the whole weakly magnetized disc wind.

\begin{figure*}[t!]
\centerline{
   \includegraphics[width=8.5cm]{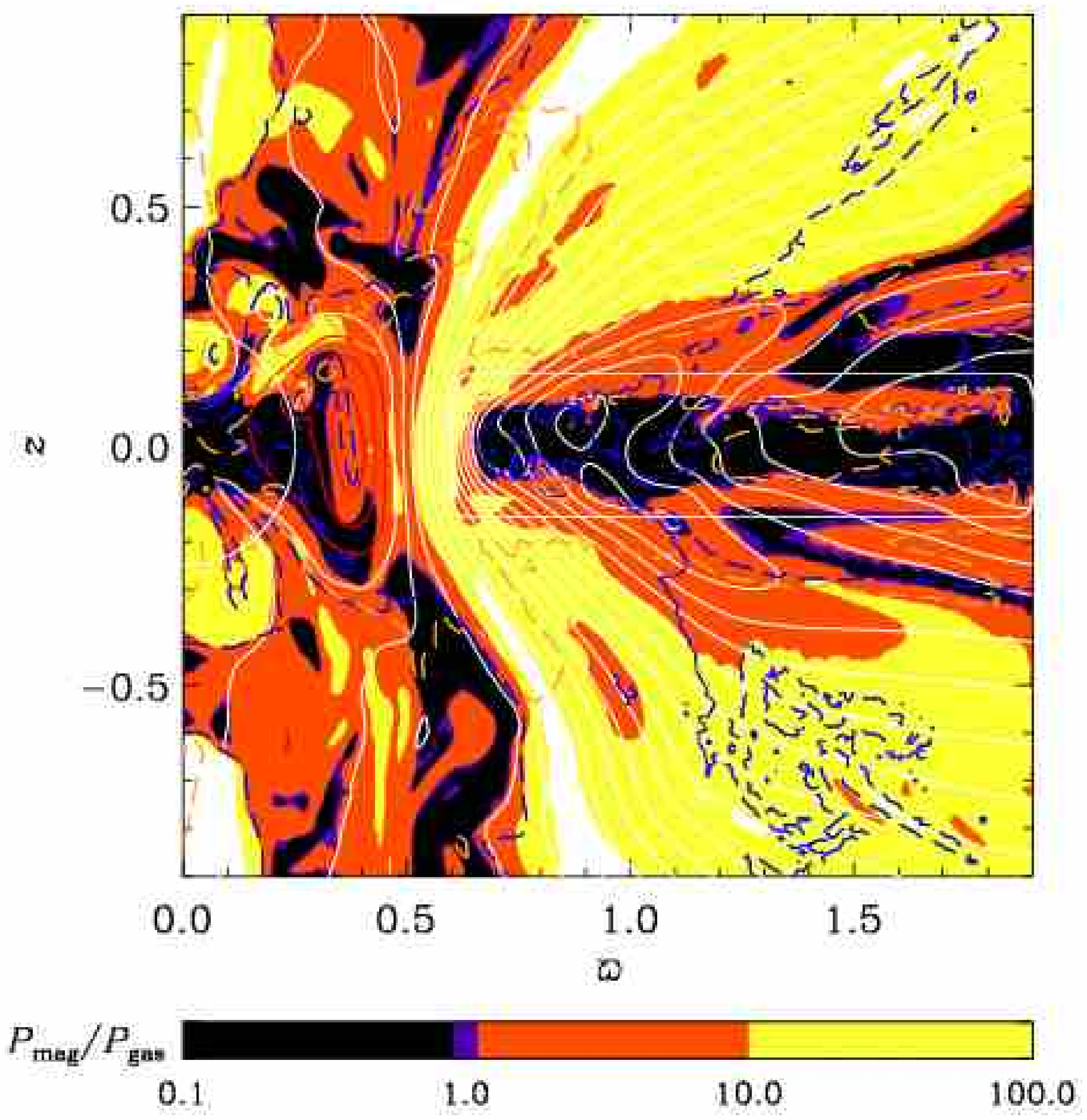}
   \includegraphics[width=8.5cm]{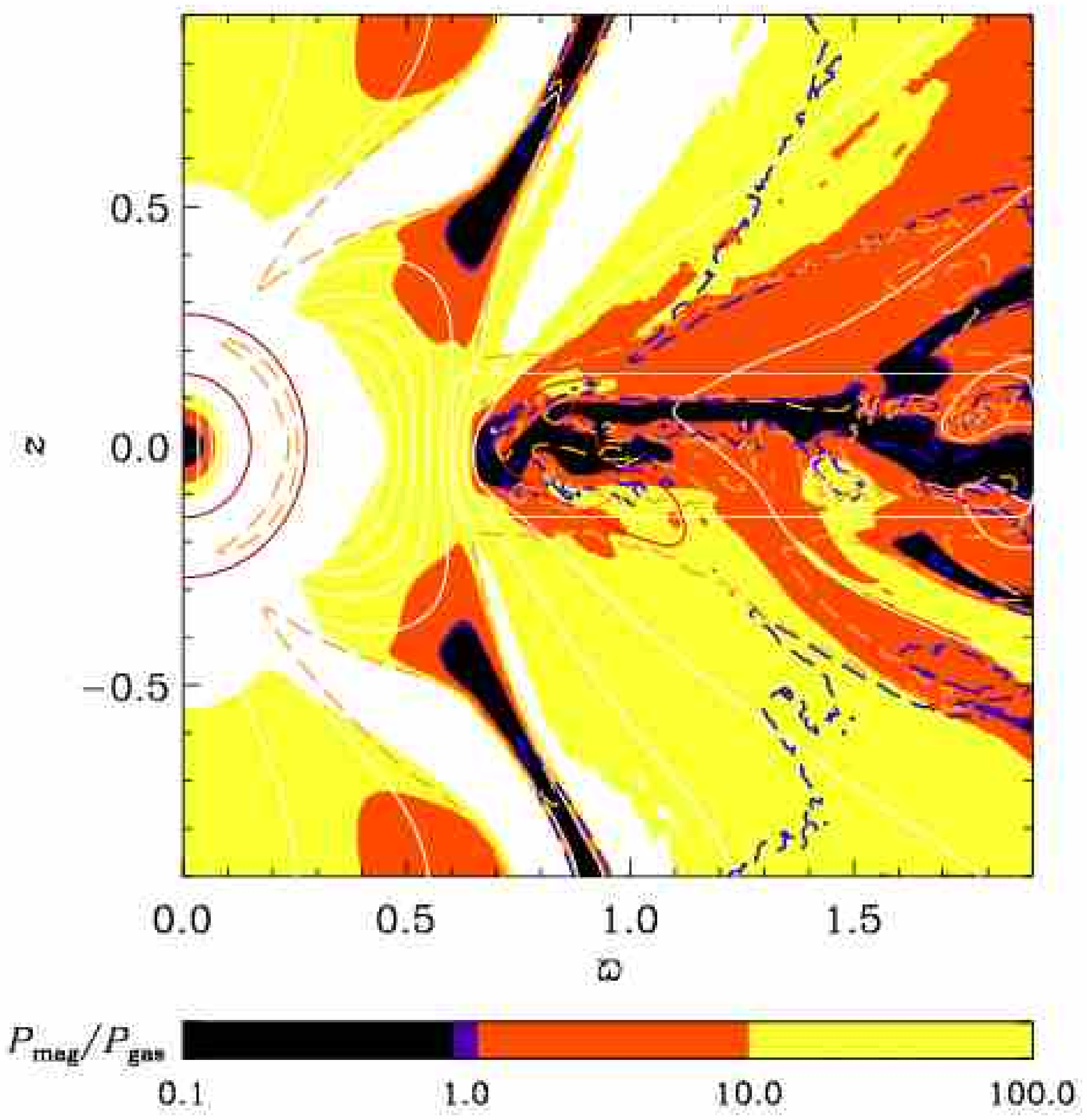}
}
\caption[]{
The inverse plasma beta
is also shown in the disc but not in the inner part of the star.
Note the region of low plasma beta also in the region of the inner stellar wind
in Model~S ({\it right hand panel}).
There, one has only the poloidal magnetic pressure; the toroidal field and
the angular velocity are very small.
Blue \& Orange dashed: Alfv\'en \& sonic surfaces, respectively.
{\it Left}: Model~Ref.
{\it Right}: Model~S, where the outer stellar wind is very similar to
the inner disc wind.
}\label{pEBbeta1_R3_strong}
\end{figure*}

The nature of the stellar wind depends strongly on the field configuration
of the stellar magnetic field.
This can be seen in Fig.~\ref{pnew_pEBforce-159.85860-ori-R6b} which
shows a time snapshot of a model where there
is only a stellar dynamo; the disc dynamo is turned off.
Above and below the disc, there are narrow structures at about $45^\circ$
angle where magneto-centrifugal driving is important.
A similar structure is also present in the model with an anchored dipole;
see the right hand panel of Fig.~\ref{pEBforce_R3_strong}.
Here, the magnetic field can act as an efficient lever arm that is
required for acceleration.
All magneto-centrifugally accelerated winds are supersonic
but the Alfv\'en surface is further away.
Angular momentum is transported along field lines (originating
from the disc) that make an angle away from
the vertical axis that exceeds $30^\circ$ (Blandford \& Payne 1982).

In Fig.~\ref{pEBbeta1_R3_strong} we plot the inverse plasma beta,
$\beta^{-1} \equiv P_{\rm mag} / P_{\rm gas} = |B|^2/(2 \mu_0 p)$, where
the symbol $P_{\rm gas} = p$ has been used.
It is noticeable that in the stellar dipole model, the inner stellar wind
has very low plasma beta but no poloidal component of the Lorentz force
along the field lines. This is because the magnetic pressure due to the
{\it poloidal} field exceeds the gas pressure.

\begin{figure}[b!]
\centerline{
   \includegraphics[width=8.5cm]{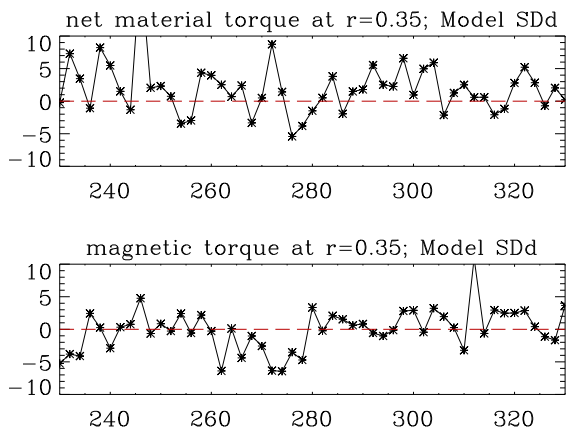}
}
\caption[]{
Model~SDd.
$T_{\rm mat,net}(r)$ and $T_{\rm mag}(r)$ at fixed spherical radius $r=0.35$
(close to the star)
for various times.
The mean amplitude of the net material torque is comparable
to that of the
magnetic torque,
but both torques are highly
time-dependent and change sign in an irregular fashion.
Time is normalized, with $302$ and $305$ corresponding to
the times of Fig.~\ref{pnew_p_R5}: $464$ days and $469$ days, respectively.
}\label{times_torques_R5}
\end{figure}

\begin{figure*}[t!]
\centerline{
   \includegraphics[width=8.5cm]{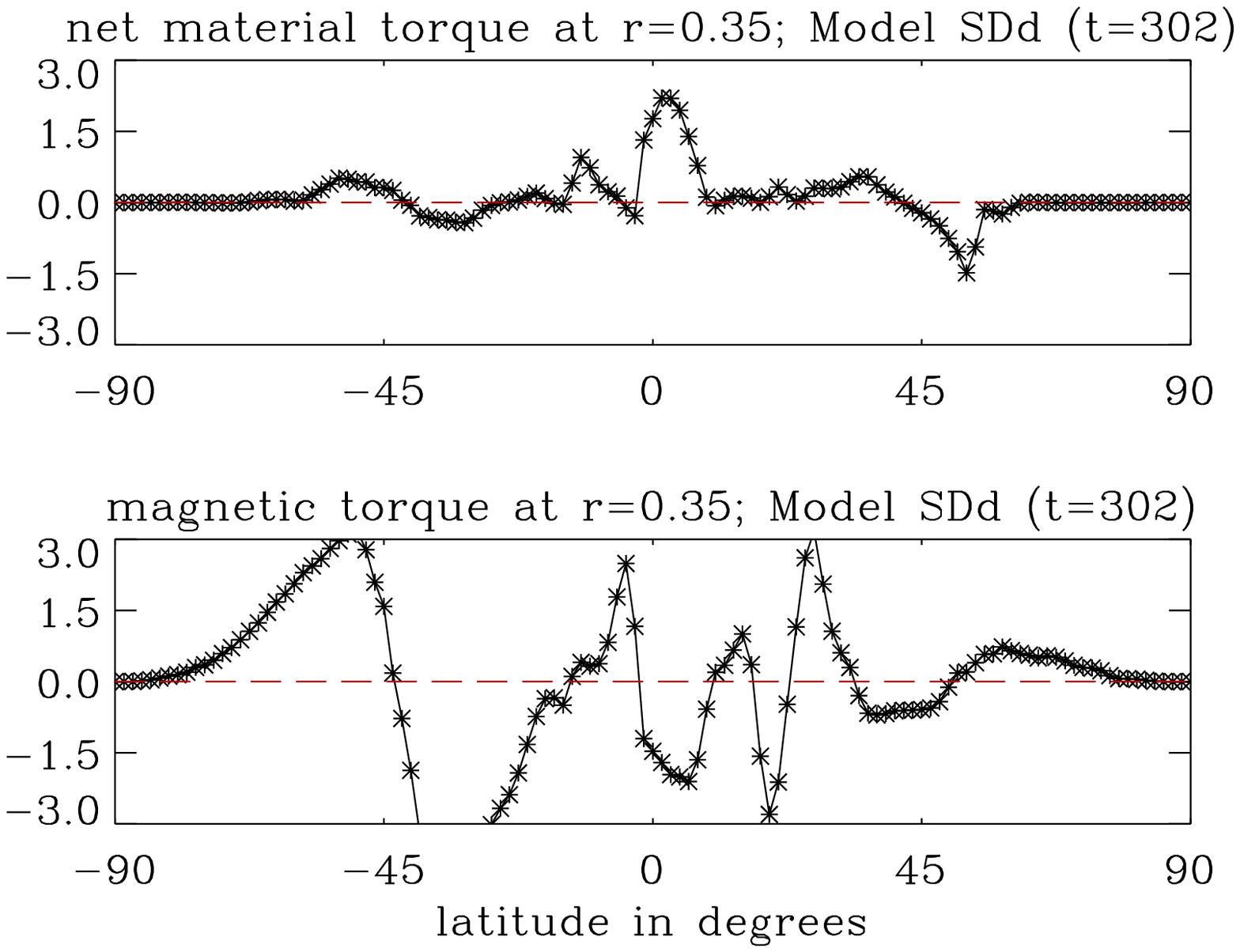}
   \includegraphics[width=8.5cm]{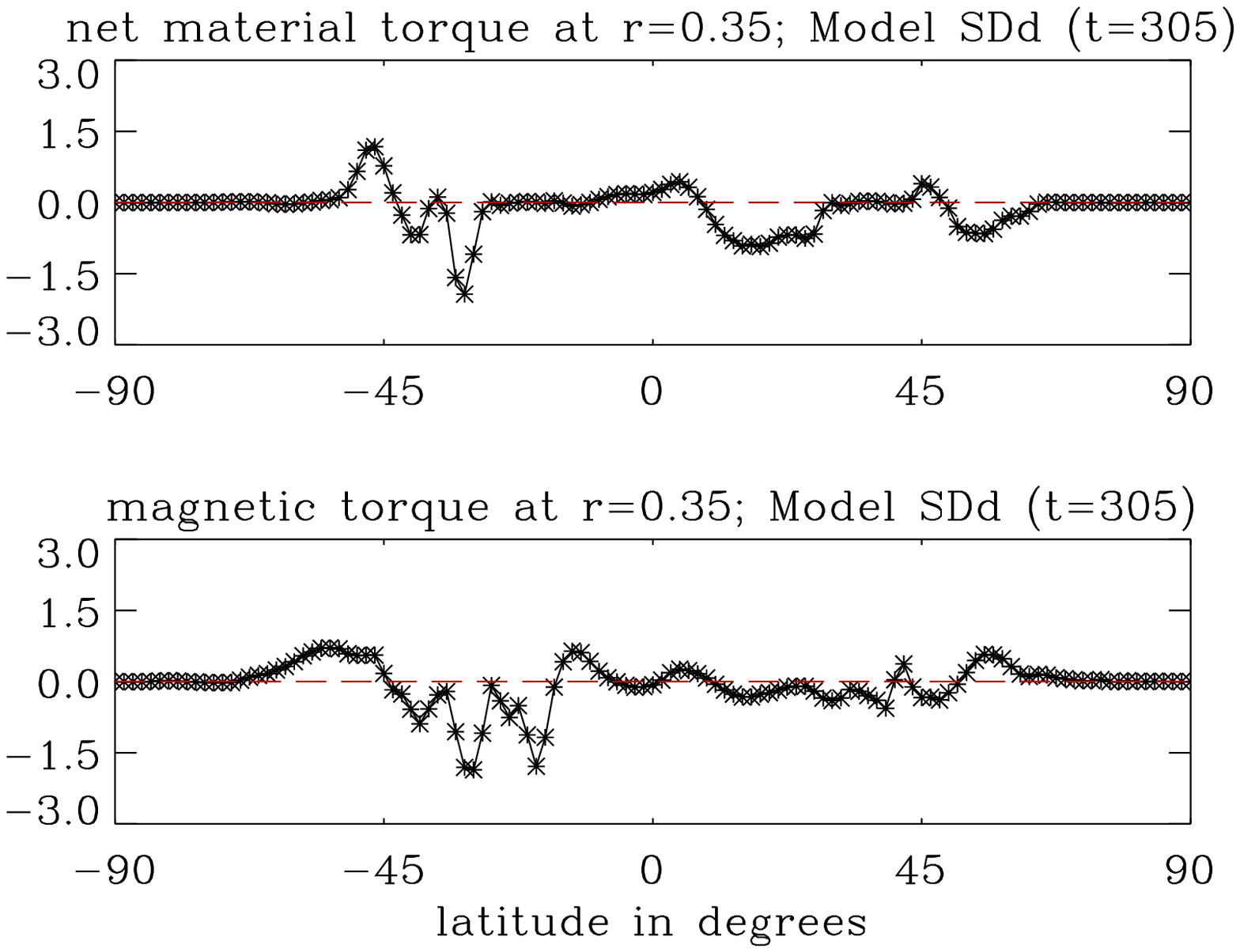}
}
\caption[]{
Model~SDd.
$\tilde{T}_{\rm mat,net}(r,\Theta)$ and $\tilde{T}_{\rm mag}(r,\Theta)$
at fixed spherical radius $r=0.35$ (close to the star) as functions
of latitude at two different times,
$t\approx 464$ days ({\it left}: connected state) and
$t\approx 469$ days ({\it right}: disconnected state); cf.\ Fig~\ref{pnew_p_R5}.
Net material torque and magnetic torque are comparable in amplitude.
{\it Left}:
At this time, the latitudinally integrated net material torque
($T_{\rm mat,net}(r=0.35)$)
is positive and the latitudinally integrated magnetic torque
($T_{\rm mag,net}(r=0.35)$)
is negative.
Low-latitude accretion leads to a material spin-up.
Higher-latitude stellar wind leads to a material spin-down but also to a
material spin-up in places.
{\it Right}:
At this time, $T_{\rm mat,net}(r=0.35) < 0$ and $T_{\rm mag,net}(r=0.35) < 0$.
There is no accretion of disc matter,
and the stellar wind generally leads to a material spin-down.
}\label{Theta_torques_R5}
\end{figure*}

In summary, it turns out that in the stellar dynamo models,
the overall stellar wind is driven by magnetic and gas pressure.
This wind is hot, dense, very fast and rotating with a period $P_{\rm rot}$
of a few days.
However, in places where field lines assume a configuration such that they can
act as a lever arm, magneto-centrifugal acceleration becomes possible.
The magnetic field geometry of the stellar field generated
by the dynamo in the star is complex with an irregular temporal behavior,
and configurations favorable for magneto-centrifugal acceleration are
assumed locally for short times; the exact behavior is sensitive to the
stellar dynamo parameters.
In the dipole model, the stellar wind is clearly structured. Here
the inner stellar wind is mostly driven by poloidal magnetic pressure.
This wind is hot, dense, slow and rotating with $P_{\rm rot} \approx$ 10~days.
The outer stellar wind is magneto-centrifugally
accelerated and very similar to the inner disc wind
(cooler, less dense, fast and rotating with $P_{\rm rot} \approx$ 3~days).
The magneto-centrifugally accelerated regions of the winds generally correspond
to the coolest wind regions and have temperatures between $10^4$ and $10^5$
Kelvin. In the dipole model, the outer stellar wind and the inner disc wind
are separated by a hot current sheet of about $10^6\K$.

As remarked before, the imposed stellar dipole field yields very little
toroidal field in the plasma above the star, giving a poloidal
component of the Lorentz force along the field lines only in a narrow strip.
A stellar dynamo produces a more irregular field,
and there can be a significant gradient in the toroidal magnetic field,
giving rise to the possibility of magneto-centrifugal driving
of cooler and less dense wind structures.

\subsection{Stellar spin-up or spin-down?} \label{updown}

In order to address the question of spin-up versus spin-down of the star,
we need to calculate the sum of the various torques exerted on the star,
i.e.\ the {\it net} material torque
($T_{\rm mat,net}$),
the magnetic torque
($T_{\rm mag}$) and the viscous torque.
The {\it net} material torque is the {\it difference} in angular
momentum flux toward the star, carried by matter, when comparing the angular
velocity of the accreting or outflowing matter, $\Omega$, and the
{\it effective} angular velocity of the stellar surface, $\Omega_*=L_*/I_*$.
Here, $L_*$ is the total stellar angular momentum, $I_* = M_* \varpi_*^2$ is
the star's effective moment of inertia, and $\varpi_*^2$ is the effective
lever arm that is, for reasons that become evident in a moment, defined as
a weighed average of the form
\EQ
\varpi_*^2\equiv
\langle\varpi^2\rangle=\left.
\oint\varrho\uu_{\rm pol}\varpi^2\cdot\dd\SSS
\right/\oint\varrho\uu_{\rm pol}\cdot\dd\SSS,
\EN
where the integrals are taken over the stellar surface.
This defines $\Omega_*$, which replaces the intuitive definition of
$\Omega_*$ as the stellar surface rotation rate (see Sect.~3.8 of vRB04
and the appendix therein).
We assume that the effective lever arm is constant in time,
which means that during
accretion or mass loss of the star matter is redistributed in such a way
that $\varpi_*^2$ is constant.
This is done because in the model the stellar interior is insufficiently
resolved and
any variation of $\varpi_*^2$ would be artificial.
It is therefore important that our spin-up or spin-down calculations
are not affected by this.

\begin{figure*}[t!]
\centerline{
 \includegraphics[width=8.5cm]{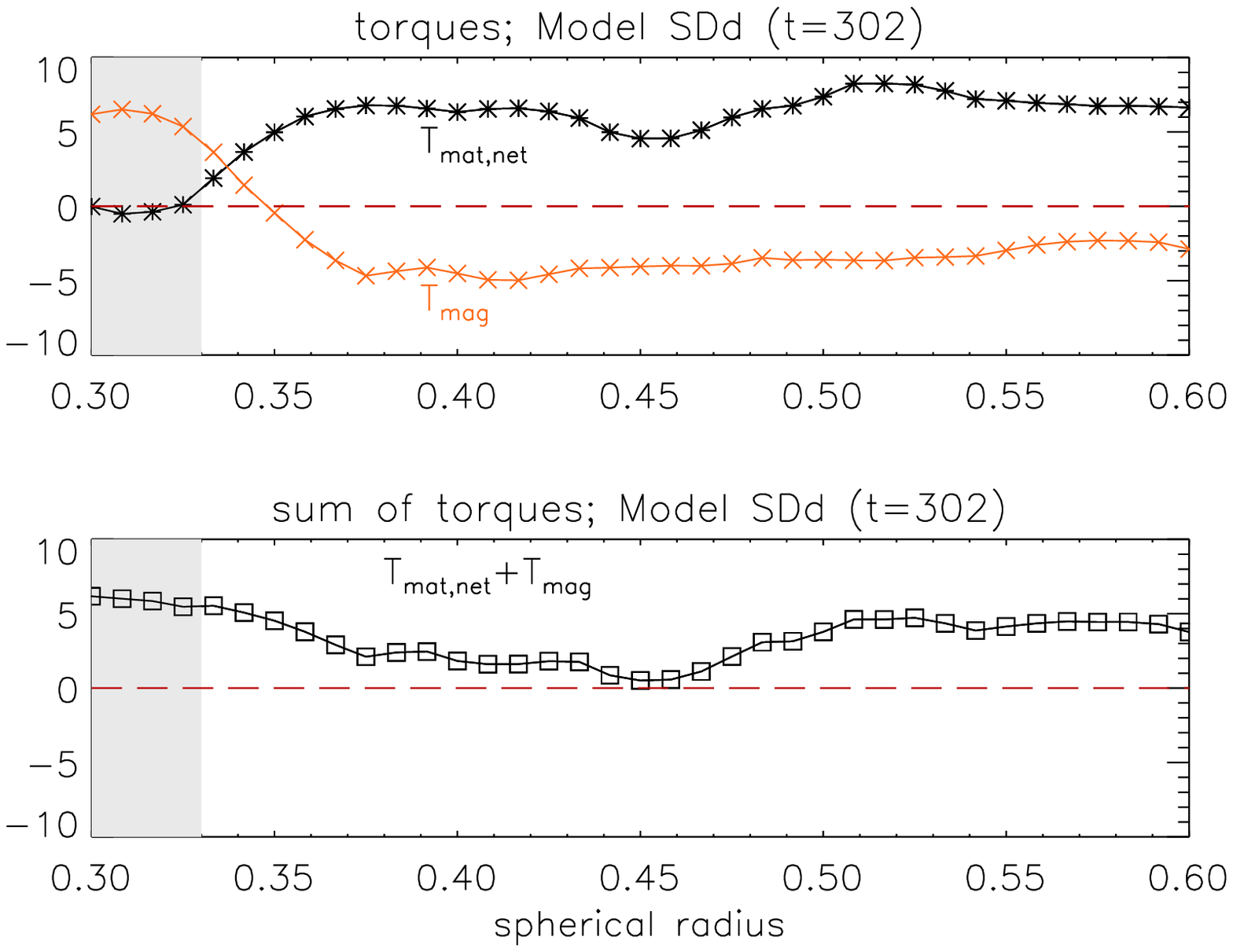}
 \includegraphics[width=8.5cm]{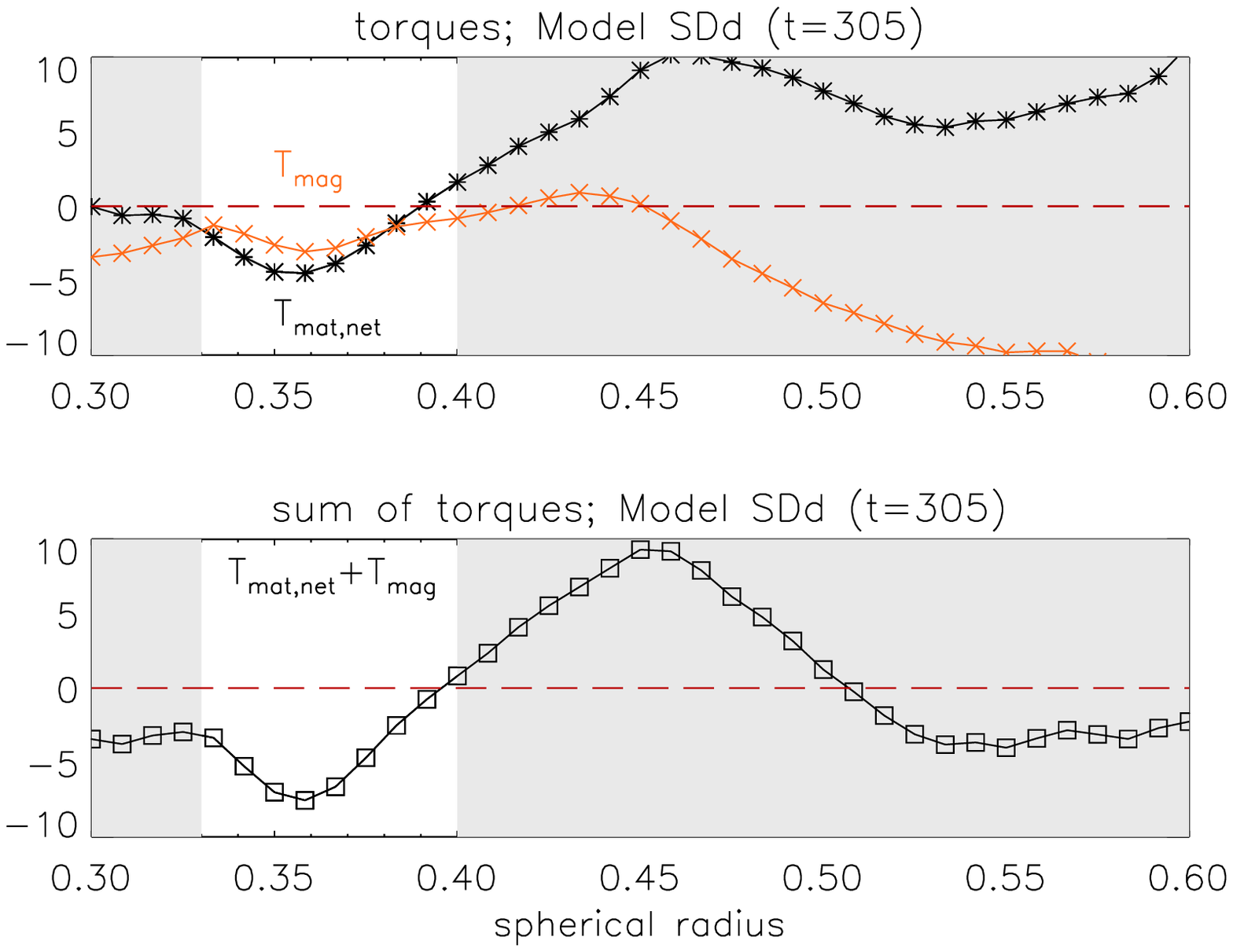}
}
\caption[]{
Model~SDd.
Radial profiles of the latitudinally integrated torques (\ref{Tacc}) and (\ref{Tmag})
at the times of Fig.~\ref{Theta_torques_R5}, $t\approx 464$ days ({\it left}:
connected state) and $t\approx 469$ days ({\it right}: disconnected state).
}\label{ppltorques_R5}
\end{figure*}

Given the definition $\Omega_*=L_*/I_*$, we can calculate its rate of change as
\EQ
\dot{\Omega}_*=\dot{L}_*/I_*-L_*\dot{I}_*/I_*^2.
\EN
Since we have assumed $\varpi_*^2=\mbox{const}$, we have
$\dot{I}_*=\dot{M}_*\varpi_*^2$.
To determine the sign of $\dot{\Omega}_*$ we consider the sign of
\EQ
I_*\dot{\Omega}_*=\dot{L}_*-L_*\dot{M}_*/M_*,
\label{IOmegadot}
\EN
where $\dot{L}_*$ is taken to be the sum of material and magnetic torques;
the viscous torque is neglected because we found that it is much smaller
than the other torques.
Using $L_*=\Omega_*M_*\varpi_*^2$
and $\dot{M}_*=-\oint\varrho\uu_{\rm pol}\cdot\dd\SSS$,
the second term on the right hand side of
\Eq{IOmegadot} can be written as
\EQ
-L_*\dot{M}_*/M_*=-\Omega_*\varpi_*^2\dot{M}_*
=\oint\Omega_*\varrho\uu_{\rm pol}\varpi^2\cdot\dd\SSS.
\EN
This term can be subsumed into the material torque by defining a
{\it net} material torque as (cf.\ vRB04)
\begin{equation}
T_{\rm mat,net}(r) = r\int_0^\pi\tilde{T}_{\rm mat,net}(r,\Theta)\,{\rm d}\Theta
\label{Tacc}
\end{equation}
with
\begin{equation}
\tilde{T}_{\rm mat,net}(r,\Theta)
= -2\pi \varpi^3 \varrho u_r (\Omega - \Omega_*).
\label{Ttildeacc}
\end{equation}
Here we have used spherical polar coordinates, $(r,\Theta,\varphi)$,
so $\varpi=r\sin\Theta$ and $\dd S=2\pi r\varpi\dd\Theta$.
The magnetic torque is defined correspondingly,
\begin{equation}
T_{\rm mag}(r) = r\int_0^\pi\tilde{T}_{\rm mag}(r,\Theta)\,{\rm d}\Theta,
\label{Tmag}
\end{equation}
where
\begin{equation}
\tilde{T}_{\rm mag}(r,\Theta)
=  2\pi \varpi^2 B_r B_\varphi/\mu_0.
\label{Ttildemag}
\end{equation}
Both radial and latitudinal torque profiles will be discussed.
The sum of both magnetic and net material torques gives
\EQ
I_*\dot{\Omega}_*=T_{\rm mat,net}(r_*) + T_{\rm mag}(r_*),
\label{spin}
\EN
and its sign determines whether the star spins up or down.
Positive means stellar spin-up (and, since $\varpi_*^2$ is constant in
time, gain in mean specific angular momentum of the star,
$l_*=\varpi_*^2\Omega_*$), whilst negative means stellar spin-down
(and loss in mean specific angular momentum of the star).

In the following we
first turn to the dynamo models, in particular to Model~SDd.
In Fig.~\ref{times_torques_R5} we show,
at a fixed spherical radius $r=0.35$ (close to the star),
the latitudinally integrated torques
$T_{\rm mat,net}(r)$ and $T_{\rm mag}(r)$
for various times.
Both torques are highly time-dependent and change sign in an irregular fashion.
It turns out that close to the star, the mean amplitude of net material torque is
comparable to that of the magnetic torque.

\begin{figure*}[t!]
\centerline{
 \includegraphics[width=8.5cm]{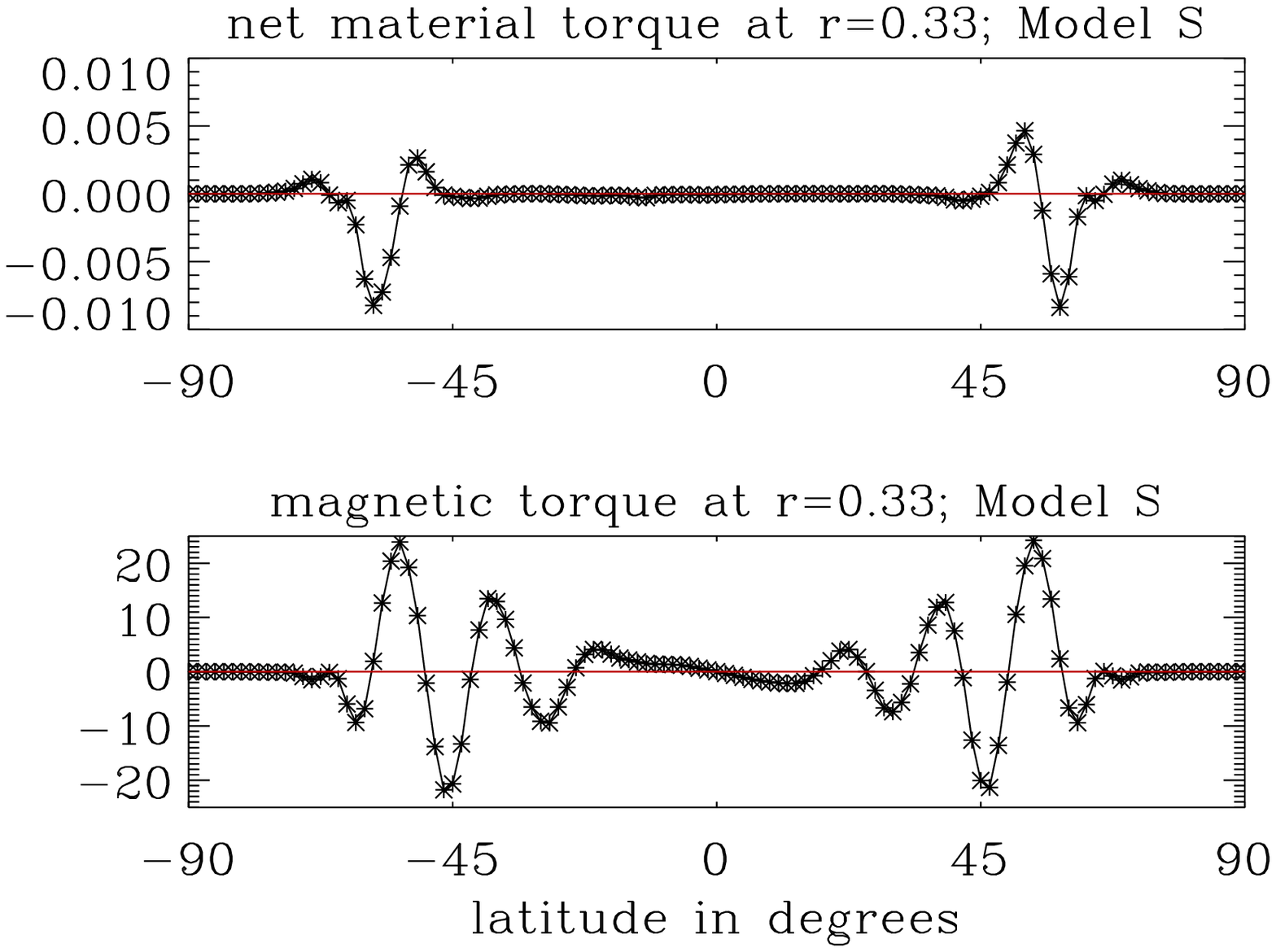}
 \includegraphics[width=8.5cm]{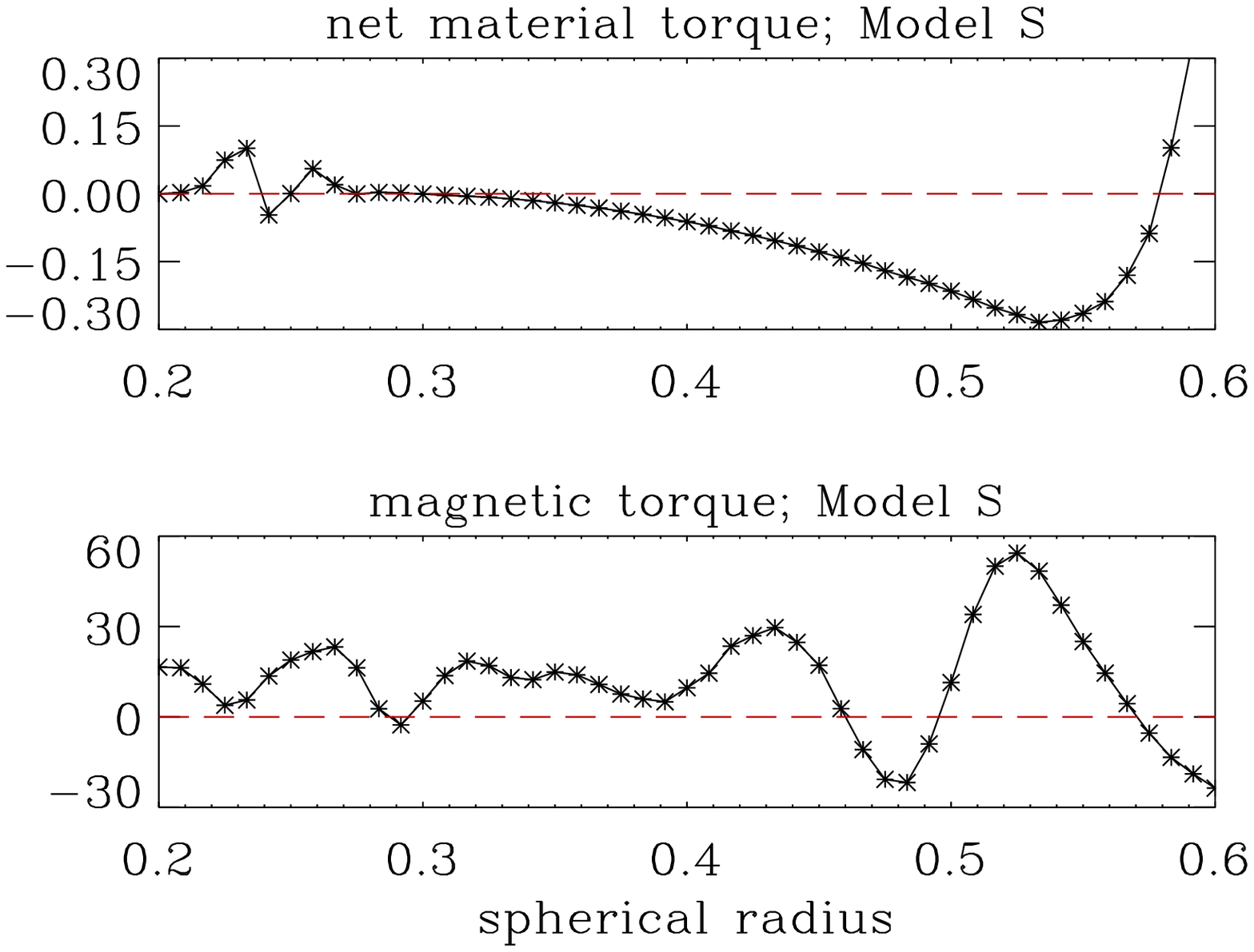}
}
\caption[]{
Latitudinal ({\it left column}) and radial ({\it right column}) profiles of the
torques for Model~S at $t\approx 158$ days.
For the latitudinal profiles we have plotted
the torques (\ref{Ttildeacc}) and (\ref{Ttildemag}) at $r=0.33$,
and for the radial profiles we have plotted the latitudinally integrated
torques (\ref{Tacc}) and (\ref{Tmag}).
}\label{torques-VAR938-strong}
\end{figure*}

At this point we need to mention that a positive (negative) net material torque
can reflect two situations: matter rotating faster (slower) than the star
is flowing toward the star, or matter flowing away from the star is leaving
(taking) rotational energy.

In Fig.~\ref{Theta_torques_R5} we plot latitudinal profiles of the
torques both during the connected state (left hand panel) and during
the disconnected state (right hand panel).
Note that during the connected state, i.e.\ when matter from the disc
falls onto the equatorial regions of the star, there is a significant
positive net material torque in a thin equatorial strip. This means
that accretion leads to a material spin-up of the star at those latitudes.
The stellar wind from higher latitudes can spin down the star or spin it up,
depending on the ability of carrying away angular momentum with it.

The magnetic field around the star
is controlled by both stellar and disc dynamos.
It is complex enough
so that the resulting magnetic torque changes sign in latitude several times
(cf.\ Torkelsson 1998).
At mid and high latitudes, the magnetic torque exerted on the star
is due to the field generated by the stellar dynamo; this stellar field
is not connected to the disc
but can be positive as well as negative.
In the equatorial regions, the disc dynamo-generated field connects
the inner disc regions to the star; here, the magnetic torque exerted
on the star is negative which means that the disc field spins the star down.
However, this magnetic
spin-down torque can barely counteract the material spin-up torque.
This is true also for the latitudinally integrated torques,
showing a total spin-up of the star in the connected state
(of Fig.~\ref{ppltorques_R5}, left hand panel) due to accreting matter,
although the magnetic torque is predominantly negative. (The stellar surface
lies effectively around $0.33$, due to a smoothed profile for the star.)

During the disconnected state, the stellar wind in total carries away
angular momentum which leads to a material spin-down of the star.
In this state, spin-down is supported by a negative magnetic torque;
however, the field is not connected to the disc at all. In Model~SDd,
the overall angular momentum transport by the stellar wind decreases further
away from the star, so that $T_{\rm mat,net}$ becomes positive while
$T_{\rm mag}$ changes sign twice. However, the stronger the stellar dynamo is,
the more efficient is the braking by the stellar wind.

\begin{figure*}[t!]
\centerline{
   \includegraphics[width=8.5cm]{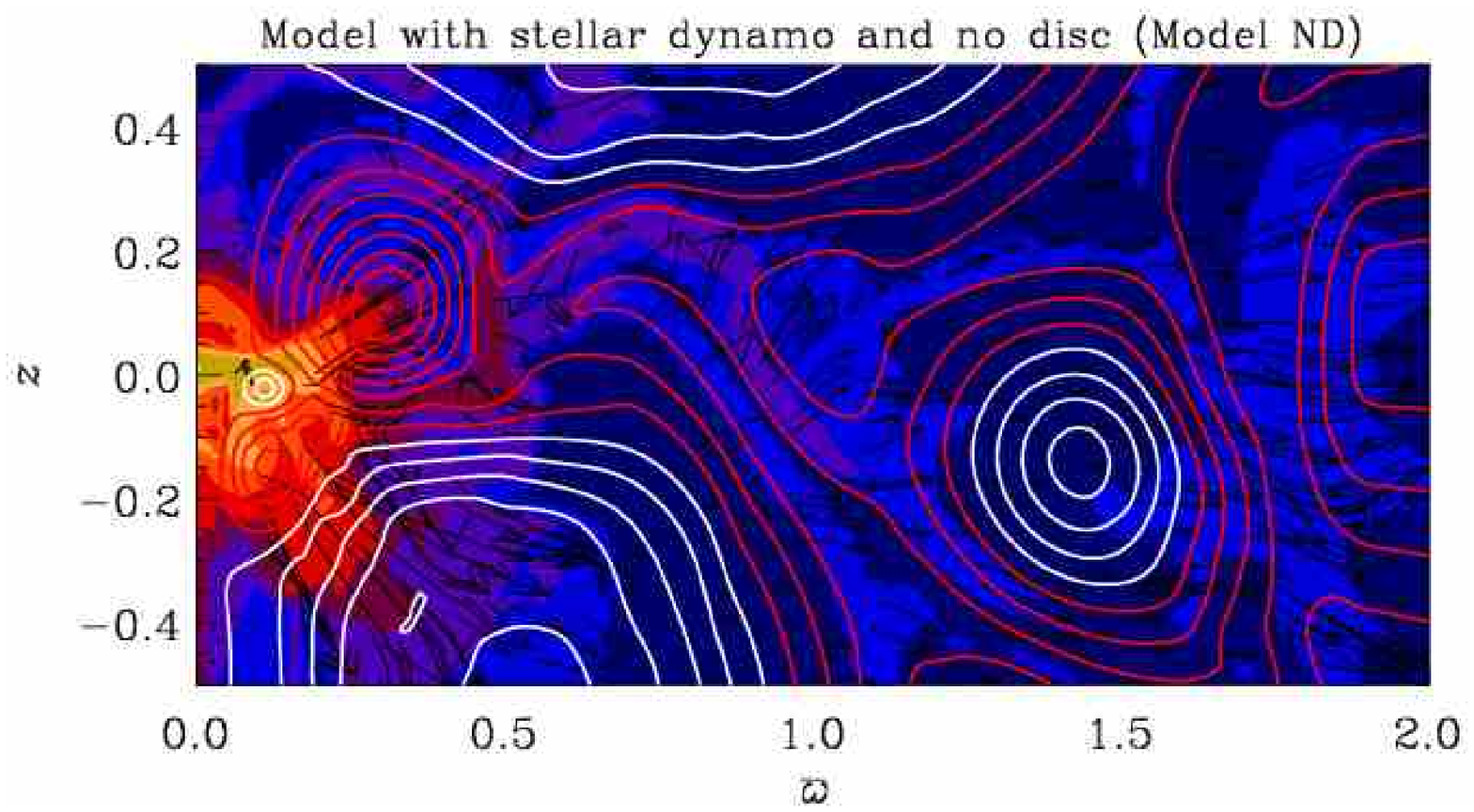}
   \includegraphics[width=8.5cm]{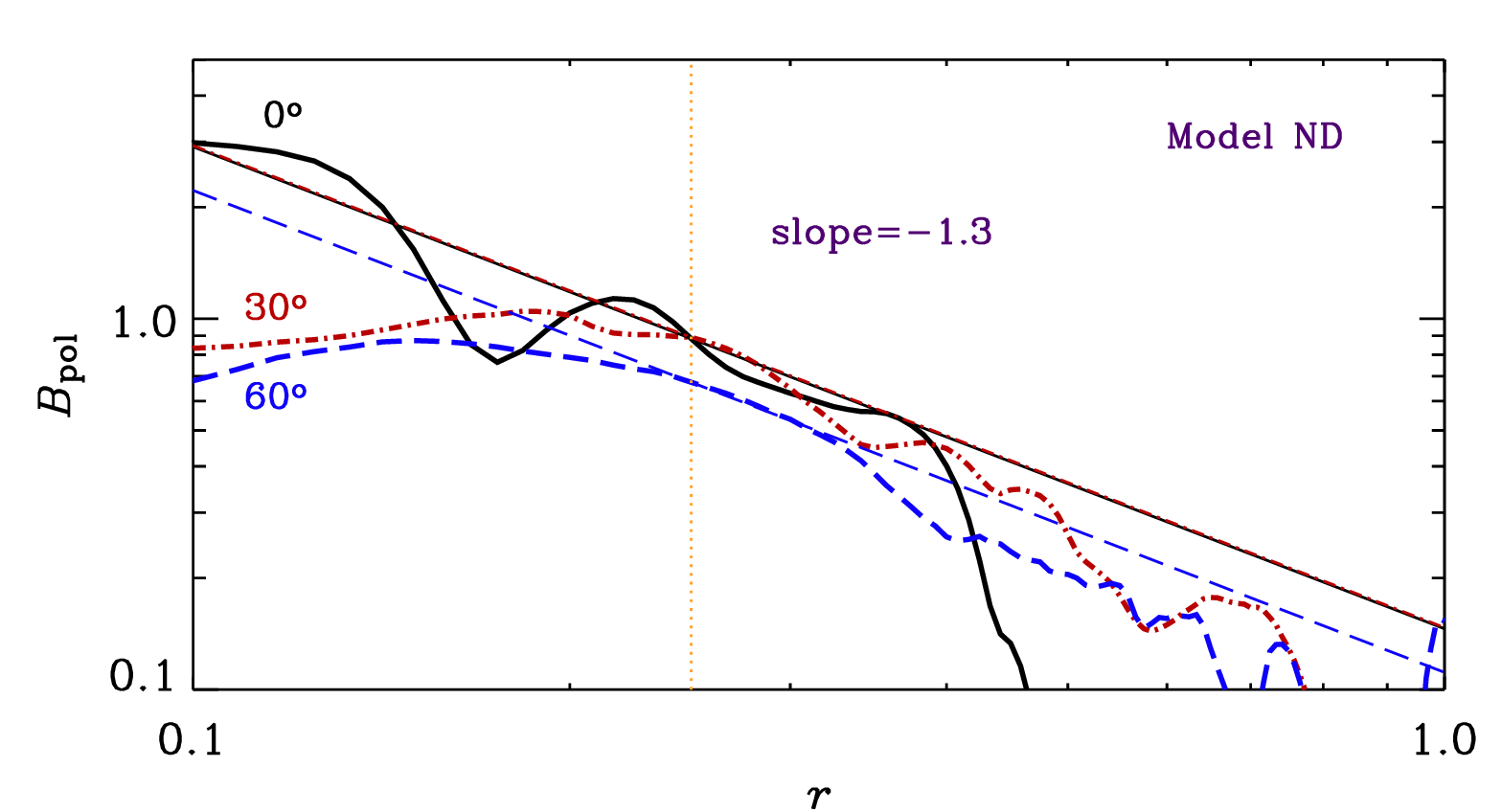}
}
\centerline{
   \includegraphics[width=8.5cm]{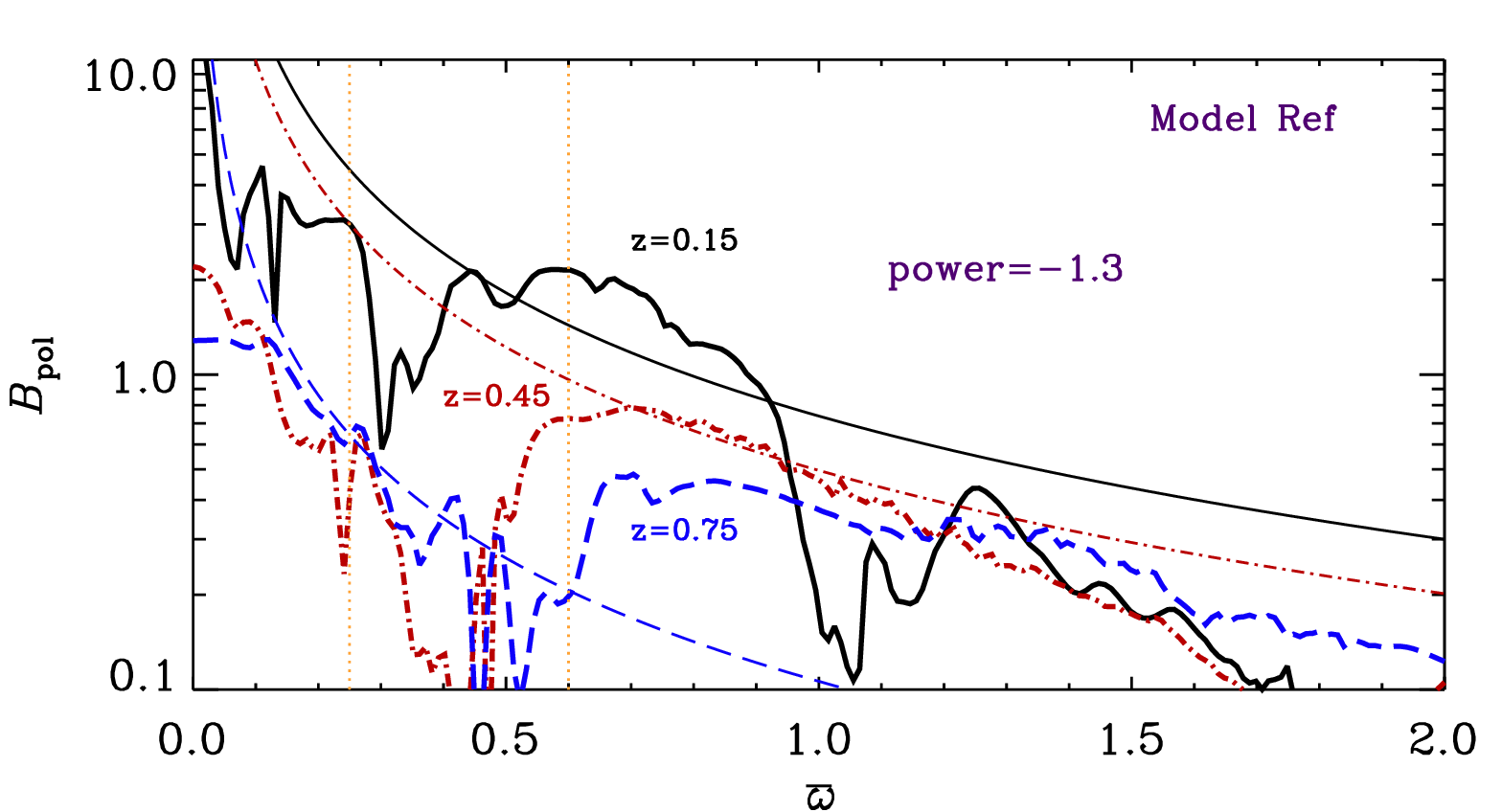}
   \includegraphics[width=8.5cm]{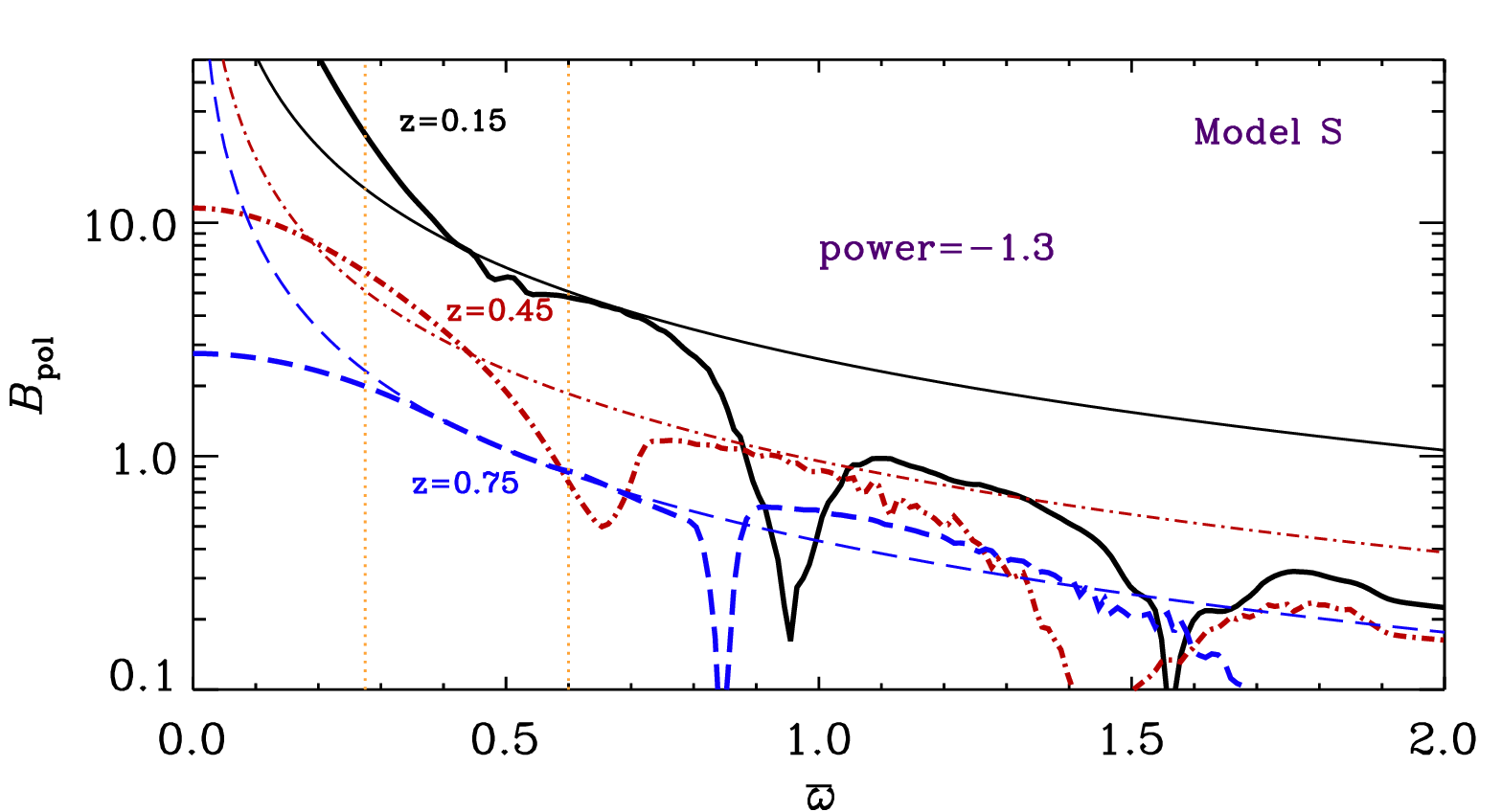}
}
\caption[]{
{\it Top row}: Model~ND at $t \approx 376$ days.
In this model, the computational domain extends to $[-2,+2]$ in $z$.
{\it Top left}:
Black arrows: azimuthally integrated poloidal momentum density vectors
(not shown in the star);
colors/gray shades: gas density.
{\it Top right}:
Poloidal magnetic field strength $B_{\rm pol}$ as a function of spherical
radius $r$
at various latitudes (0, 30 and 60 degrees), compared with fields
decaying as $B_* \, (r/r_*)^{-1.3}$.
{\it Bottom row}:
Poloidal magnetic field strength $B_{\rm pol}$ as a function of cylindrical
radius $\varpi$
at various heights ($z=0.15$, $z=0.45$ and $z=0.75$), compared with fields
decaying as $\varpi^{-1.3}$.
{\it Bottom left}: Model~Ref at $t \approx 1171$ days.
{\it Bottom right}: Model~S at $t \approx 158$ days.
The vertical lines mark the stellar radius ({\it top right and bottom row})
and the inner disc radius ({\it bottom row}).
}\label{p_B-Dip_nodiscbig}
\end{figure*}

In conclusion, stellar spin-up due to accretion cannot always be
compensated by the magnetic torque when the latter is negative.
Stellar spin-down by a stellar wind is equally important to stellar spin-down
by a magnetic field. The total (latitudinally integrated) torques
at the stellar surface are changing sign over
time scales ranging from a few days up to about 30 days, so that the star is
alternately spun up and down.

In Fig.~\ref{torques-VAR938-strong} we show the torques
for the model with the anchored stellar dipole field, Model~S.
Note that here, unlike the stellar dynamo case, the magnetic torque
exceeds the net material torque and the stellar magnetic field leads to
a spin-up of the star.
Fig.~\ref{torques-VAR938-strong} also shows that the fast outer stellar wind
-- emanating from fast-rotating latitudes with a rotation period of about
3~days -- does yield a negative net material torque at these latitudes, causing
a material spin-down of these latitudes. The major part of the stellar surface,
however, is already rotating rather slowly, with a rotation period of about
10~days. Moreover, the magnetic torque is much larger than the net
material torque and in total spins the star up.

\subsection{Collimation of stellar and disc winds}
\label{Collimation}

A formal way of determining whether or not there is collimation was
proposed by Fendt \& {\v C}emelji{\'c} (2002).
Following their proposal, we 
choose a box within each of the two winds and
define the mass load onto the stellar and disc winds as
\begin{equation}
f_{\rm load}^{\rm star} \equiv {\dot{M}_\varpi^{\rm star} \over \dot{M}_z^{\rm star}}
\equiv{\left.
       \int_{z     =0.5 }^{z     =0.7 }2\pi\varpi\varrho u_\varpi\, dz
       \right|_{\varpi=0.7}
\over  \left.
       \int_{\varpi=0.5 }^{\varpi=0.7 }2\pi\varpi\varrho u_z     \, d\varpi
       \right|_{z     =0.7}
      },
\end{equation}
and
\begin{equation}
f_{\rm load}^{\rm disc} \equiv {\dot{M}_\varpi^{\rm disc} \over \dot{M}_z^{\rm disc}}
\equiv{\left.
       \int_{z     =0.6 }^{z     =0.75}2\pi\varpi\varrho u_\varpi\, dz
       \right|_{\varpi=1.2}
\over  \left.
       \int_{\varpi=1.05}^{\varpi=1.2 }2\pi\varpi\varrho u_z     \, d\varpi
       \right|_{z     =0.75}
      }.
\end{equation}
Mass loads less than one indicate collimation while
mass loads greater than one indicate decollimation.
We have calculated these quantities for Models~Ref and S; see also
Figs~\ref{pnew_R3} and \ref{Runs_strong-103_VAR1525-VAR1540-R3}
(bottom left), respectively.
For Model~Ref, the results are slightly different during the
predominantly dipolar phase (Fig.~\ref{pnew_R3}, left hand panel, where
$f_{\rm load}^{\rm star}\approx 0.5$ and
$f_{\rm load}^{\rm disc}\approx 2...3$) and during the
predominantly quadrupolar phase (Fig.~\ref{pnew_R3}, right hand panel, where
$f_{\rm load}^{\rm star}\approx 0.7$ and
$f_{\rm load}^{\rm disc}\approx 6.5$).
For Model~S, we have $f_{\rm load}^{\rm star}\approx 0.6$ and
$f_{\rm load}^{\rm disc}\approx 0.9$
(see Fig.~\ref{Runs_strong-103_VAR1525-VAR1540-R3}, bottom left).
These results suggest that at {\it low} heights,
the stellar wind shows a stronger tendency toward
collimation than the disc wind. Collimation of the disc wind requires a
well-organized and sufficiently strong magnetic field above and below the disc.

\begin{figure*}[t!]
\centerline{
   \includegraphics[width=8.5cm]{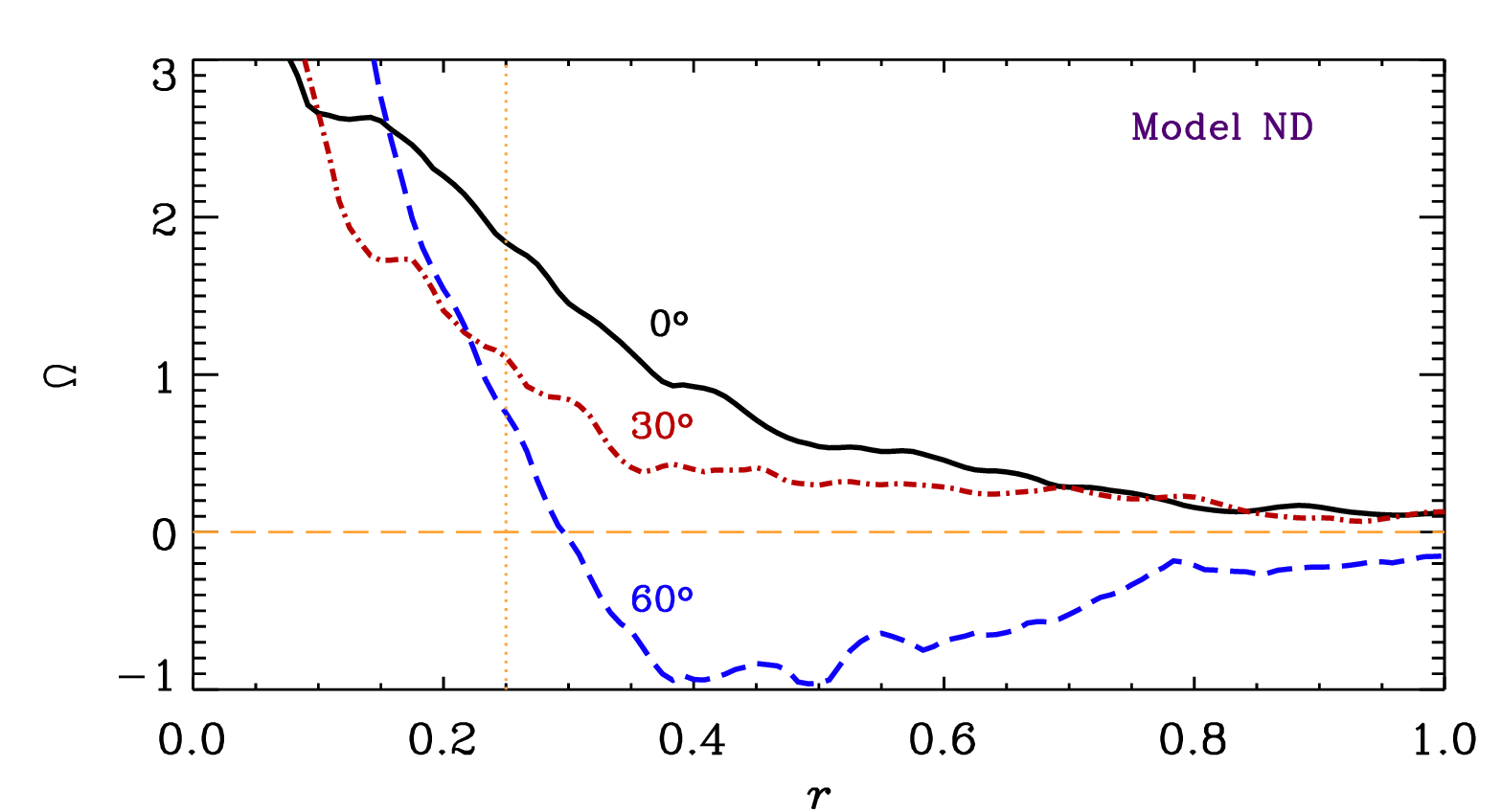}
   \includegraphics[width=8.5cm]{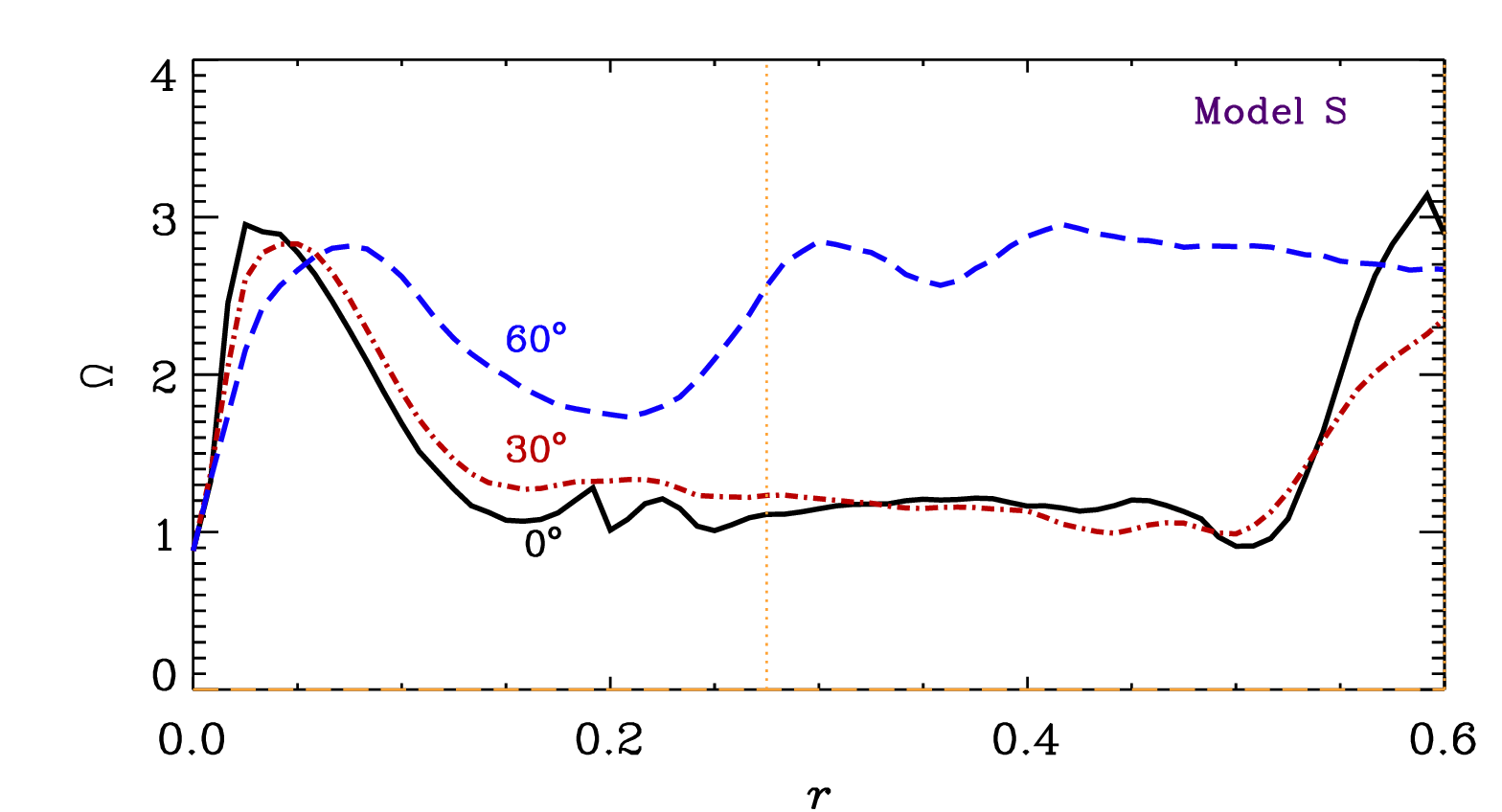}
}
\centerline{
   \includegraphics[width=8.5cm]{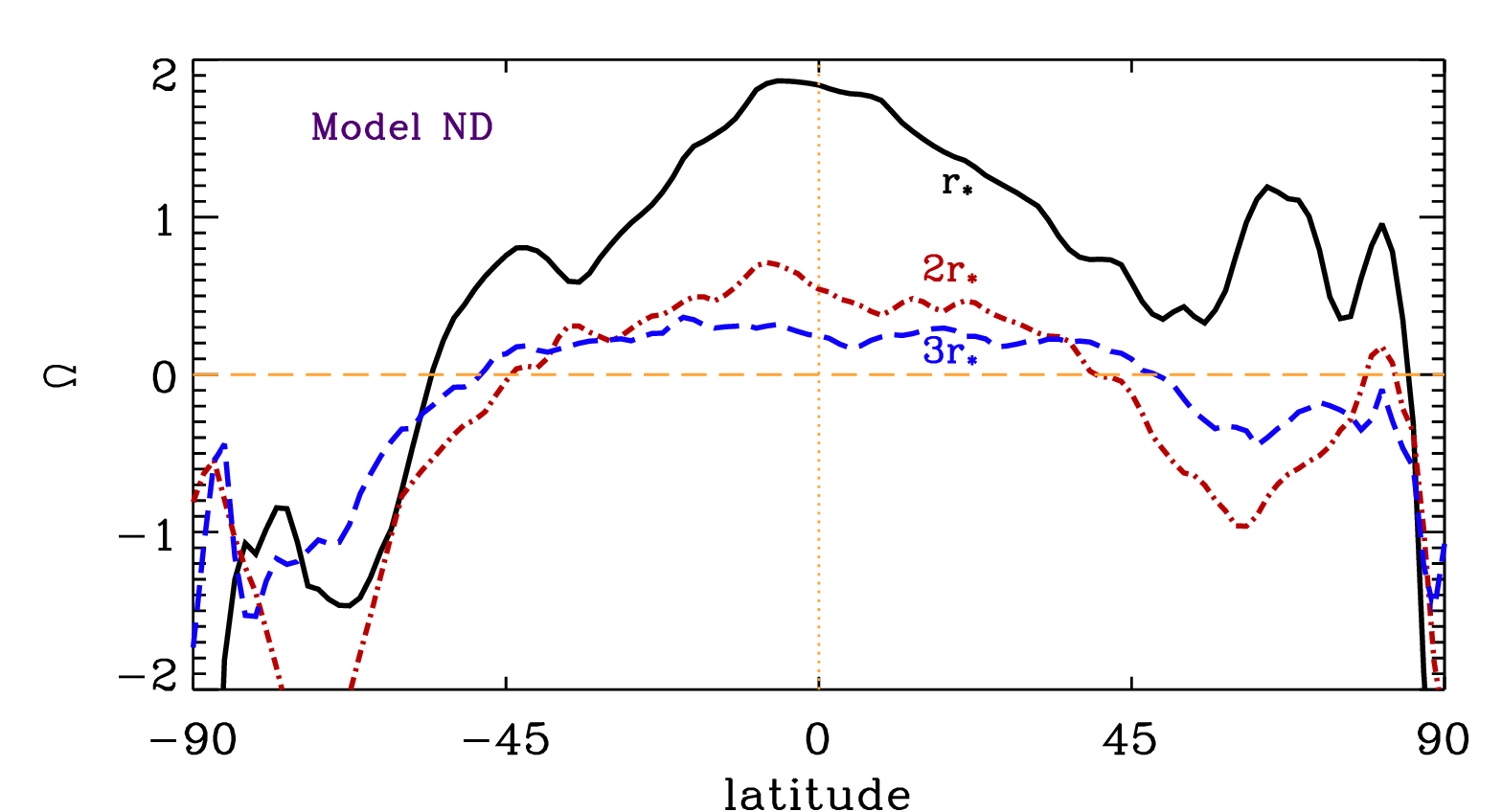}
   \includegraphics[width=8.5cm]{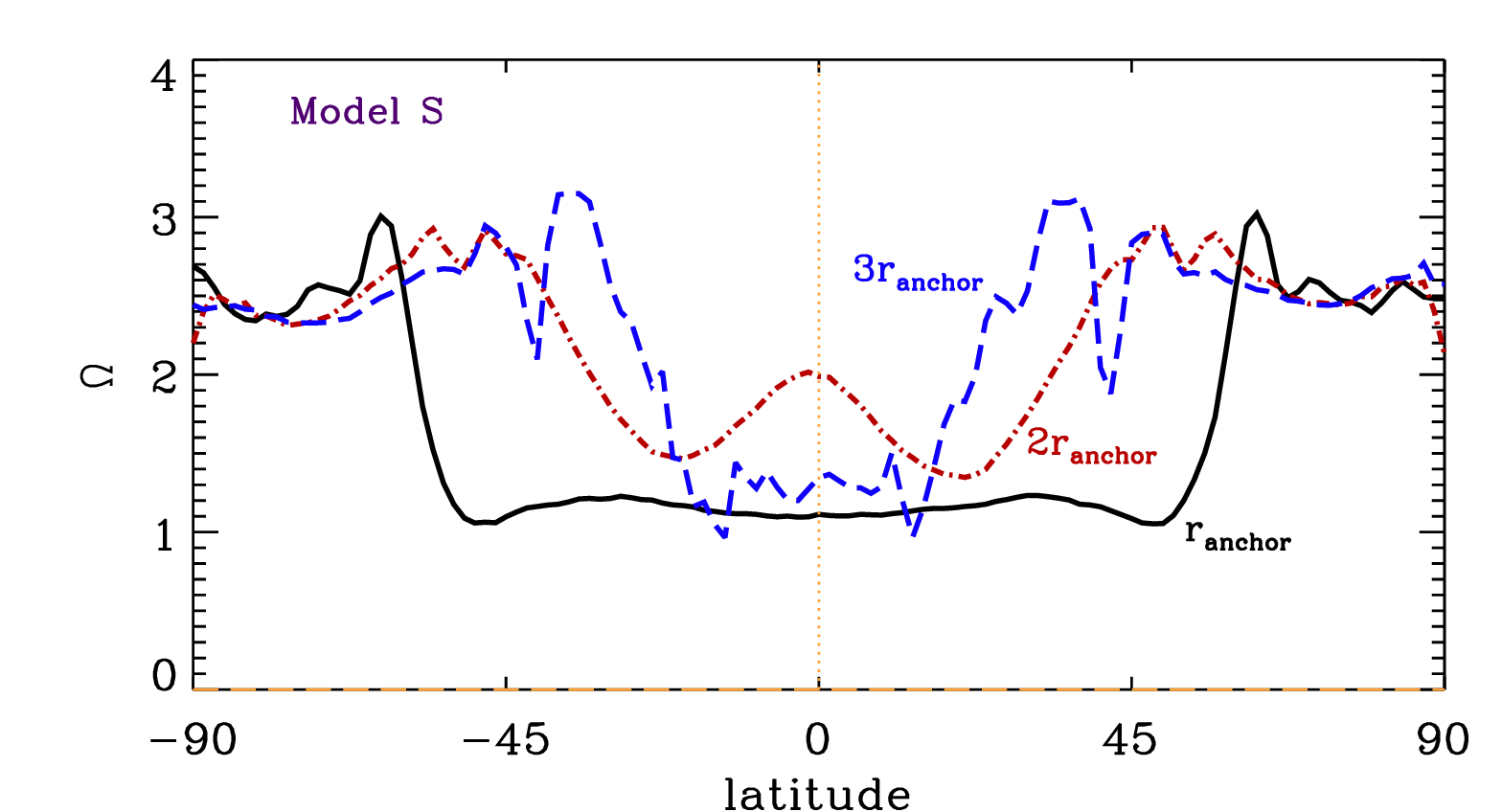}
}
\caption[]{
Note the presence of radial and latitudinal
shear in a model with stellar dynamo and no disc (Model~ND; {\it left column}),
which is compared to Model~S ({\it right column}).
(A behavior of $\Omega$ similar to that of Model~ND is found in Model~Ref.)
{\it Top row}:
Angular velocity as a function of spherical radius at various latitudes
(0, 30 and 60 degrees),
averaged over times 307 to 461 days ({\it left})
and at time $t \approx 158$ days ({\it right}), respectively.
{\it Bottom row}:
Angular velocity as a function of latitude at various spherical radii
(stellar surface, 2 times stellar surface and 3 times stellar surface),
averaged over times 307 to 461 days ({\it left})
and at time $t \approx 158$ days ({\it right}), respectively.
}\label{Om_nodiscbig}
\end{figure*}

The poloidal collimation properties of the stellar and disc winds depend on
how rapidly
the poloidal magnetic field decays with cylindrical radius
(Spruit et al.\ 1997).
The magnetic pressure from the poloidal field outside the jet
is important. Spruit et al.\ (1997) derived the criterion that
$B_{\rm pol} \propto \varpi^{-\mu}$ with $\mu \le 1.3$ is sufficient
for poloidal collimation.

For an ordinary dipole, the (poloidal) field decays with distance like $1/r^3$.
This is also true of a dynamo-generated magnetic field in a vacuum,
but here we are dealing with a dynamo in a conducting medium.
In order to determine the radial decay of a dynamo-generated
stellar field in our simulations, we consider Model~ND with no disc;
see Fig.~\ref{p_B-Dip_nodiscbig}, top row.
The plot in the upper right panel shows the (spherically)
radial decay of the poloidal
field at different latitudes.
There is no clean power law behavior but it is clear that the decay
is slower than $1/r^3$.
In fact, the decay of the poloidal field is similar to $1/r^{1.3}$ just outside
the star.

The tendency to poloidal collimation of the inner stellar wind in a stellar
dynamo model can be seen in Fig.~\ref{p_B-Dip_nodiscbig}, bottom left,
where we have plotted the poloidal magnetic field as a function of
cylindrical radius at the disc surface and in the corona.
The inner disc wind shows a tendency to poloidal collimation over a broader
range in $\varpi$ only at larger $z$.

In Model~S, the decay of the magnetosphere is $B_{\rm pol}(r) \propto 1/r^3$
except for the equatorial region where it is compressed by the disc field and
therefore decaying more slowly. The decay with cylindrical radius is
$B_{\rm pol}(\varpi) \propto 1/\varpi^{1.3}$ in both magneto-centrifugally
accelerated winds; see Fig.~\ref{p_B-Dip_nodiscbig}, bottom right.

A general feature of our models is that
our dipolar disc dynamo-generated field advected into the corona satisfies
the criterion for poloidal collimation
of Spruit et al.\ (1997) in a certain range over $\varpi$ and $z$.
Our models cannot be compared directly with M.\ v. Rekowski et al.\ (2000), whose
disc dynamo with dipolar polarity satisfies the condition everywhere at the disc
surface, because in their model there are no reversals in $B_z$ in the disc.

Figure~\ref{Om_nodiscbig}, left column, shows that in a model with stellar dynamo,
mid-latitudes rotate more slowly than the equator around the stellar surface.
There is latitudinal as well as negative radial rotational shear also outside
the star, leading
to a magnetic field that is decaying more slowly than a dipole with distance
from the star (see Fig.~\ref{p_B-Dip_nodiscbig}, top right). A similar behavior
in the angular velocity is found in Model~Ref (a model with stellar dynamo,
disc and disc dynamo).
In a model with anchored stellar magnetosphere (Fig.~\ref{Om_nodiscbig},
right column), the radial shear is about 0
around $r_{\rm anchor}$ at low latitudes but there is positive radial shear around $r_{\rm anchor}$
at higher latitudes where the fast magneto-centrifugally accelerated outer stellar
wind is launched. Around $r=r_{\rm anchor}$, these are also the latitudes that rotate
fastest.
This shows again that fast-rotating latitudes are more likely to produce
a fast wind.

\section{Conclusions}

The present work has shown that a dynamo-generated stellar field, instead
of just an anchored stellar dipole field, can render the interaction
between protostar and its surrounding disc in several ways more realistic.
First, the observed structure of protostellar fields is known to be more
complex than just a pure dipole.
Second, dynamo-generated stellar fields can contribute quite efficiently
to launching an overall fast stellar wind (up to $450\kms$ terminal velocity,
which exceeds the terminal disc wind speed that itself is about twice the
Keplerian speed of the inner disc edge).
Such behavior was never found with just an imposed stellar dipole field.
A stellar wind is equally important as the magnetic field
for spinning down the star -- especially
because the braking via the magnetic
interaction with the surrounding disc is now
found to be much less efficient than previously thought.
We also suggest that this stellar wind could contribute to the observed jets,
together with the wind from the inner part of the surrounding accretion disc.

As before in the case of an imposed central dipole field, the interaction
between star and disc tends to occur in a time-dependent fashion where
the system alternates between phases where star and disc are magnetically
connected and mass can flow onto the star, and when they are magnetically
disconnected and mass transfer is blocked.
Unsteady star--disc interaction is now becoming a prominent feature
of magnetospheric accretion onto T Tauri stars seen both in observations 
(Bouvier et al.\ 2003, 2004) as well as in
numerical simulations (Goodson et al.\ 1997, 1999,
Goodson \& Winglee 1999, Matt et al.\ 2002, vRB04).
At the same time, inefficient star--disc coupling is beginning
to be a well recognized feature when stars couple to the differentially
rotating disc (Matt \& Pudritz 2004, 2005).

However, time variability is no longer periodic but more irregular and
on small scales, which is due to the more complex magnetic field configuration.
The episodic magnetic star--disc coupling is no longer facilitated by the stellar
magnetosphere but by the disc dynamo-generated magnetic field so that
the episodic accretion flow is not a slow polar funnel flow but happens
at low latitudes along a very weak advected disc field, where it is much faster.
Stellar braking is not only due to a magnetic field but also
due to a fast overall stellar wind, which develops
in the presence of a stellar dynamo and is driven by gas and magnetic pressure,
with magneto-centrifugal acceleration at times and in places where the stellar
field geometry is appropriate.
This suggestion is supported by recent observational evidence for high
speed stellar winds from T Tauri stars (Dupree et al.\ 2005).

Extension to nonaxisymmetry has been made for some dynamo models,
and the results will be reported in a forthcoming paper.
Possible steps to verify the dynamo models further could be
to relax the constraints imposed by
the adoption of mean-field theory and by
the restriction to a piecewise polytropic model.
Another serious restriction that should be relaxed in future models is the
assumption of time-independent profiles for the shapes of star and disc.
Ideally, such profiles should be calculated in a self-consistent manner.
In this paper, we have focused on the wind launching mechanisms and
accretion processes. Studying the shaping and collimation of the winds
further away from the star--disc system, requires a bigger computational domain.

\begin{acknowledgements}
We thank Eric Blackman for useful suggestions.
This work was supported in part by a Sch\"onberg Fellowship (B.v.R.)
in Uppsala (Sweden).
B.v.R.\ thanks NORDITA for hospitality.
Use of the supercomputer SGI 3800 in Link\"oping and of the PPARC supported
supercomputers in St Andrews and Leicester is acknowledged. This research was
conducted using the resources of High Performance Computing Center North
(HPC2N).
The Danish Center for Scientific Computing is acknowledged
for granting time on the Linux cluster in Odense (Horseshoe).
\end{acknowledgements}

\appendix

\section{The piecewise polytropic model}
\label{Sec-cool-hot}

As described in our earlier papers,
we model a cool, dense disc embedded in a hot, rarefied corona by prescribing
contrasts of the specific entropy $s$ such that $s$ is smaller within the disc
and larger in the corona. For a perfect gas this implies
$p=K\varrho^\gamma$ (in dimensionless form), with the polytrope
parameter $K\equiv e^{s/c_v}$ being a function of position.
The same idea is also applied to the star.
In the models considered here,
we have an intermediate value for the specific entropy within the star,
resulting in a relatively cool and dense star.
We prescribe the polytrope parameter to be unity in the corona and
less than unity in the disc and in the star, so we put
\begin{equation}
e^{s/c_p}
= 1-(1{-}\beta_{\rm disc})\xi_{\rm disc}-(1{-}\beta_{\rm star})\xi_{\rm star},
\label{Kdef}
\end{equation}
where $\xi_{\rm disc}$ and $\xi_{\rm star}$
are explained in Sect.~2 of von Rekowski et al.\ (2003).
The free parameters, $0<\beta_{\rm disc},\beta_{\rm star}<1$, control the
specific entropy contrasts between the disc and corona, and between the
star and corona, respectively. With prescription (\ref{Kdef}),
$\beta_{\rm disc}=1$ means that the disc is absent.

In the absence of the stellar and disc dynamos,
our initial state is a hydrostatic equilibrium with no poloidal velocity, where
we assume an initially non-rotating, hot and rarefied corona that is supported
by the pressure gradient.
We solve the vertical and radial balance equations, as described in
von Rekowski et al.\ (2003).
Since we model a disc that is cool and dense, the disc is mainly centrifugally
supported, and as a result it is rotating slightly sub-Keplerian.
The resulting stellar rotation is differential, and the initial stellar surface
rotation periods depend on the chosen stellar radius.

The temperature ratio between disc and corona is roughly $\beta_{\rm disc}$.
Assuming pressure equilibrium between disc and corona,
and $p\propto\rho T$ for a perfect gas, the corresponding density ratio is then
$\beta_{\rm disc}^{-1}$.
Thus, the entropy contrast between disc and corona chosen here
($\beta_{\rm disc} = 0.005$), leads to
density and inverse temperature ratios of $200:1$ between disc and corona.

A rough estimate for the initial toroidal velocity, $u_{\varphi0}$,
in the midplane of the disc
follows from the hydrostatic equilibrium as
$u_{\varphi0}\approx\sqrt{1-\beta_{\rm disc}}\vec{\uu}_{\rm K}$.
For $\beta_{\rm disc}=0.005$, the toroidal velocity is within 0.25\%
of the Keplerian speed.

\section{Physical units of our variables} \label{NonDim}

The resulting velocity unit is $[\vec{\uu}]=c_{\rm s0} = 10^2\km\s^{-1}$,
the length unit is $[\vec{\rr}]=GM_*/[\vec{\uu}]^{2}\approx0.1\AU$,
the time unit is $[t]=[\vec{\rr}]/[\vec{\uu}]\approx1.536~{\rm days}$,
the unit for the angular velocity is $[\Omega]=[\vec{\uu}]/[\vec{\rr}]
\approx 7.5\times10^{-6}\s^{-1}$,
the unit for the rotation period is $[P_{\rm rot}] \approx 10~{\rm days}$,
the unit for the kinematic viscosity and magnetic diffusivity
is $[\nu]=[\eta]=[\vec{\uu}][\vec{\rr}]\approx 1.5\times 10^{19}\cm^{2}\s^{-1}$,
the unit for specific entropy is $[s]=c_p=\gamma/(\gamma-1){\cal R}/\mu
\approx 3.5\times10^8\cm^2\s^{-2}\K^{-1}$,
the unit for specific enthalpy is $[h] = [\vec{\uu}]^{2} = 10^4\km^2\s^{-2}$,
the temperature unit is $[T] = [h]/[s] \approx 3\times10^5\K$,
the density unit is
$[\varrho]=\Sigma_0/[\vec{\rr}]\approx7.5\times10^{-13}\g\cm^{-3}$,
the mass unit is
$[M] = \Sigma_0 [\vec{\rr}]^{2} \approx 2 \times 10^{24} \g
\approx 10^{-9} \,M_\odot$,
the pressure unit is
$[p]=(\gamma-1)/\gamma[\varrho][h]\approx 30\g\cm^{-1}\s^{-2}$,
the magnetic field unit is
$[\vec{\BB}]=[\vec{\uu}](4\pi[\varrho])^{1/2}\approx30\G$
(it is $\mu_0=4\pi$),
and the unit for the magnetic vector potential is
$[\vec{\AAA}]=[\vec{\BB}] [\vec{\rr}] \approx 4\times10^{13}\G\cm$.
Further, the unit for the mass accretion rate and mass loss rates due to the
stellar and disc winds is
$[\dot{M}]=\Sigma_0[\vec{\uu}][\vec{\rr}]\approx 2\times10^{-7}M_\odot\yr^{-1}$,
the unit for the torques is $[{\cal T}] = GM_* \Sigma_0 [\vec{\uu}] [t]
\approx 2 \times 10^{38}\erg$,
and the unit for the luminosity is $[{\cal L}] = GM_* \Sigma_0 [\vec{\uu}]$
$= [{\cal T}]/[t]
\approx 1.5 \times 10^{33}\erg \s^{-1}$. [From this follows that even for a very
low mass accretion rate of $\dot{M}_{\rm accr} = 10^{-10} M_\odot\yr^{-1}$ and a
protostellar radius of $r_* = 5 R_\odot$, one finds an accretion luminosity
that is as high as
${\cal L}_{\rm accr} = GM_* \dot{M}_{\rm accr} / r_* \approx 2.5\times 10^{30}
\erg \s^{-1}$.]

\newcommand{\yapj}[3]{ #1, {ApJ,} {#2}, #3}
\newcommand{\yass}[3]{ #1, {Ap\&SS,} {#2}, #3}
\newcommand{\ypasj}[3]{ #1, {Publ. Astron. Soc. Japan,} {#2}, #3}
\newcommand{\yana}[3]{ #1, {A\&A,} {#2}, #3}
\newcommand{\ymn}[3]{ #1, {MNRAS,} {#2}, #3}
\newcommand{\ybook}[3]{ #1, {#2} (#3)}
\newcommand{\ynat}[3]{ #1, {Nat,} {#2}, #3}
\newcommand{\ypr}[3]{ #1, {Phys. Rev.} {#2}, #3}
\newcommand{\yjgr}[3]{ #1, {JGR,} {#2}, #3}
\newcommand{\ysph}[3]{ #1, {Sol. Phys.,} {#2}, #3}
\newcommand{\yproc}[5]{ #1, in {#3}, ed. #4 (#5), p.~#2}

\vfill\bigskip\noindent\tiny\it{
$ $Id: paper.tex,v 1.120 2005/09/10 12:23:03 gitta Exp $ $}
\end{document}